\documentclass[reqno]{amsart}

\usepackage{booktabs}
\usepackage{IEEEtrantools}
\usepackage{amssymb,latexsym,amsfonts,amsmath}
\usepackage{graphicx}
\usepackage{mathrsfs}
\usepackage{dsfont}

\usepackage {url}
\usepackage[implicit=false]{hyperref}

\usepackage{tikz}
\usetikzlibrary{calc,shapes,arrows}

\usepackage{algorithm}
\usepackage{algorithmic}

\usepackage{xspace}

\usepackage{enumerate}   
\usepackage{mdframed,lipsum}
\usepackage[many]{tcolorbox}
\usetikzlibrary{calc}
\tcbuselibrary{skins}

\newtcolorbox{resp}[1][]{%
	enhanced jigsaw,%
	colback=gray!5!white,%
	colframe=gray!80!black,%
	size=small,%
	boxrule=1pt,%
	halign title=flush center,%
	coltitle=black,%
	breakable,%
	drop shadow=black!50!white,%
	attach boxed title to top left={xshift=1cm,yshift=-\tcboxedtitleheight/2,yshifttext=-\tcboxedtitleheight/2},%
	minipage boxed title=3cm,%
	boxed title style={%
		colback=white,%
		size=fbox,%
		boxrule=1pt,%
		boxsep=2pt,%
		underlay={%
			\coordinate (dotA) at ($(interior.west) + (-0.5pt,0)$);
			\coordinate (dotB) at ($(interior.east) + (0.5pt,0)$);
			\begin{scope}[gray!80!black]
				\fill (dotA) circle (2pt);
				\fill (dotB) circle (2pt);
			\end{scope}
		}%
	},%
	#1%
}

\newtheorem{theorem}{Theorem}[section]
\newtheorem{lemma}[theorem]{Lemma}

\newtheorem{corollary}[theorem]{Corollary}

\newtheorem{definition}[theorem]{Definition}
\newtheorem{example}[theorem]{Example}

\newtheorem{remark}[theorem]{Remark}
\newtheorem{assumption}{Assumption}
\numberwithin{equation}{section}
\newtheorem{open}{Open Problem}

\newcommand{\R}{{\mathbb{R}}}

\newcommand{\N}{{\mathbb{N}}}

\newcommand{\EE}{\mathds{E}}
\newcommand{\PP}{\mathds{P}}

\newcommand{\AP}{AP} 
\newcommand{\alphabeth}{\Sigma_{\textsf{a}}}
\newcommand{\word}{\omega}
\newcommand{\Lab}{\mathsf{L}}

\newcommand{\until}{\mathbin{\sf U}}

\newcommand{\nex}{\mathord{\bigcirc}}
\newcommand{\trans}{\mathsf{t}}

\begin{document}
	
\begin{abstract}
Stochastic hybrid systems have received significant attentions as a relevant modelling framework describing many systems, from engineering to the life sciences:  
they enable the study of numerous applications, 
including transportation networks, biological systems and chemical reaction networks, smart energy and power grids, and beyond. 
Automated verification and policy synthesis for stochastic hybrid systems
can be inherently challenging: this is due to the heterogeneity of their dynamics (presence of continuous and discrete components), the presence of uncertainty, 
and in some applications the large dimension of state and input sets. 
Over the past few years, a few hundred articles have investigated these models, 
and developed diverse and powerful approaches to mitigate difficulties encountered in the analysis and synthesis of such complex stochastic systems. 
In this survey, we overview the most recent results in the literature and discuss different approaches, including  
\emph{(in)finite abstractions}, 
\emph{verification and synthesis for temporal logic specifications}, 
\emph{stochastic similarity relations},  
\emph{(control) barrier certificates}, 
\emph{compositional techniques}, and a selection of results on 
\emph{continuous-time stochastic systems};  
we finally survey recently developed \emph{software tools} that implement the discussed approaches. 
Throughout the manuscript we discuss a few open topics to be considered as potential future research directions:
we hope that this survey will guide younger researchers through a comprehensive understanding of the various challenges, tools, and solutions in this enticing and rich scientific area. 
\end{abstract}

\title[Automated Verification and Synthesis of Stochastic Hybrid Systems: A Survey]{Automated Verification and Synthesis of Stochastic Hybrid Systems: A Survey}

\author{Abolfazl Lavaei$^{1}$}
\author{Sadegh Soudjani$^{2}$}
\author{Alessandro Abate$^{3}$}
\author{Majid Zamani$^{4,5}$}
\address{$^1$Institute for Dynamic Systems and Control, ETH Zurich, Switzerland}
\email{alavaei@ethz.ch}
\address{$^2$School of Computing, Newcastle University, United Kingdom}
\email{Sadegh.Soudjani@newcastle.ac.uk}
\address{$^3$Department of Computer Science, University of Oxford, United Kingdom}
\email{aabate@cs.ox.ac.uk}
\address{$^4$Department of Computer Science, University of Colorado Boulder, USA}
\address{$^5$Department of Computer Science, LMU Munich, Germany}
\email{majid.zamani@colorado.edu}
\maketitle

\section{Introduction}

Stochastic hybrid systems (SHS) 
concern complex dynamical models combining both digital-computation elements and physical components, tightly interacting with each other in feedback interconnections. 
SHS models thus comprise discrete dynamics modelling computational components including hardware and software, 
and continuous dynamics that model the physical system.  
Due to their broad real-world applications, such as
(air) traffic networks, transportation systems, energy networks, process engineering, biological systems, and robotic manufacturing, (cf.~\cite{hu2004modeling,hespanha2005stochastic,prandini2008application,singh2010stochastic,vargas2018elucidating}, to name a few), 
over the past few years SHS have gained remarkable attention in the areas of control theory, formal verification, applied mathematics, and performance evaluation, among others.   
SHS applications have become more complex, with more digital components (e.g., for computation and communication) that interact with physical (analog) parts: 
this tight interaction causes major difficulties in designing and analyzing these complex systems. 
Accordingly, the ability to handle the interaction between continuous and discrete dynamics is a prerequisite for providing a rigorous formal framework for formal verification and synthesis of SHS.   

Grown first within the area of hybrid systems and of stochastic control, 
SHS have been first comprehensively presented and widely discussed in the books in \cite{blom2006stochastic} and in \cite{cassandras2018stochastic}.  
The historical roots in hybrid systems that underpin SHS research have brought to an inter-disciplinary look at these models, 
with emphasis split between classical dynamical analysis and control synthesis on the one hand, 
as well as computability and formal verification around rich, high-level specifications on the other.
Automated verification and policy synthesis for SHS around high-level temporal requirements, \emph{e.g.,} those expressed as (linear) temporal logic formulae~\cite{pnueli1977temporal}, is the core emphasis of this survey.
Given a temporal property of interest for a dynamical model, formal verification is concerned to soundly check whether the desired specification is satisfied. 
If the underlying model is stochastic, 
the goal translates in formally quantifying the probability of satisfying the property of interest. 
A synthesis problem instead concerns dynamical models with the presence of control inputs: 
the goal is to formally design a controller (also known in different areas as policy, or strategy, or scheduler), 
which is by and large a state-feedback architecture, to enforce the property of interest. 
This procedure is also called ``correct-by-construction control design", 
since every step in the controller synthesis procedure comes with a formal guarantee.  
In a stochastic setting, the key objective is to synthesize a controller that optimizes (e.g., maximizes) the probability of satisfying the given specification. 
As a result of their intrinsic soundness, 
formal methods approaches do not require any costly, exhaustive, and possibly unsuccessful post-facto validation, 
which is needed in many safety-critical, real-world applications.

The intrinsic complexity of SHS models, 
resulting from the aforementioned interaction of discrete and continuous components, as well as from the presence of uncertainty that is modelled via probability terms, makes it in general difficult - if at all possible - to obtain analytical results in their formal verification or for control synthesis tasks. Hence, verification and policy synthesis for SHS are generally addressed by techniques that either leverage model (finite) approximations, or the use of sufficient conditions for analysis. Accordingly, we categorize these two classes of approaches as (i) discretization-based and (ii) discretization-free techniques. 

\subsection{Discretization-based Techniques}\label{subsec:DBT}

\begin{figure}[ht]
	\begin{center}
		\includegraphics[width=0.5\linewidth]{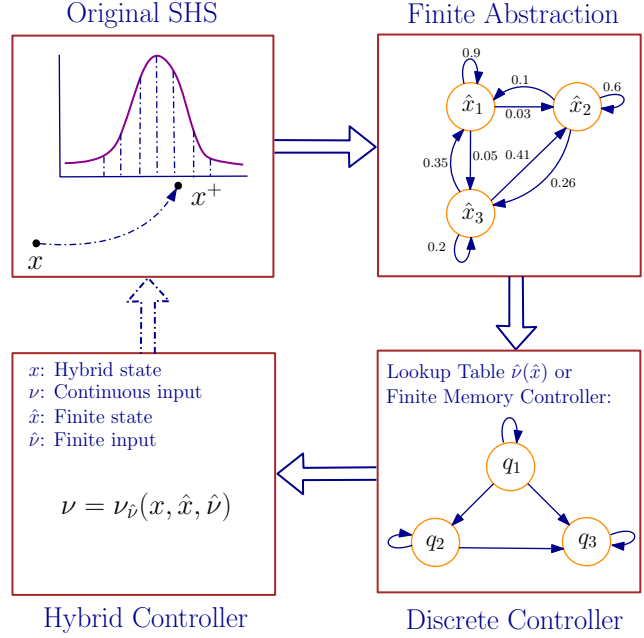} 
		\caption{Illustration of the procedure underlying discretization-based techniques based on finite abstractions. 
			The discrete controller can be a static lookup table or a dynamic controller (with finite memory).} 
		\label{Fig1}
	\end{center}
\end{figure}

In the analysis of SHS, it is often the case that quantities of interest, such as value functions, 
or the characterization of optimal policies, 
are in general not available in a closed (explicit, analytical) form. 
Therefore, a suitable approach for analyzing SHS is to approximate given (``concrete'') SHS models by simpler ones endowed with finite state spaces, also known as ``finite abstractions''.\footnote{An alternative to the described approach is to generate simpler models that are still uncountable, as approximations of concrete SHS. As examples, these could be obtained by linearizing the (continuous) dynamics, by disregarding additive noise terms, or by reducing the dimension of the concrete models. We place less emphasis on this alternative set of approaches, since they are not in general automated, requiring instead manual solutions which are not generally applicable to SHS (as much as discretization-based techniques are), and since they seldom come with the guarantees that are instead typical of discretization-based techniques.} 
Finite abstractions of SHS are often in the form of Markov decision processes (MDP), 
where each discrete state corresponds to a set of continuous states of the concrete SHS model (similarly for inputs). 
In practice, such finite abstractions can be generated by partitioning state/input sets of the concrete models given some discretization parameters. If the underlying SHS is autonomous, \emph{i.e.,} without control input, the finite MDP is then reduced to a finite Markov chain (MC).
The discrete dynamics of the finite abstractions are similarly obtained from those of the concrete continuous models (cf. Fig.~\ref{Fig1}).  
Since the obtained abstractions are finite, 
many algorithmic machineries from computer science~\cite{baier2008principles} are directly applicable to perform analysis, model checking, 
or to synthesize controllers maximizing rewards or enforcing complex properties, including those expressed as temporal logic formulae (to be discussed later). 
A crucial step related to these discretization-based techniques is to provide formal guarantees on the obtained abstractions, 
so that the verification or synthesis results on abstract models can be formally carried over to the original SHS: 
this is a key feature that characterizes the overall approach.
Discretization-based techniques using finite abstractions are schematically illustrated in Fig.~\ref{Fig1}. As it can be observed, the original SHS is first approximated by a finite abstraction with discrete state and input sets. Then a discrete controller, in the form of a static lookup table or a dynamic controller (with finite memory), is synthesized over the constructed finite abstraction. Finally, the discrete controller is refined back over the original SHS via a \emph{hybrid} interface map that contains states of both original and abstract systems, and the discrete input.

\begin{remark} 
	We remark a fundamental difference between the discussed discretization-based techniques for abstractions, 
	which are focused on formally simplifying (SHS) models into abstract models that are amenable to be verified or subject to synthesis tasks; 
	and standard approaches in literature that resort to (e.g., spatial) discretization to provide numerical implementations of algorithms for analysis or synthesis.  
	The latter approaches deal with numerical solutions for quantities of interest, such as value functions or optimal policies.  
	Beyond this fundamental difference, note that the latter approaches often do not come with correctness guarantees.
\end{remark} 

\subsection{Discretization-free Techniques} 

\begin{figure}[h!]
	\begin{center}
		\includegraphics[width=0.5\linewidth]{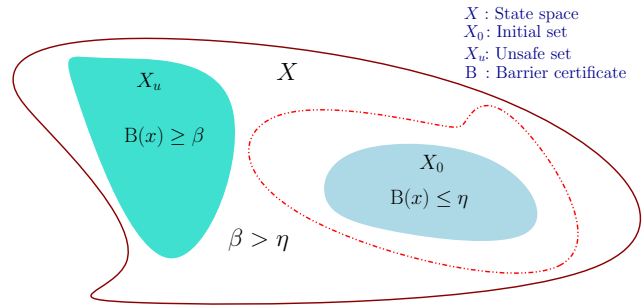} 
		\caption{Discretization-free techniques can study probabilistic safety based on the construction of control barrier certificates. The (red) dashed line denotes the level set $ \text{B}(x) = \eta$.} 
		\label{Barrier}
	\end{center}
\end{figure}

The techniques discussed in the setting of finite abstractions rely on the discretization (that is, partitioning, or gridding) of state and input/action sets; 
consequently, they can suffer from an issue known as the \emph{curse of dimensionality}: 
the complexity of constructing the abstraction grows exponentially with the state/input dimension of the SHS. 
This critical challenge motivates the development of discretization-free approaches, 
such as those based on \emph{(control) barrier certificates}.
These approaches, 
which have been recently introduced (over the last $15$ years) for verification and/or controller synthesis of complex SHS, 
should again provide ``sufficient results'' for the analysis and/or synthesis over the given SHS models.
Barrier certificates are Lyapunov-like functions defined over the state space of the system and satisfying a set of inequalities on both the function itself and the one-step transition (or the infinitesimal generator along the flow) of the system. 
An appropriate level set of a barrier certificate can separate an unsafe region from all system trajectories starting from a given set of initial conditions (cf. Fig.~\ref{Barrier}) with some probability lower bound. Consequently, the existence of such a function provides a formal probabilistic certificate for system safety. Notice that although barrier certificates are natively employed to ensure the safety of a SHS model, they have also been recently used in the literature to enforce alternative properties, such as temporal requirements (cf. Section~\ref{CBC}).

It is worth mentioning that there exist other discretization-free techniques for analysis and controller synthesis in relevant literature, which are mainly based on optimization approaches, such as model predictive control (cf. Subsection~\ref{Optimization}). For instance, stochastic model predictive control (SMPC)~\cite{mesbah2016stochastic} is a widely investigated setup which, however, is not aimed at providing the formal guarantees on verification and controller synthesis of complex SHS on which this survey focuses. Model-order reductions are alternative types of discretization-free techniques, which originate from control literature in the frequency domain, the main goal of which is to establish a closeness relation between the transfer function of the original system and that of its reduced-order model: this is attained by providing a bound on the $\mathcal H_2$ norm of the error between transfer functions at given frequencies \cite{cheng2017reduction,yu2019,yu2021h_2}. In Section~\ref{Infinite} instead, we incorporate model-order reductions in the time domain within the development of  \emph{infinite-abstraction} techniques, which can thus handle high-level logic properties on model's trajectories, such as safety, reachability, etc. - this contrasts with the mentioned classical reduction techniques in the frequency domain, which are by and large exclusively developed for the analysis of input-output behaviour and stability.

In this survey paper, we discuss recent approaches grounded on both discretization-based and -free techniques. 
We should mention that the main focus of this survey is on  \emph{discrete-time, continuous-space} stochastic hybrid systems, 
whereas we dedicate only one section (Sec.~\ref{CTSS}) to the otherwise interesting framework of \emph{continuous-time, continuous-space} models, 
where we overview the corresponding major theoretical results. 
We should also stress that much of the presented work builds on the extensive theoretical and algorithmic background of \emph{finite-space} Markov models, 
which is not overviewed here in view of length limitations:  
we refer the interested reader to~\cite{baier2008principles,kwiatkowska2011prism} for informative overviews.    

\subsection{Different Types of Closeness Guarantees} 

Earlier, we have emphasized the importance of providing ``formal'' abstractions:
in this survey we discuss four different types of closeness guarantees (or error bounds) between original SHS and their finite abstractions, 
as introduced next. 
These guarantees allow to perform computations over the abstract models, and to formally refine them over the concrete SHS.  

\begin{definition}\label{CG}
	Let $\Sigma$ be a concrete SHS and $\widehat \Sigma$ be its abstraction. 
	For a given specification,
	the probabilistic closeness between $\Sigma$ and $\widehat\Sigma$ is defined according to one of the following:
	\begin{tcolorbox}[enhanced,
		standard jigsaw,
		boxrule=0.5pt,
		opacityback=0,
		]
		\begin{enumerate}[(i)]
			\item the difference between probabilities of satisfaction of specifications over the original system $\Sigma$ and its corresponding abstraction $\widehat \Sigma$ (cf. equation~\eqref{Pro1} or~\eqref{Pro1_2}); 
			\item the probability of the difference between the output trajectories of $\Sigma$ and $\widehat \Sigma$ being less than a given threshold (cf. equation~\eqref{Pro2});
			\item the expectation (moment) of the difference between output trajectories of original system $\Sigma$ and those of its abstraction $\widehat \Sigma$ (cf. equation~\eqref{Pro3});
			\item the probability of satisfaction of logic properties over the abstract system $\widehat \Sigma$ is either lower- or upper-bounds the satisfaction probability over original system $\Sigma$ (cf. equations~\eqref{Pro4} and~\eqref{Pro5}).
		\end{enumerate}
	\end{tcolorbox}
\end{definition}

It is worth mentioning that the proposed probabilistic closeness bounds in Definition~\ref{CG} can be employed for abstractions that can be either specification-guided (i, iv) or specification-free (ii, iii). In general, abstractions that are specification-dependent are potentially less conservative as they are a-priori tailored to some given specifications. In comparison, specification-free abstractions are more general since, their corresponding closeness guarantees hold for classes of properties of interest, however this comes at the cost of an increase in their computational complexity. We shall further discuss closeness guarantees corresponding to these two types of abstractions in Sections~\ref{SSR} and~\ref{Infinite}.

\subsection{Contributions and Organization of this Survey}

This paper provides the first survey of literature on automated formal verification and synthesis of stochastic hybrid systems (SHS).  
While trying to be comprehensive, we focus on the most recent and sharpest results in the literature, and discuss related approaches in various sections in coarser detail. 
Besides the selection of the most relevant articles, 
this survey is intended to help researchers to gain an overall understanding of the many challenges and solution strategies related to the formal verification and the control synthesis of SHS, 
as well as the associated software tools that have been developed to support the theory. We discuss approaches in relevant literatures via both discretization-based and -free techniques, 
categorizing them over four different closeness guarantees between the concrete SHS and their abstractions, according to Definition~\ref{CG}. 
We employ a running example and discuss approaches under the lens of \emph{(i) time complexity}, and \emph{(ii) memory requirements}.
We also discuss many open problems throughout this survey paper.

We remark that although the survey paper in~\cite{teel2014stability} also covers \emph{stochastic} hybrid systems, 
its main focus is on stability analysis: 
different notions of stability are overviewed, including Lyapunov, Lagrange, asymptotic stability, and recurrence analysis.
In contrast, here we focus on formal verification and synthesis goals, defined around complex properties including those expressed as temporal logic formulae (simple instances are safety and reachability specifications), as well as more general properties expressed via omega-regular languages~\cite{baier2008principles}. 
In addition, we zoom in on \emph{algorithmic solutions} for verification and synthesis of SHS against temporal properties. An overview of the main developments in the area of stochastic model predictive control (SMPC) is in~\cite{mesbah2016stochastic}: these results focus on constrained, optimal control synthesis, however they are not natively aimed at providing the formal guarantees on verification and controller synthesis of complex SHS, which are the core focus of our survey.

This survey paper is structured as follows. In Section~\ref{NPM}, we formalize the models under study (syntax and semantics) and present preliminaries and main notations from control theory and computer science, which are widely employed throughout the survey. 
In Section~\ref{SSR}, we present one of the pivotal theorems of the article, elaborating on different types
of closeness guarantees between a discrete-time SHS and its abstraction.  
We discuss in depth the required assumptions, and present tools to compute such guarantees. 
Corresponding results on stochastic similarity relations to connect the probabilistic behavior of concrete models to that of their abstractions are also presented in the same section. 
Building on these notions of relations amongst models, work on the construction of \emph{infinite} abstractions for SHS is discussed in Section~\ref{Infinite}, and corresponding results on the construction of \emph{finite} abstractions are studied in Section~\ref{Finite}. We also discuss existing abstraction algorithms, together with the assumptions and details underpinning them.

In Section~\ref{CBC}, we first formally present the definition of control barrier certificates, as a discretization-free approach, for the analysis and synthesis of SHS. We then present another main theorem of this survey, which allows to quantify an upper bound on the probability that the given system reaches an unsafe region over both finite and infinite time horizons. We also briefly survey results on optimization-based methods for the analysis of SHS, as alternative discretization-free approaches that are, however, not core to this survey. Temporal logic verification and synthesis of SHS are studied in Section~\ref{TLVS}. Section~\ref{Network} is devoted to compositional techniques as a potential direction for mitigating the curse of dimensionality. We present the definition of subsystem, together with the formal definition of interconnected systems. We then discuss the main compositionality results, based on two different techniques from literature. 

Results for \emph{continuous-time} SHS are briefly presented in Section~\ref{CTSS}. 	
Section~\ref{SIM_SHS} is dedicated to surveying sample- and simulation-based analysis of SHS.  
Software tools on verification and synthesis of SHS are discussed in Section~\ref{RST}. In Section~\ref{CORD}, we summarize the existing analysis methods and highlight relevant directions for future research. In particular, we discuss a few open problems including ``formal analysis of SHS via learning and data-driven approaches", ``formal synthesis of partially-observed SHS", ``secured-by-construction controller synthesis for SHS", ``(mixed)-monotonicity of SHS", ``compositional construction of interval Markov processes", ``compositional controller synthesis for SHS", and ``potential extensions of software tools".

\section{Notations, Preliminaries, and Models}\label{NPM}

The sets of non-negative and positive integers are denoted by $\mathbb N := \{0,1,2,\ldots\}$ and $\mathbb N_{\ge 1} := \{1,2,3,\ldots\}$, respectively. Moreover,
the symbols $\mathbb R$, $\mathbb R_{>0}$, and $\mathbb R_{\ge 0}$ denote, respectively, the sets of real, positive and nonnegative real numbers. For any set $X$ we denote by $2^X$ the power set of $X$, namely the set of all subsets of $X$.
Given $N$ vectors $x_i \in \mathbb R^{n_i}$, $n_i\in \mathbb N_{\ge 1}$, and $i\in\{1,\ldots,N\}$, we use $x = [x_1;\ldots;x_N]$ to denote the corresponding column vector of dimension $\sum_i n_i$.
We denote by $\Vert\cdot\Vert$ and $\Vert\cdot\Vert_2$ the infinity and Euclidean norms, respectively. Given any $a\in\mathbb R$, $\vert a\vert$ denotes the absolute value of $a$. Symbols $\mathds{I}_n$, $\mathbf{0}_n$, and $\mathds{1}_n$ denote the identity matrix in $\mathbb R^{n\times{n}}$ and the column vector in $\mathbb R^{n\times{1}}$ with all elements equal to zero and one, respectively. Given a matrix
$P = \{p_{ij}\} \in \mathbb R^{n\times n}$, we denote the trace of
$P$ by $\textsf{Tr}(P)$, where $\textsf{Tr}(P) = \sum_{i=1}^{n}p_{ii}$. We denote the disjunction ($\vee$) and conjunction ($\wedge$) of Boolean functions $f:\Gamma\rightarrow \{0,1\}$ over a (possibly infinite) index set $\Gamma$ by $\underset{\alpha\in\Gamma}\vee f(\alpha)$ and $\underset{\alpha\in\Gamma}\wedge f(\alpha)$, respectively.
Given functions $f_i:X_i\rightarrow Y_i$, for any $i\in\{1,\ldots,N\}$, their Cartesian product $\prod_{i=1}^{N}f_i:\prod_{i=1}^{N}X_i\rightarrow\prod_{i=1}^{N}Y_i$ is defined as $(\prod_{i=1}^{N}f_i)(x_1,\ldots,x_N)=[f_1(x_1);\ldots;f_N(x_N)]$.
Given sets $X$ and $Y$, a relation $\mathscr{R}\subseteq X \times Y$ is a subset of the Cartesian product $X \times Y$ that relates $x \in X$ with $y \in Y$ if $(x, y) \in \mathscr{R}$, which is equivalently denoted by $x\mathscr{R}y$. A function $\gamma: \mathbb R_{\ge 0}\rightarrow \mathbb R_{\ge 0}$, is said to be a class $\mathcal{K}$ function if it is continuous, strictly increasing, and $\gamma(0)=0$. A class $\mathcal{K}$ function $\gamma(\cdot)$ is said to be a class $\mathcal{K}_{\infty}$ if $\gamma(r) \rightarrow \infty$ as $r\rightarrow\infty$. A continuous function $\beta: \mathbb R_{\ge 0}\times \mathbb R_{\ge 0}\rightarrow\mathbb R_{\ge 0}$ is said to belong to class $\mathcal{KL}$ if, for each fixed $t$, the map $\beta(r,t)$ belongs to class $\mathcal{K}$ with respect to $r$, and for each fixed nonzero $r$, the map $\beta(r,t)$ is
decreasing with respect to $t$, and $\beta(r,t)\rightarrow 0$ as $t\rightarrow\infty$.

We consider a probability space $(\Omega,\mathcal F_{\Omega},\mathds{P}_{\Omega})$,
where $\Omega$ is the sample space, $\mathcal F_{\Omega}$ is a sigma-algebra on $\Omega$ comprising subsets of $\Omega$ as events, and $\mathds{P}_{\Omega}$ is a probability measure that assigns probabilities to events. We assume that the random variables introduced and discussed in this article are measurable functions of the form $X:(\Omega,\mathcal F_{\Omega})\rightarrow (S_X,\mathcal F_X)$ (relevant literature contains details supporting these claims).
As such, any random variable $X$ induces a probability measure on  its space $(S_X,\mathcal F_X)$ as $Prob\{A\} = \mathds{P}_{\Omega}\{X^{-1}(A)\}$ for any $A\in \mathcal F_X$.
We often directly discuss the probability measure on $(S_X,\mathcal F_X)$ without explicitly mentioning the underlying probability space and the function $X$ itself.

A topological space $\mathcal S$ is called a Borel space if it is homeomorphic to a Borel subset of a Polish space (\emph{i.e.,} a separable and completely metrizable topological space).
Examples of a Borel space are Euclidean spaces $\mathbb R^n$, its Borel subsets endowed with a subspace topology as well as hybrid state spaces  \cite{APLS08}. 
Any Borel space $\mathcal S$ is assumed to be endowed with a Borel sigma-algebra, which is denoted by $\mathcal B(\mathcal S)$. We say that a map $f : \mathcal S\rightarrow Y$ is measurable whenever it is Borel measurable. 

\subsection{Discrete-Time Stochastic Hybrid Systems}
In this survey, we consider stochastic hybrid systems models in discrete time (dt-SHS), first introduced in \cite{bcAAPLS06,APLS08}, and defined formally as follows.

\begin{definition}\label{SHS}
	A discrete-time stochastic hybrid system (dt-SHS) is characterized by the tuple
	\begin{equation}\label{Eq:SHS}
		\Sigma=(\mathcal Q,n,X,U,T_{\mathsf x},Y,h), \textrm{	where}
	\end{equation}
	\begin{itemize}
		\item $\mathcal Q:= \{q_1,\dots,q_p\}$ for some $p \in \mathbb N_{\ge 1}$,
		represents the discrete-state space;
		\item $n: \mathcal Q \rightarrow \mathbb N_{\ge 1}$ assigns to each discrete state value $q \in \mathcal Q$ the dimension of the continuous-state space $\mathbb R^{n(q)}$;
		\item $X\subseteq \cup_{q\in\mathcal Q}\{q\}\times \mathbb R^{n(q)}$ is a Borel space as the hybrid-state space of the system. We denote by $(X, \mathcal B (X))$ the measurable space, with $\mathcal B (X)$  being  the Borel sigma-algebra over the state space;
		\item $U\subseteq \mathbb R^m$ is a Borel space as the input space of the system;
		\item $T_{\mathsf x}:\mathcal B(X)\times X\times U\rightarrow[0,1]$ 
		is a conditional stochastic kernel that assigns to any $x \in X$, and $\nu\in U$, a probability measure $T_{\mathsf x}(\cdot | x,\nu)$
		on the measurable space $(X,\mathcal B(X))$. This stochastic kernel specifies probabilities over executions $\{x(k),k\in\mathbb N\}$ of the hybrid system, 
		such that for any set $\mathcal{A} \in \mathcal B(X)$ and any $k\in\mathbb N$,
		\begin{align}\notag
			\mathbb P (x(k+1)\in \mathcal{A}\big| x(k),\nu(k))= \int_\mathcal{A} T_{\mathsf x} (\mathsf{d}x(k+1)\big|x(k),\nu(k)); 
		\end{align}
		\item  $Y\subseteq \mathbb R^{q}$ is a Borel space as the output space of the system;
		\item  $h:X\rightarrow Y$ is a measurable function as the output map that takes a state $x\in X$ to its output $y = h(x)$.
	\end{itemize}
\end{definition}	

An example of dt-SHS is discussed in the running example and equation~\eqref{RE}.

\begin{figure}[ht]
	\begin{center}
		\includegraphics[width=8.3cm]{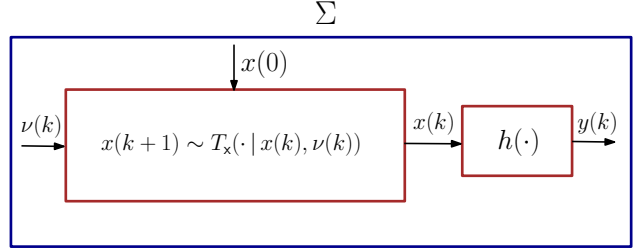}
		\caption{A schematic representation of a dt-SCS $\Sigma$.}
		\label{Fig_gMDP}
	\end{center}
\end{figure}

This definition is general and describes numerous applications. 
As this general structure of the state space can be notationally heavy,  
for the scope of this survey and for the sake of an easier presentation, 
we will introduce definitions, algorithms, and theorems based on a specific class of SHS with a single discrete mode (\emph{i.e.,} $X\subseteq \mathbb R^{n}$),  
called discrete-time stochastic control systems (dt-SCS)~\cite{meyn1993markov,hernandez1996discrete}. 
We emphasize that broadly the notions and approaches underlying the proposed results can be generalized to SHS endowed with {\em hybrid} state spaces.
A schematic representation of dt-SCS $\Sigma$ is shown in Fig.~\ref{Fig_gMDP}.

As argued in \cite{kallenberg1997foundations}, any dt-SCS endowed with a stochastic transition kernel $T_{\mathsf x}$ as in Definition \ref{SHS} can be \emph{equivalently} represented by a dt-SCS with pair $(f,\varsigma)$,
as formalized next. Note that this alternative representation is more common in control theory. 
It is often easier to show specific results of this paper based on the alternative representation.

\begin{definition}\label{Def:1}
	A discrete-time stochastic control system (dt-SCS) is represented by the tuple
	\begin{equation}\label{Eq:1}
		\Sigma=(X,U,\varsigma,f,Y,h), \textrm{   where}
	\end{equation}
	\begin{itemize}
		\item $X\subseteq \mathbb R^n$ is a Borel space as the state space of the system; 
		\item $U\subseteq \mathbb R^m$ is a Borel space as the input space of the system;
		\item $\varsigma$ is a sequence of independent and identically distributed (i.i.d.) random variables from a sample space $\Omega$ to the measurable space $(\mathcal{V}_\varsigma,\mathcal F_\varsigma)$
		\begin{equation*}
			\varsigma:=\{\varsigma(k):(\Omega,\mathcal F_\Omega)\rightarrow (\mathcal{V}_\varsigma,\mathcal F_\varsigma),\,\,k\in\N\}; 
		\end{equation*}
		\item $f:X\times U\times \mathcal{V}_{\varsigma} \rightarrow X$ is a measurable function characterizing the state evolution of the system;
		\item  $Y\subseteq \mathbb R^{q}$ is a Borel space as the output space of the system;
		\item  $h:X\rightarrow Y$ is a measurable function as the output map.
	\end{itemize}
	
	For given initial state $x(0)\in X$ and input sequence $\nu(\cdot):\mathbb N\rightarrow U$, the evolution of the state of the dt-SCS $\Sigma$ can be written, with $k\in\mathbb N$,  as
	\begin{equation}\label{Eq:2}
		\Sigma\!:\left\{\hspace{-1.5mm}\begin{array}{l}x(k+1)=f(x(k),\nu(k),\varsigma(k)),\\
			y(k)=h(x(k)).\\
		\end{array}\right.
	\end{equation}
	We denote by $\mathbb U$ the collection of input sequences $\{\nu(k):\Omega\rightarrow U,\,\,k\in\N\}$, in which $\nu(k)$ is independent of $\varsigma(z)$ for any $k,z\in\mathbb N$ and $z\ge k$. For any initial state $a\in X$, and input $\nu(\cdot)\in\mathbb U$, the random sequences $x_{a\nu}:\Omega \times\N \rightarrow X$, and $y_{a\nu}:\Omega \times \N \rightarrow Y$ that satisfy \eqref{Eq:2} are respectively called the \emph{solution process} and the \emph{output trajectory} of $\Sigma$ under an input $\nu$ and an initial state $a$. System $\Sigma$ is said to be finite if $ X$ and $U$ are finite sets, and infinite otherwise.
\end{definition}
{\bf Running Example.}  To help the reader gain a better understanding of the details in this survey paper, we present a simple yet interesting running example. 
We tailor the models and apply the results presented in this survey to a temperature regulation problem for a room equipped with a heater.
The model of this case study is borrowed from~\cite{fehnker2004benchmarks,meyer},
but modified by including an additive noise, which is intended to capture the effect of uncertain weather- or user-dependent factors. 
The evolution of the temperature $T(\cdot)$ over time can be described by the following dt-SCS:
\begin{equation}\label{RE}
	\Sigma\!:\left\{\hspace{-1.5mm}\begin{array}{l}{T}(k+1)=a(k){T}(k)+\gamma T_{h}\nu(k)+ \theta T_{e}+R\varsigma(k),\\
		y(k)=T(k),\end{array}\right.
\end{equation}
where the signal $a(k) := (1-\theta-\gamma\nu(k))$ depends on the input $\nu(k)$, $R = 0.6$ is the noise coefficient, and $\theta = 0.4$, and
$\gamma = 0.5$ are factors that affect the rate of heat conduction between the external environment and the room, 
and between the heater and the room. The parameter $T_{e}=-1\,^\circ C$ is the outside temperature, $T_h=50\,^\circ C$ is the heater temperature, 
and $y$ is the (sensed, observed) output of the system, which in this instance corresponds to the temperature itself. Finally, $\varsigma$ is assumed to be i.i.d. with a normal distribution having zero mean and a covariance equal to 1.

The model in~\eqref{RE} can be alternatively (and equivalently) characterized via the tuple in~\eqref{Eq:1}: 
here $X,U$ are subsets of the real numbers, $f(x(k),\nu(k),\varsigma(k)) = a(k){T}(k)+\gamma T_{h}\nu(k)+ \theta T_{e}+R\varsigma(k)$, 
and the output map $h$ is identity (accordingly, the output space $Y = X$). 
Note that this system is a very special instance of SHS in~\eqref{Eq:SHS} endowed with a single discrete mode, where the conditional stochastic kernel $T_{\mathsf x}$ is a normal distribution with mean $a(k){T}(k)+\gamma T_{h}\nu(k)+ \theta T_{e}$ and covariance $R^2$.
Alternatively, if the input $\nu(k)$ is assumed to be a finite-valued function of the state, 
e.g. a binary function switching upon hitting the boundaries of a temperature interval, 
then the model can be interpreted as a two-mode SHS. 

\begin{figure}
	\centering
	\includegraphics[width=7.3cm]{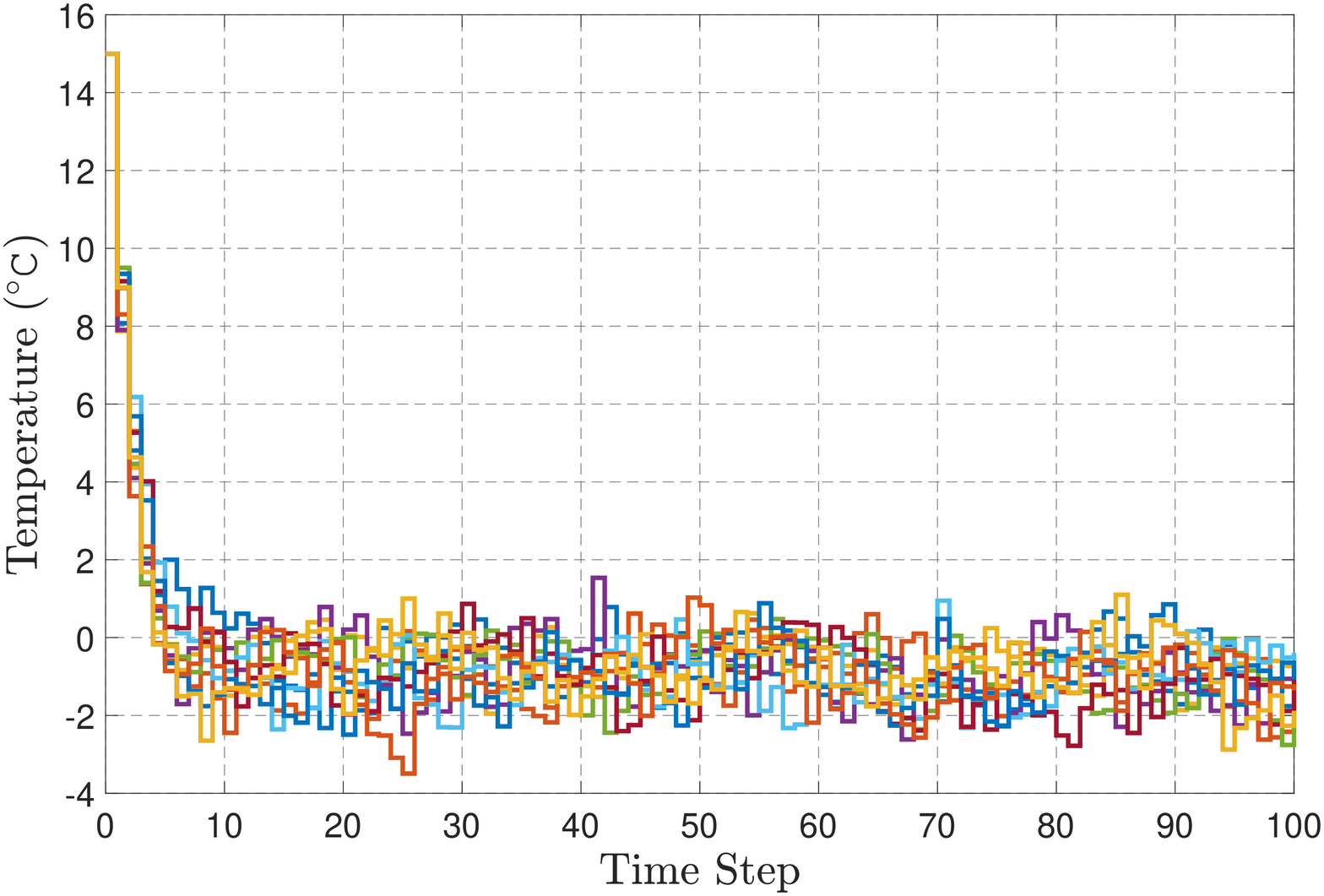}
	\includegraphics[width=7.3cm]{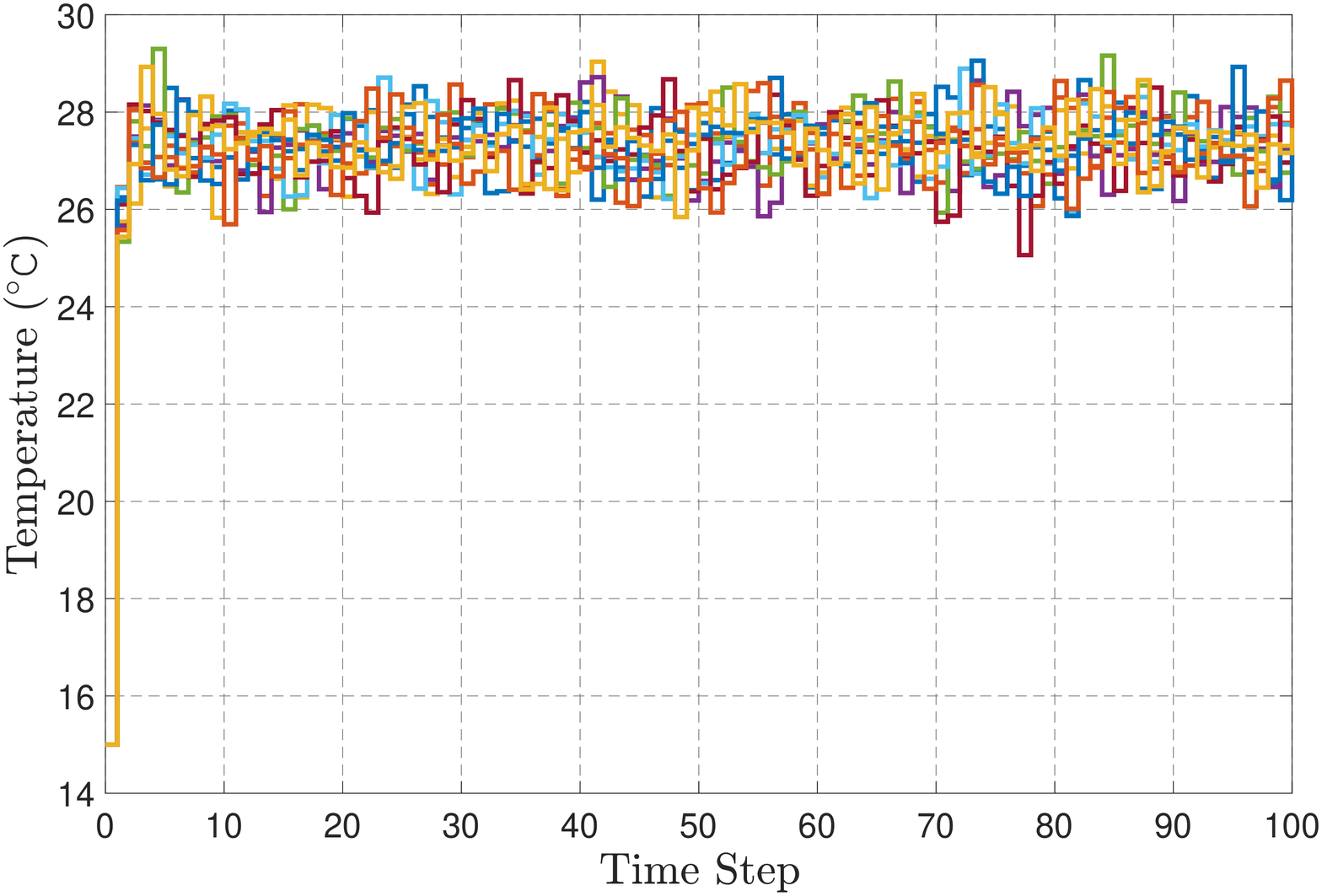}
	\caption{State trajectories, generated for the running example with $10$ different noise realizations within the finite time horizon $T_d=100$ from an initial condition $x_0 = 15$,  and with $\nu = 0$ (top, heating off) and $\nu = 1$ (bottom, heating on), respectively.}
	\label{Fig_Running}
\end{figure}

In order to provide some intuitions on the evolution of temperature, we plot in Fig.~\ref{Fig_Running} the state trajectories of the running example with $10$ different noise realizations within the finite time horizon $T_d=100$ from an initial condition $x_0 = 15$ and with inputs $\nu = 0$ and $\nu = 1$.$\hfill\square$

Given the dt-SCS model in \eqref{Eq:1}, we introduce \emph{Markov policies} as follows.

\begin{definition}\label{Def:2}
	A Markov policy for the dt-SCS $\Sigma$ in \eqref{Eq:1} is a sequence
	$\mu = (\mu_0,\mu_1,\mu_2,\ldots)$ of universally measurable stochastic kernels $\mu_n$ \cite{BS96},
	each defined on the input space $U$ given $X$ and such that for all $x(n)\in X$, $\mu_n(U(x(n))\,\big|\,x(n))=1$.
	The class of all Markov policies is denoted by $\mathcal M_p$.
\end{definition} 

Informally, Markov policies are history-independent and the control input taken at the current time instance is selected, possibly randomly, with a distribution that depends only on the current state.

\subsection{Relations between Models}

We now define the notion of incremental input-to-state stability for $\Sigma$, as a pivotal assumption that will allow some of the results, in particular to provide closeness guarantees between output trajectories of concrete system $\Sigma$ and its abstraction $\widehat \Sigma$, as per~\emph{(iii)} and~\emph{(iv)} in Definition~\ref{CG}.

\begin{definition}\label{Def:4}
	A dt-SCS $\Sigma=(X, U,\varsigma, f, Y,h)$ is called \emph{incrementally input-to-state stable ($\delta$-ISS)} if there exists a function $S: X\times X\to \mathbb{R}_{\geq0}$  such that $\forall x, x'\in X$, $\forall \nu,\nu'\in U$, the following two inequalities hold: 	
	\begin{align}\label{Eq:6}
		\underline{\alpha}(\Vert x-x'\Vert ) \leq S(x,x')\leq \overline{\alpha} (\Vert x-x'\Vert ),
	\end{align}	
	and	
	\begin{align}\label{Eq:7}
		\mathbb{E} \Big[S(f(x,\nu,\varsigma),f(x',\nu',\varsigma))\,\big|\,x,x',\nu,\nu'\Big]-S(x,x')\leq-{\bar\kappa}(S(x,x'))+ \rho(\Vert \nu-\nu'\Vert),
	\end{align}
	for some $\underline{\alpha}, \overline{\alpha}, {\bar\kappa}\in \mathcal{K}_{\infty}$,and $\rho\in\mathcal{K}_\infty\cup\{0\}$.
\end{definition}
Later we will show how one can use the $\delta$-ISS property to bound the distance between two solution processes starting from different initial conditions and under different input trajectories. 

We now define the notion of stochastic simulation functions (SSF) between $\widehat\Sigma$ and $\Sigma$, 
which allows to provide closeness guarantees between the output trajectories of the two models, 
as per~\emph{(ii)} in Definition~\ref{CG}. 

\begin{definition}\label{SSF}
	Consider two dt-SCS
	$\Sigma =(X,U,\varsigma,f,Y,h)$ and
	$\widehat\Sigma =(\hat X,\hat U,\varsigma,\hat f,\hat Y,\hat h)$.
	A function $V:X\times\hat X\to\R_{\ge0}$ is
	called a \emph{stochastic simulation function} (SSF) from $\widehat\Sigma$  to $\Sigma$ if
	\begin{itemize}
		\item $\exists \alpha\in \mathcal{K}_{\infty}$ such that
		\begin{eqnarray}\notag
			\forall x\in X,\forall \hat x\in\hat X,\quad \alpha(\Vert h(x)-\hat h(\hat x)\Vert)\le V(x,\hat x),
		\end{eqnarray}
		\item $\forall x\in X,\hat{x}\in\hat X,\hat{u}\in\hat U$, $\exists u\in U$ such that
		\begin{align}\label{eq:V_dec}
			\EE \Big[V(f(x,u,\varsigma),\hat{f}(\hat x,\hat u,\varsigma))\,\big|\,x,\hat{x},u, \hat{u}\Big]\leq-\kappa V(x,\hat{x})
			+\rho_{\mathrm{ext}}(\Vert\hat u\Vert)+\psi,
		\end{align}
		for some $0<\kappa<1$, $\rho_{\mathrm{ext}} \in \mathcal{K}_{\infty}\cup \{0\}$, and $\psi \in\R_{\ge 0}$.
	\end{itemize}
	We denote by $\widehat\Sigma\preceq\Sigma$
	if there exists an SSF $V$ from $\widehat\Sigma$ to $\Sigma$, and call the system $\widehat\Sigma$ an abstraction of the concrete (original) system $\Sigma$. Note that $\widehat{\Sigma}$ may be finite or infinite, depending on the cardinality of the sets $\hat X$ and $\hat U$. 
\end{definition}

Informally, stochastic simulation functions are Lyapunov-like functions defined over the Cartesian product of the state spaces of two models, 
which relate their output trajectories and indeed guarantee that their mismatch (namely the difference between their outputs) remains within some guaranteed error bounds. 
This mismatch can be conceived as the abstraction error if one model is obtained as the simplification of a given concrete model.  
In particular, since SHS are in general complex and intractable, stochastic simulation functions are beneficial to connect the probabilistic behavior of concrete SHS to that of their abstractions: in particular, by providing closeness guarantees between output trajectories of two systems via the established stochastic simulation functions, one can perform formal analysis over the simplified abstractions and transfer the obtained results back over the original SHS.

\begin{remark}\label{Interface}
	The second condition in Definition~\ref{SSF} implies the existence of a function $\nu=\nu_{\hat \nu}(x,\hat x,\hat \nu)$ for the satisfaction of \eqref{eq:V_dec}. 
	This function is called the ``interface function" and will be used to refine a synthesized policy $\hat\nu$ for $\widehat\Sigma$ to a policy $\nu$ for $\Sigma$ (cf. Fig. \ref{Fig1}), and will be discussed later in Sections~\ref{Infinite} and~\ref{Finite}.  
\end{remark}

\subsection{Temporal Logic Specifications}\label{LTL}
Formal specifications provide a rigorous and unambiguous formalism to express formal requirements over models. 
A common way to describe such formal requirements is utilizing specifications expressed as automata or in a temporal logic, \emph{e.g.,} formulae expressed in linear temporal logic (LTL)~\cite{pnueli1977temporal}.Let us start with some basic properties. Consider the dt-SCS in~\eqref{Eq:1} and measurable sets $\mathsf A,\mathsf B\subset Y$, named respectively ``safe'' and ``target'' set. 
(Later, in Def. \ref{LTL_Def}, we shall encompass these sets through a labelling function $\Lab$.) 
We define the bounded-horizon safety property as $\square^{\le T_d}\mathsf A$, indicating that all output trajectories $\{y(k)\}_{k\ge 0}$ start from
the safe set $\mathsf A$ and remain inside it over the finite-time horizon $k\in [0,T_d]$.  
Similarly, we say that an output trajectory $\{y(k)\}_{k\ge 0}$ reaches a target set $\mathsf B$ within the discrete time interval $[0,T_d]\subset \mathbb N$, if there exists a $k\in [0,T_d]$ such that $y(k)\in \mathsf B$: this bounded-horizon reachability property is denoted by $ \lozenge^{\le T_d}\{y\in \mathsf B\}$ or briefly $\lozenge^{\le T_d}\mathsf B$. Extending the above requirements to infinite horizons by $T_d\rightarrow \infty$, we denote the corresponding safety and reachability properties as $\square\mathsf A$ and $\lozenge \mathsf B$, which are colloquially said  ``always $\mathsf A$'' and ``eventually $\mathsf B$'', respectively. Additionally, we define \emph{reach-avoid} specifications by the formula  $\mathsf A\until\mathsf B$, requiring the output trajectories to reach the target set $\mathsf B$ while remaining in the safe set $\mathsf A$ - this property is also known as \emph{constrained reachability}. More generally, all the described basic properties can be reframed as specifications in LTL.

We formally define syntax and semantics of linear temporal logic (LTL) as follows. 

\begin{definition}\label{LTL_Def}
	Consider a set of atomic propositions $\AP$ and the alphabet $\alphabeth := 2^{\AP}$. 
	Let $\word=\word(0),\word(1),\word(2),\\\ldots$ be an infinite word, that is, a string composed of letters from $\alphabeth$ (i.e., $\word(i)\in\alphabeth, \forall i\in\N$).
	We are interested in those atomic propositions that are relevant to the dt-SCS via a measurable labeling function $\Lab$ from the (continuous) output space to the (finite) alphabet 
	as
	$\Lab:Y\rightarrow \alphabeth$. Informally, the output space is ``tagged with labels'' that are relevant to the specifications of interest.
	Accordingly, output trajectories  $\{y(k)\}_{k\geq 0}\in Y^{\mathbb N}$ of the dt-SCS 
	can be readily mapped to the set of infinite words $\alphabeth^{\mathbb N}$, as
	\[\word=\Lab(\{y(k)\}_{k\geq0}):=\{\word\in \alphabeth^{\mathbb N}\,|\,\word(k)= \Lab(y(k)) \}.\]
	We define the LTL syntax \cite{baier2008principles} as 
	\begin{equation*}
		\varphi ::= \operatorname{ true} \mid p \mid \neg \varphi \mid \varphi_1 \wedge \varphi_2 \mid
		\nex \varphi  \mid
		\varphi_1\until\varphi_2.
	\end{equation*}
	Given a trace $\word=\word(0),\word(1),\word(2),\ldots$, let us denote the suffix of $\word$ starting from $\word(i)$ by 
	$$(\word,i) = \word(i),\word(i+1),\word(i+2),\ldots.$$
	We denote by $(\word,i)\models \varphi$ when the LTL formula $\varphi$ is true on the suffix $(\word,i)$.
	This satisfaction is defined inductively as follows:
	\begin{itemize}
		\item $(\word,i)\models \textsf{true}$;
		\item $(\word,i)\models p$, for $p\in AP$ iff $p\in\word (i)$;
		\item $(\word,i)\models \neg\varphi$ iff $(\word,i)\not\models\varphi$;
		\item $(\word,i)\models \varphi_1\wedge\varphi_2$ iff $(\word,i)\models\varphi_1$ and $(\word,i)\models\varphi_2$;
		\item $(\word,i)\models \nex \varphi$ iff $(\word,i+1)\models\varphi$;
		\item $(\word,i)\models  \varphi_1\until\varphi_2$ iff for some $j$ such that $i\leq j$, we have $(\word,j)\models\varphi_2$, and for all $k$ s.t. $i\leq k<j$, we have $(\word,k)\models\varphi_1$.
	\end{itemize}
	Formula $\varphi$ is true on $\word$, denoted by $\word\models \varphi$, if and only if $(\word,0)\models \varphi$. 
\end{definition}
Based on the above operators, we can also introduce other formulae, obtained via propositional or temporal manipulations. 
These can encode simple properties, such as the mentioned reachability and safety specifications, or more complicated requirements, obtained by composing arbitrary numbers of operators.
For instance, 
$\varphi_1 \vee\varphi_2$,
$\lozenge\varphi$, and  
$\square\varphi$
have semantics
\begin{itemize}
	\item \emph{$(\word,i)\models \varphi_1\vee\varphi_2$ iff $(\word,i)\models\varphi_1$ or $(\word,i)\models\varphi_2$;} 
	\item \emph{$(\word,i)\models  \lozenge\varphi$ iff for some $j$ such that $i\leq j$, we have $(\word,j)\models\varphi$;}
	\item \emph{$(\word,i)\models  \square\varphi$ iff for all $j$ such that $i\leq j$, we have $(\word,j)\models\varphi$.}
\end{itemize}
Later, we shall be interested in quantifying the likelihood of verifying given LTL formulae by Markov models, such as MDPs or dt-SCS. 
Clearly, this requires reasoning about measurability of events associated to the LTL specifications introduced above; 
however, we shall not delve into measure-theoretical  issues in the present survey.

We now introduce a fragment of LTL properties known as \emph{syntactically co-safe} linear temporal logic (scLTL) \cite{KupfermanVardi2001}. 
scLTL properties are popular since their satisfaction can be sufficiently witnessed by finite-length traces.

\begin{definition}\label{def:scLTL}
	An scLTL over a set of atomic propositions $\AP$ is a fragment of LTL such
	that the negation operator ($\neg$) only occurs before atomic propositions, 
	and it is characterized by the following grammar:
	\begin{equation*}
		\varphi ::=  \operatorname{true} \,|\, p \,|\, \neg p \,|\,\varphi_1\wedge \varphi_2\,|\,\varphi_1 \vee \varphi_2\,|\, \nex \varphi \,|\, \varphi_1\until \varphi_2\,|\, \lozenge \varphi,
	\end{equation*} 
	with $p\in \AP$.
	The semantics of satisfaction follows from that of LTL.
\end{definition}
Another enticing aspect about scLTL is that it can be alternatively expressed by means of simple finite-state  automata~\cite{KupfermanVardi2001,Calin17}. 
This means that the set of words satisfying a given scLTL formula can be equivalently expressed as the set of words that are accepted by a proper (not necessarily unique) finite-state automaton. 
More specifically, we introduce a class of models known as deterministic finite-state automata (DFA).
\begin{definition}\label{DFA_Original}
	A DFA is a tuple $\mathcal A = \left(Q_{\ell},q_0,\mathsf{\Sigma}_{\textsf{a}},F_{\textsf{a}},\trans\right)$, where
	$Q_{\ell}$ is a finite set of locations (states),
	$q_0\in Q_{\ell}$ is the initial location,
	$\mathsf{\Sigma}_{\textsf{a}}$ is a finite set (a.k.a. alphabet),
	$F_{\textsf{a}}\subseteq Q_{\ell}$ is a set of accepting locations, and
	$\trans: Q_{\ell}\times\mathsf{\Sigma}_{\textsf{a}}\rightarrow Q_{\ell}$ is a transition function.
\end{definition}
Consider a set of atomic propositions $\AP$ and the alphabet $\mathsf{\Sigma}_{\textsf{a}} := 2^{\AP}$. 
A finite word composed of letters of the alphabet, \emph{i.e.,} $\word_f = (\word_f(0),\ldots,\word_f(n))\in \mathsf{\Sigma}_{\textsf{a}}^{n+1}$, is accepted by a DFA $\mathcal A$ if there exists a finite run $q =(q(0),\ldots,q(n + 1))\in Q_{\ell}^{n+2}$ such that $q(0) = q_0$,
$q(i + 1) =\trans(q(i),\word_f(i))$ for all $0\le i\le n$, and $q(n+1)\in F_{\textsf{a}}$. The accepted language of $\mathcal A$, denoted $\mathcal L(\mathcal A)$, is the set of all words accepted by $\mathcal A$. For every scLTL property $\varphi$, cf. Definition \ref{def:scLTL}, there exists a DFA $\mathcal A_{\varphi}$ such that
\begin{equation}\notag
	\mathcal L_f(\varphi) = \mathcal L(\mathcal A_{\varphi}),
\end{equation}
where $\mathcal L_f$ denotes the set of all words associated to an scLTL formula $\varphi$. In some parts of this article, we focus on the computation of probability of $\omega_f \in\mathcal L(\mathcal A_\varphi)$ over bounded intervals. In other words, we fix a time horizon $T_d$ and compute $\mathbb P(\omega_f(0)\omega_f(1)\ldots\omega_f(T_d)\in\mathcal L(\mathcal A_\varphi)~\text{s.t.}~ |\word_f|\le T_d+1)$, with $|\word_f|$ denoting the length of $\word_f$.

\begin{figure}[ht]
	\begin{center}
		\includegraphics[width=7cm]{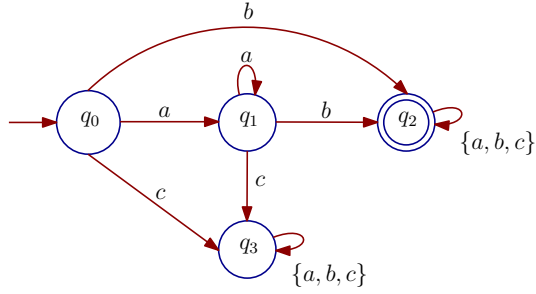}
		\caption{DFA $\mathcal A_{\varphi}$ of the reach-avoid specification $(a\,\until b)$.}
		\label{DFA}
	\end{center}
\end{figure}

The following example, borrowed from~\cite{lavaei2018CDCJ}, provides an automaton associated with a reach-avoid specification
\begin{example}\label{illustration}
	Consider two measurable sets $\mathsf A,\mathsf B\subset Y$ as the safe and target sets, respectively. We present the DFA for the specification $(\mathsf A\,\until\mathsf B)$
	which requires the output trajectories to reach the target set $\mathsf B$ while remaining in the safe set $\mathsf A$. Note that we do not assume these two sets are disjoint. Consider the set of atomic propositions $AP = \{\mathsf A,\mathsf B\}$ and the alphabet $\mathsf{\Sigma}_{\textsf{a}} = \{\emptyset,\{\mathsf A\},\{\mathsf B\}, \{\mathsf A,\mathsf B\}\}$. Define the labeling function as
	\begin{equation*}
		\Lab(y) =
		\begin{cases}
			\{\mathsf A\}=:a & \text{ if }\,\, y\in \mathsf A\backslash \mathsf B,\\
			\{\mathsf B\}=:b & \text{ if }\,\,y\in \mathsf B,\\
			\emptyset=:c & \text{ if } \,\,y\notin\mathsf A\cup \mathsf B.
		\end{cases}
	\end{equation*}
	As can be seen from the above definition of the labeling function $\Lab$, it induces a partition over the output space $Y$ as
	\begin{equation*}
		\Lab^{-1}(a) = \mathsf A\backslash \mathsf B,\quad \Lab^{-1}(b) = \mathsf B,\quad \Lab^{-1}(c) = Y\backslash(\mathsf A\cup \mathsf B).
	\end{equation*}
	Note that we have indicated the elements of $\,\mathsf{\Sigma}_{\textsf{a}}$ with lower-case letters, for ease of notation. 
	The specification $(\mathsf A\,\until\mathsf B)$ can be equivalently written as $(a\,\until b)$ with the associated DFA depicted in Figure~\ref{DFA}.
	This DFA has the set of locations $Q_{\ell}=\{q_0,q_1,q_2,q_3\}$, an initial location $q_0$, and an accepting location $F_{\textsf{a}} =\{q_2\}$.
	Thus output trajectories of a dt-SCS $\Sigma$ satisfy the specification $(a\,\until b)$ if and only if their associated words are accepted by this DFA.
\end{example}

Generalizing beyond scLTL (and corresponding DFAs), 
often we are interested in infinite paths through a system and thus in infinite words. 
$\omega$-regular languages generalize the definition of regular languages to encompass sets of infinite-length words.
Correspondingly, $\omega$-regular properties are specifications expressed via $\omega$-regular languages and $\omega$-automata are finite-state models that can accept them.
Two of the most commonly used automata to express $\omega$-regular properties
are \emph{B{\"u}chi} and \emph{Rabin} automata~\cite{baier2008principles}. 
Indeed, it can be shown that non-deterministic\footnote{In a deterministic automaton, each transition is \emph{uniquely} determined by its source state and input symbol, whereas a non-deterministic automaton does not abide by this requirement. Non-determinism in B{\"u}chi automata is required to encompass LTL.} \emph{B{\"u}chi} automata (NBA) encompass the entire LTL.
\begin{definition}\label{NBA_Original}
	An NBA is a tuple $\mathcal A = \left(Q_{\ell},q_0,\mathsf{\Sigma}_{\textsf{a}},F_{\textsf{a}},\trans\right)$, where
	$Q_{\ell}$ is a finite set of locations,
	$q_0 \subseteq Q_{\ell}$ is the initial location,
	$\mathsf{\Sigma}_{\textsf{a}}$ is the finite alphabet,
	$F_{\textsf{a}} \subseteq Q_{\ell}$ is a set of accept locations, and
	$\trans: Q_{\ell}\times\mathsf{\Sigma}_{\textsf{a}}\rightarrow 2^{Q_{\ell}}$ is a transition relation.
\end{definition} 
An infinite word composed of letters of the alphabet, \emph{i.e.,} $\word = (\word(0),\word(1)\ldots)\in \mathsf{\Sigma}_{\textsf{a}}^{\omega}$,
is accepted by the NBA $\mathcal A$ if there exists an infinite run $q =(q(0),q(1),\ldots)\in Q_{\ell}^{\omega}$ such that $q(0) \in q_0$, $q(i + 1) \in \trans(q(i),\word(i))$ for all $0\le i$, and $q(i)\in F_{\textsf{a}}$ infinitely often.  Let us remark that the \emph{non-deterministic} feature of NBA is necessary to express $\omega$-regular properties, and in particular the set of LTL specifications. 
Alternatively, later in this survey, we shall mention limit-deterministic \emph{B{\"u}chi} automata. Similarly, deterministic \emph{Rabin} automata can be utilized, which employ a different (and more involuted) acceptance semantics. However, for the sake of space we avoid to detail them and instead refer the readers to \cite{baier2008principles} for a comprehensive discussion.

For probabilistic models, 
properties of interest can be expressible in a different logic, which encompasses a probabilistic operator in its syntax and is such that its satisfaction is defined over states (branching semantics), as opposed to the case of LTL specifications whose satisfaction is defined over trajectories (linear semantics).
Probabilistic computation tree logic (PCTL)~\cite{ciesinski2004probabilistic} can be introduced as follows.
\begin{definition}
	The syntax of (PCTL) formulae is defined recursively using the following operators:
	\begin{eqnarray*}
		&& \varphi ::= \operatorname{true} \mid p \mid \varphi \wedge \varphi \mid \neg \varphi \mid \mathbb{P}_{\sim p}[\psi], \\
		&& \psi ::= \nex \varphi \mid \varphi\, \until \varphi, 
	\end{eqnarray*}
	where $p \in AP$ and $\sim \in \{<, \leq, \geq, >\}$, and $p \in [0,1]$. 
	The semantics for $\nex$ and for $\until$ are as before, and the semantics for $\operatorname{true}$, $p$, $\wedge$, and $\neg$ are also identical except for being defined with reference to a state $s = \omega(0)$, instead of the first element of a path $\omega$. 
	The satisfaction semantics for the expression $s \models \mathbb{P}_{\sim p}[\psi]$
	is defined as follows: 
	\begin{equation*}
		\Pr(\{\word \in \mathsf{\Sigma}_{\textsf{a}}^{\omega} ~\vert~ \word(0) = s \text{ and } \word,0 \models \psi\}) \sim p, 
	\end{equation*}
	where $\Pr$ is the probability distribution over the infinite paths through $Y$ induced by the stochastic dynamics.
\end{definition}

Discussion of characterization and computation of basic PCTL specifications, such as safety and reachability, for dt-SCS is introduced in \cite{bcAAPLS06} and generalized in \cite{APLS08}, 
is applied to invariance in \cite{DBLP:conf/cdc/PolaP06}, 
and extended to reach-avoid properties in  \cite{DBLP:journals/automatica/SummersL10}. The characterization of properties in these works is based on a dynamic programming recursion,  which is generalized in \cite{DBLP:conf/hybrid/RamponiCSL10}, which connects the characterization of PCTL to dynamic programming. Further, \cite{tmka2013,tkachev2017quantitative} generalizes this work to regular and $\omega$-regular properties, respectively, leveraging DFA and  deterministic NBA models from above, 
and a product construction further discussed in Section \ref{TLVS}.  

\begin{remark}\label{measurable} 
	Note that it has been shown in relevant literature that all random variables discussed in this survey are measurable functions of the form $X:(\Omega,\mathcal F_{\Omega})\rightarrow (S_X,\mathcal F_X)$. Accordingly, all specifications discussed in this survey encompass measurable events under the system dynamics, so  that one can properly assign probabilities to those events~\cite[Proposition 2.3]{vardi1985automatic}. 
\end{remark}

\begin{remark}
	Let us remark that literature on SHS deals with diverse types of logical specifications, including safety, reachability, reach-avoid, co-safe linear temporal logic (scLTL), bounded linear temporal logic (BLTL)~\cite{maler2004monitoring},
	signal temporal logic (STL)~\cite{maler2008checking}, probabilistic computation tree logic (PCTL), and metric interval temporal logic (MITL)~\cite{maler1995timing,maler2005real}.
	However, for the sake of better readability, we mainly focus on simpler, more common requirements that are widely employed in the literature and in the practice, including  safety, reachability, reach-avoid, as well as LTL and PCTL properties.
\end{remark}

\section{Stochastic Similarity Relations for Abstractions}\label{SSR} 
In view of the generality of SHS and of the properties of interest, closed-form solutions for verification and synthesis problems, concerning the expression of value functions or of synthesized feedback policies, are in general not available explicitly, and thus require to be computed numerically. 
In this instance, an effective approach is to approximate a given SHS model by simpler abstract ones, for example models with lower dimensionality, simpler dynamics, or even a finite state space. 
In order to render this approximation ``formal,'' it is desirable to provide guarantees on this approximation step, so that the analysis and/or the synthesis on the derived abstract models can be translated back to the original SHS. 
As anticipated earlier, stochastic similarity relations can indeed be employed to relate the probabilistic behavior of a concrete model (e.g., a given SHS) to that of its abstractions. 
They can be framed as stochastic simulation and bisimulation relations, in either exact or approximate form.  

In the following, we present four theorems that summarize several results from relevant literature, 
and thus provide the four different types of closeness guarantees, as introduced earlier, between a concrete SHS and a derived abstraction. 
First, with focus on dt-SCS, we present bounds on the difference between the probability of satisfaction of logic properties over a given system $\Sigma$ and its corresponding finite abstraction $\widehat \Sigma$ \cite{SSoudjani,tmka2013}, as introduced in Sec. \ref{subsec:DBT}. This type of probabilistic closeness requires a Lipschitz continuity assumption on the stochastic kernel of the dt-SCS, as in the following. 

\begin{definition}\label{Lipt}
	The dt-SCS in Definition~\ref{Def:1} is Lipschitz continuous if the stochastic kernel $T_{\mathsf x}$ admits a density function $t_s(\bar x | x,u)$ satisfying the following inequality for a some constant $\mathscr{\bar H}\ge 0$:
	\begin{align}
		\label{Lipschitz1}
		\vert t_s(\bar x\,|\, x,u) - t_s(\bar x \,|\, x',u')\vert 
		\leq \mathscr{\bar H}(\Vert x -x'\Vert + \Vert u -u'\Vert),
	\end{align}
	for all $x,x',\bar x \in X$ and all $u, u' \in U$.
	If the policy for dt-SCS is given as $\nu: X\rightarrow U$, we define the Lipschitz constant of the stochastic kernel by $\mathscr{H}$\!, where
	\begin{align}
		\label{Lipschitz2}
		\vert t_s(\bar x\,|\, x,\nu(x)) - t_s(\bar x \,|\, x',\nu(x'))\vert 
		\leq \mathscr{H}\Vert x -x'\Vert,
	\end{align}
	for all $x,x',\bar x \in X$.
\end{definition}

Now, we have all the ingredients to introduce the first approximation theorem, which is related to~\emph{(i)} in Definition~\ref{CG}. Note that a finite abstraction $\widehat\Sigma$ is obtained from the original system $\Sigma$ by first constructing finite partitions of state and input sets, and then selecting arbitrary ``representative points'' as abstract states and inputs. Transition probabilities in the finite abstraction $\widehat\Sigma$ are computed accordingly (cf. Section~\ref{Finite}, Algorithm~\ref{algo:MC_app}).
\begin{theorem}
	Let $\Sigma=(X,U,\varsigma,f, Y, h)$ be a continuous-space dt-SCS and $\widehat\Sigma=(\hat X, \hat U,\varsigma,\hat f, \hat Y, \hat h)$ be its finite abstraction. 
	Assume that the original system $\Sigma$ is Lipschitz continuous, as per Definition~\ref{Lipt}.
	For a given logic specification $\varphi$, and for any policy $\hat\nu(\cdot)\in\mathbb{\hat U}$ that preserves the Markov property for the closed-loop $\widehat\Sigma$ (i.e., system $\widehat \Sigma$  fed by input $\hat\nu(\cdot)$, which is denoted by $\widehat\Sigma_{\hat \nu}$), the probabilistic closeness between two systems is as follows:
	\begin{tcolorbox}[enhanced,
		standard jigsaw,
		boxrule=0.5pt,
		opacityback=0,]
		\begin{equation}\label{Pro1}
			|\PP(\Sigma_{\hat \nu}\vDash\varphi) - \PP (\widehat\Sigma_{\hat \nu}\vDash\varphi)|\le \lambda_1,
		\end{equation}
	\end{tcolorbox}
	
	with $\lambda_1 := T_d \delta \mathscr{H}\mathscr{L}_b$, where $T_d$ is the finite-time horizon, $\delta$ is the state discretization parameter, $\mathscr{H}$ is the Lipschitz constant of the stochastic kernel $T_{\mathsf x}$ under policy $\hat\nu$ as in~\eqref{Lipschitz2}, and $\mathscr{L}_b$ is the Lebesgue measure of the state space.
	Moreover,
	the difference between the optimal probabilities of satisfying a given LTL specification $\varphi$ by the two models is bounded by
	\begin{tcolorbox}[enhanced,
		standard jigsaw,
		boxrule=0.5pt,
		opacityback=0,]
		\begin{equation}\label{Pro1_2}
			\big|\sup_{\nu}\mathbb P(\Sigma_{\nu}\vDash\varphi) - \sup_{\hat\nu}\mathbb P(\widehat\Sigma_{\hat \nu}\vDash\varphi)\big|\le \bar\lambda_1,
		\end{equation}
	\end{tcolorbox}
	with $\bar \lambda_1 := T_d \delta \mathscr{\bar H}\mathscr{L}_b$, where $\mathscr{\bar H}$ is the Lipschitz constant of the stochastic kernel $T_{\mathsf x}$ over the state $x$ and input $\nu$ as in~\eqref{Lipschitz1}. Furthermore, for the optimal policy $\hat \nu^*$ that maximizes the satisfaction probability of $\varphi$ for the abstraction $\widehat\Sigma$, we have
	\begin{tcolorbox}[enhanced,
		standard jigsaw,
		boxrule=0.5pt,
		opacityback=0,]
		\begin{equation}\label{Pro1_1}
			\big|\mathbb P(\Sigma_{\hat\nu^*}\vDash\varphi) - \mathbb P(\widehat\Sigma_{\hat \nu^*}\vDash\varphi)\big|\le 2\bar\lambda_1.
		\end{equation}
	\end{tcolorbox}
\end{theorem}

\begin{remark}
	Note that the  Lebesgue measure $\mathscr{L}_b$
	(informally, the ``volume'') of the set of interest (within the state space) appears in the error formula, as per~\eqref{Pro1}-\eqref{Pro1_1}, 
	which makes them meaningful over bounded domain and, possibly, quite conservative. 
	There exist techniques based on an adaptive and sequential gridding scheme (e.g.,~\cite{SA13}) that mitigate both shortcomings.   
\end{remark}

\begin{remark}
	Let us remark that, in general, the construction of the abstract system $\widehat \Sigma$ is performed in a way that allows a proper interpretation of the concrete specification $\varphi$ on the abstract model - as such, the abstraction will be property-dependent. In the closeness bounds~\eqref{Pro1}-\eqref{Pro1_1}, the specification $\varphi$ is defined over the state space of both $\Sigma$ and $\widehat \Sigma$. 
\end{remark}

\begin{remark}\label{Lipschitz}
	For a dt-SCS $\Sigma$ with linear dynamics $x(k+1) = Ax(k)+ B\nu(k)+\varsigma(k)$, where $A= [a_{ij}]$, $B= [b_{ij}]$, and $\varsigma(k)$ is
	i.i.d.\ for $k=0,1,2,\ldots$ with a normal distribution having zero mean and covariance matrix
	$\operatorname{diag}(\sigma_1,\ldots,\sigma_n)$, one can obtain $\mathscr{H} = \sum_{i,j}\dfrac{2|a_{ij}|}{\sigma_i\sqrt{2\pi}}$ and $\mathscr{\bar H} = \sum_{i,j}\dfrac{2|a_{ij}|}{\sigma_i\sqrt{2\pi}} + \sum_{i,j}\dfrac{2|b_{ij}|}{\sigma_i\sqrt{2\pi}}$. We refer the interested reader to~\cite{SA13} for the computation of Lipschitz constant $\mathscr{H}$ and $\mathscr{\bar H}$ in the general case. 
\end{remark}

In the next theorem, we present a condition such that the probabilistic distance between output trajectories of $\Sigma$ and $\widehat \Sigma$ is less than a given threshold, which is related to~\emph{(ii)} in Definition~\ref{CG},
as proposed in~\cite{lavaei2017compositional}. Let us note that this condition can work with either a finite abstraction or with an infinite abstraction with a lower-dimensional state space, which for instance can be constructed by means of a linear transformation of the state space, obtained with a rectangular matrix (cf. Theorem~\ref{IAbs} and Figure~\ref{Fig1_2}).
\begin{theorem}
	Let $\Sigma=(X,U,\varsigma,f, Y, h)$ be a continuous-space dt-SCS and $\widehat\Sigma=(\hat X, \hat U,\varsigma,\hat f, \hat Y, \hat h)$ be its abstraction, which can be either with a lower dimension or defined over a finite state set.
	Suppose there exists a SSF $V:X\times\hat X\to\R_{\ge0}$ from $\widehat\Sigma$  to $\Sigma$ as in Definition~\ref{SSF}. For any input trajectory $\hat\nu(\cdot)\in\mathbb{\hat U}$ that preserves the Markov property for the closed-loop $\widehat\Sigma$, and for any random variables $a$ and $\hat a$ as the initial states of dt-SCS $\Sigma$ and $\widehat \Sigma$, respectively, one can construct an input trajectory $\nu(\cdot)\in\mathbb{U}$ for $\Sigma$ through an interface function associated with $V$ (cf. Def.~\ref{SSF})
	such that:
	
	\begin{tcolorbox}[enhanced,
		standard jigsaw,
		boxrule=0.5pt,
		opacityback=0,]
		\begin{equation}\label{Pro2}
			\PP\Big\{\sup_{0\leq k\leq T}\Vert y_{a\nu}(k)-\hat y_{\hat a \hat\nu}(k)\Vert\geq\varepsilon\,|\,[a;\hat a]\Big\}\leq\lambda_2,
		\end{equation}
	\end{tcolorbox}
	where,
	\begin{align}\notag
		\lambda_2 :=\begin{cases}
			1-(1-\frac{V(a,\hat a)}{\alpha(\varepsilon)})(1-\frac{\hat\psi}{\alpha(\varepsilon)})^{T}, & ~~~\text{if}~\alpha(\varepsilon)\geq\frac{\hat\psi}{1-\kappa},\\
			(\frac{V(a,\hat a)}{\alpha(\varepsilon)})\kappa^T+(\frac{\hat\psi}{(1-\kappa)\alpha(\varepsilon)})(1-\kappa^T), & ~~~\text{if}~\alpha(\varepsilon)<\frac{\hat\psi}{1-\kappa},
		\end{cases}
	\end{align}
	with $\hat\psi\geq \rho_{\mathrm{ext}}(\Vert\hat \nu\Vert_{\infty})+\psi$, where $\alpha\in \mathcal{K}_{\infty}, 0< \kappa <1$,  $\rho_{\mathrm{ext}} \in \mathcal{K}_{\infty}\cup \{0\}$, and $\varepsilon, \psi \in\R_{\geq 0}$ as introduced in Def.~\ref{SSF}.
\end{theorem}

\begin{remark}
	Note that the closeness bounds in~\eqref{Pro1}-\eqref{Pro1_1} are specification-dependent (cf. presence of formula $\varphi$ in the statements), whereas the provided closeness guarantee in~\eqref{Pro2} is specification-free. As a result, we observe that on the one hand the closeness bound in~\eqref{Pro2} can be considered to be more general that in~\eqref{Pro1}-\eqref{Pro1_1}, however on the other it is likely to come at the cost of being more complex to compute and of being less tight. 
\end{remark}

In order to establish the presented closeness guarantee between output trajectories of $\Sigma$ and $\widehat \Sigma$ (as per~\eqref{Pro2}), 
some conditions are required~(\emph{cf.}~\cite{lavaei2018CDCJ,zamani2016approximations}), 
namely asking that the concrete model $\Sigma$ is \emph{incrementally input-to-state stable} ($\delta$-ISS), as per Definition~\ref{Def:4}. This relates to the nature of the guarantee, pertaining models' trajectories. 
In contrast, notice that the closeness guarantee in~\eqref{Pro1} does not require original systems to be $\delta$-ISS: instead, only the Lipschitz continuity of the associated stochastic kernels is required for such guarantee~\cite{SAM15}. Accordingly, the nature of the obtained guarantee is different.

On the other hand, the abstraction error presented in~\eqref{Pro1} depends on the Lipschitz constants of the stochastic kernel, 
and the error grows to infinity when the standard deviation of the noise goes to zero, which is not the case for~\eqref{Pro2}. 
Thus, whilst different in nature, the bound in~\eqref{Pro2} can practically outperform that in~\eqref{Pro1} for noises with a small standard deviation, as long as the $\delta$-ISS assumption is satisfied by the original model. In addition, recent works~\cite{SIAM17,HS19,lavaei2019NAHS1} have proposed a closeness guarantee as a version of~\eqref{Pro1} by establishing an approximate probabilistic relation between $\Sigma$ and $\widehat\Sigma$ based on a notion called $\delta$-lifting.
The proposed framework is based on constructing an $\varepsilon$-expansion 
or $\varepsilon$-contraction of the set of interest (cf. $\varepsilon$ in~\eqref{Pro2}) over the abstract system. Accordingly, the probability of satisfaction computed over the modified sets on the abstract system provides upper and lower bounds for the probability of satisfaction on the original model.

We now present a result, related to condition~\emph{(iv)} in Definition~\ref{CG}, where the probability of satisfaction of a temporal logic property over the abstract system $\widehat \Sigma$ is either a lower or upper bound for the probability of property satisfaction over the original system $\Sigma$.

\begin{theorem}
	Let $\Sigma=(X,U,\varsigma,f, Y, h)$ be a continuous-space dt-SCS  and $\widehat\Sigma=(\hat X, \hat U,\varsigma,\hat f, \hat Y, \hat h)$ be its finite abstraction. For a given LTL specification $\varphi$, and for any policy $\hat\nu(\cdot)\in\mathbb{\hat U}$ that preserves the Markov property for the closed-loop $\widehat\Sigma$, one can construct a policy of $\nu(\cdot)\in\mathbb{U}$ for $\Sigma$
	such that:
	\begin{tcolorbox}[enhanced,
		standard jigsaw,
		boxrule=0.5pt,
		opacityback=0,]
		\begin{equation}\label{Pro4}
			\PP (\widehat\Sigma_{\hat \nu}\vDash\varphi) \leq\PP(\Sigma_{\nu}\vDash\varphi).
		\end{equation}
	\end{tcolorbox}
\end{theorem}

\begin{remark}
	As \eqref{Pro4} provides a lower bound for the probability of satisfaction over $\Sigma$, it is mainly useful when one is interested in maximizing the satisfaction probability. 
	Conversely, if the goal is to minimize the probability of satisfaction, one would want to search for an upper bound of the satisfaction probability. Such an upper bound can be quantified from~\eqref{Pro4} using the negation of the specification (i.e., $\neg\varphi$) as the following:  
	\begin{equation}\label{Pro5}
		\PP(\Sigma_{\nu}\vDash\varphi) \leq 1 - \PP (\widehat\Sigma_{\hat \nu}\vDash\neg\varphi).
	\end{equation}
\end{remark}

One can employ the same approach as in~\cite[Section 6]{lavaei2018CDCJ}
and transfer the proposed closeness bound of~\eqref{Pro2} to~\eqref{Pro1} to any specification that can be accepted by a deterministic finite automaton  (DFA)~\cite{KupfermanVardi2001}. In particular, any LTL property $\varphi$ over the concrete system can be seen as the union of events over the product output space (these events can be shown to be measurable --- cf. Remark~\ref{measurable}).
For instance, for a given safe set $\mathbb{S}\subseteq Y$,
the safety property over a finite-time horizon $T$ is a subset of $ Y^{T+1}$ (with $Y^{T+1} = \prod_{i=0}^{T}Y$) indicated by the set $\mathbb{S}^{T+1}$.
For all measurable events $\mathsf{A} \subset Y^{T+1}$, one can construct an $\epsilon$-expansion and $\epsilon$-contraction of $\mathsf{A}$ over the abstract model within a given finite-time horizon $T$, as 
\begin{align*}
	\mathsf A^{\epsilon} &:=\{\{ y(k)\}_{0:T}\in Y^{T+1}\,\big|\,\exists \{\bar y(k)\}_{0:T}\in \mathsf A~\text{s.t.} \max_{k\le T}\|\bar y(k)-y(k)\|\leq\epsilon\},\\
	\mathsf A^{-\epsilon} &:= \{\{ y(k)\}_{0:T}\in Y^{T+1}\,\big|\,\forall\{ \bar y(k)\}_{0:T}\in Y^{T+1}\backslash\mathsf A,~ \max_{k\le T}\|\bar y(k)-y(k)\|>\epsilon\},
\end{align*}
where $\{ y(k)\}_{0:T} = [y(0);\dots;y(T)]$,
whose probabilities of satisfaction give respectively upper and lower bounds for the probability of satisfaction in the concrete domain with some quantified error bounds in the form of~\eqref{Pro1}.

For the sake of completeness, 
let us remark that closeness conditions in \emph{(iii)} in Definition \ref{CG} will be covered in Section 9, in the context of continuous-time SHS. 

\subsection{Literature on Similarity Relations for Stochastic Models}

There has been substantial work in the area of Formal Methods on different types of stochastic similarity relations, 
which are employed to relate the probabilistic behavior of a concrete model to that of its abstraction and have been more recently studied for continuous-space models~\cite{pp09,abate2013approximation}. 
Early on, similarity relations over finite-state stochastic systems via exact notions of probabilistic bisimulation relations have been introduced in~\cite{larsen1991bisimulation}. Leveraging probabilistic transition systems as the underlying semantic model, the article shows how a testing algorithm can distinguish, with a probability arbitrarily close to one, between processes that are not bisimilar. 
Similarity relations over finite-state probabilistic models via exact probabilistic simulation relations are also presented in~\cite{segala1995probabilistic}. 
In general, similarities are based on simulation or bisimulation relations, and can be either exact or approximate. 
Whenever the relation between a concrete model and its abstraction is symmetric, it is called ``bisimulation relation."  
Exact simulation relations require the outputs of related systems to be exactly the same, while approximate simulation relations relax this requirement by allowing the outputs to differ up to a given error term~\cite{baier2008principles,tabuada2009verification,Calin17}.

Admittedly, exact bisimulation relations raise very strong requirements amongst models, 
and in practice very limited classes of models can admit abstractions with those types of relations \cite{desharnais2008approximate,d2012robust}. 
This is particularly true for continuous-space models \cite{abate2013approximation}.  
Similarity relations of probabilistic models via approximate versions of probabilistic (bi)simulation relations are provided in~\cite{desharnais2008approximate}. 
The proposed framework is based on two-player games: the existence of a winning strategy for one of the players induces the $\epsilon$-(bi)simulation, 
and furthermore letting $\epsilon$ = 0 gives back the exact notion.
The paper also proposes a polynomial time algorithm to compute a derived metric, where the 
distance between states $s$ and $t$ is defined as the smallest $\epsilon$ such that $s$ and $t$ are $\epsilon$-equivalent.

An approximate probabilistic bisimulation relation for discrete-time Markov chains is proposed in~\cite{d2012robust}. The provided scheme exploits the structure and properties of the approximate probabilistic bisimulation and leverages the mathematical framework of Markov set-chains~\cite{hartfiel2006markov} (related to Interval MC in Section \ref{IMDPs}) in order to provide a quantified upper bound on a metric over probabilistic realizations for labeled Markov chains. 
It is shown that the existence of an approximate probabilistic bisimulation relation implies the preservation of robust PCTL formulae.

Similarity relations for models with general, uncountable state spaces have also been proposed in the more recent literature~\cite{pp09,abate2013approximation}.
These relations can depend on stability requirements, 
on model's dynamics via martingale theory~\cite{hall2014martingale}, 
or on contractivity analysis~\cite{zamani2014symbolic}. 
Notably, the work in \cite{zamani2014symbolic} argues that every stochastic control system satisfying a probabilistic variant of incremental input-to-state stability ($\delta$-ISS), and for every given precision $\varepsilon>0$, a finite-state transition system can be constructed that is $\varepsilon$-approximately bisimilar to the original stochastic control system. It also provides a closeness bound between the $\delta$-ISS stochastic control system and its bisimilar finite abstraction (cf. closeness in~\eqref{Pro3}).

Similarity relations of dt-SCS via approximate (bi)simulation relations are proposed in~\cite{desharnais2004metrics}, in which the relations enforce structural abstractions of a model by exploiting continuity conditions on its probability laws. 
Approximation metrics of stochastic processes, in particular Markovian processes in discrete time evolving on general state spaces (which are again domains with infinite cardinality and endowed with proper measurability and metric structures), are based on the notion of probabilistic bisimulation.

Labelled Markov processes (LMP) as probabilistic versions of labelled transition systems with continuous state spaces are widely discussed in~\cite{pp09} and related to dt-SCS. This book covers basic probability and measure theory on continuous state spaces and then develops the theory of LMPs. The main topics covered are bisimulation, the logical characterization of bisimulation, metrics and approximation theory. 

Probabilistic model checking of dt-SCS via finite approximate bisimulations is proposed in~\cite{abate2014probabilistic}. The paper considers
notions of (exact and approximate) probabilistic bisimulation and proposes a technique to compute an approximate probabilistic bisimulation of a dt-SCS, 
where the resulting abstraction is characterized as a finite-state Markov chain. 

\begin{figure}
	\includegraphics[width=8.5cm]{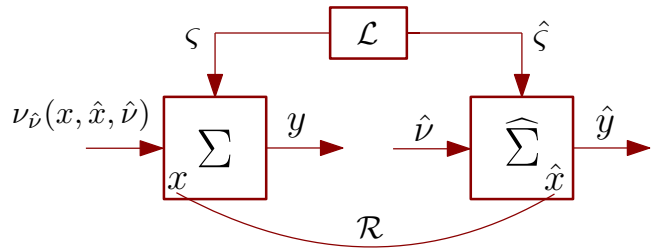}
	\caption{Notion of \emph {lifting} for specifying the similarity between a dt-SCS $\Sigma$ and its abstraction $\widehat\Sigma$.}
	\label{Fig7}
\end{figure}

A notion of approximate similarity relation based on ``lifted'' probability measures is presented in~\cite{SIAM17,lavaei2019NAHS1}, 
which is inspired by notions of similarity relations proposed in~\cite{segala1995probabilistic} for finite-state systems.  
The provided relation, underpinned by the use of metrics, allows in particular for a useful trade-off between deviations over probability distributions on states, 
and metric-based distances between model outputs. 
This new relation is inspired by a notion of simulation developed for finite-state models, and can be effectively employed over dt-SCS for both verification and synthesis purposes. 
The work also quantifies the distance in probability between the original system and its abstraction as a version of the closeness guarantee proposed in~\eqref{Pro1}. The notion of lifting  for specifying the similarity between a dt-SCS $\Sigma$ and its abstraction $\widehat\Sigma$ is schematically shown in Fig.~\ref{Fig7}. The relation $\mathcal{R}$ connects states of the two dt-SCS, and $\mathcal{L}$
specifies the relation between the two noises. The interface function $\nu_{\hat \nu}(x,\hat x,\hat \nu)$ is used for refining a policy from the abstract system to the concrete one.

These notions and results are then generalized in~\cite{haesaert2018temporal} to a larger class of temporal properties ((bounded) probabilistic reachability problems and co-safe LTL specifications)
and in~\cite{HS19} to synthesize policies for a robust satisfaction of these properties, with applications in building automation systems~\cite{HCA17}. An extension of these results to \emph{networks} of dt-SCS is presented in~\cite{lavaei2019NAHS1}, and will be discussed in more detail in Section~\ref{Network}.  

A notion of approximate probabilistic trace equivalences for both finite-state Markov processes and dt-SCS, and its relation to approximate probabilistic bisimulation, is presented in~\cite{bian2017relationship}. The proposed framework induces a tight upper bound on the approximation between finite-horizon traces, as expressed by a total variation distance. This bound can be employed to relate the closeness in satisfaction probabilities over bounded linear-time properties, 
much as in \eqref{Pro1}, 
and allows for probabilistic model checking of concrete models via their abstractions. 

An approach for computing probabilistic bisimilarity distances for finite-state probabilistic automata
has been proposed in~\cite{bacci2019computing}. The work proves that the bisimilarity distance bounds the difference in the maximal (or minimal) probability of two states to satisfy any arbitrary $\omega$-regular properties (\emph{i.e.,} namely, the notion is specification-independent).  
As expected, since the proposed results should hold for any arbitrary $\omega$-regular specification, 
it can be much more conservative and difficult to be fulfilled or checked, 
compared to establishing the previously mentioned guarantees for a given specification. 

We raise the following open challenge.
\begin{resp}
	\begin{open}
		Let $\Sigma_1$ and $\Sigma_2$ be two dt-SCSs. Develop an approach for computing the probabilistic bisimilarity distance between $\Sigma_1$ and $\Sigma_2$, satisfying any $\omega$-regular specification $\varphi$, as follows:
		\begin{equation}\notag
			|\PP(\Sigma_1\vDash\varphi) - \PP (\Sigma_2\vDash\varphi)|\le \lambda_3, \quad\text{$\forall \varphi$}.
		\end{equation} 
	\end{open}
\end{resp}\medskip

It is worth concluding this section emphasizing again that establishing stochastic similarity relations is crucial to connect the probabilistic behavior of an original SHS, which can be complex, to that of its abstraction. Consequently, by providing closeness guarantees between the output trajectories of two systems via the established stochastic similarity relations, one can perform formal analysis over the simpler abstraction and transfer the obtained results back to the original SHS.

\section{Infinite Abstractions}\label{Infinite}

\begin{figure} 
	\begin{center}
		\includegraphics[width=0.5\linewidth]{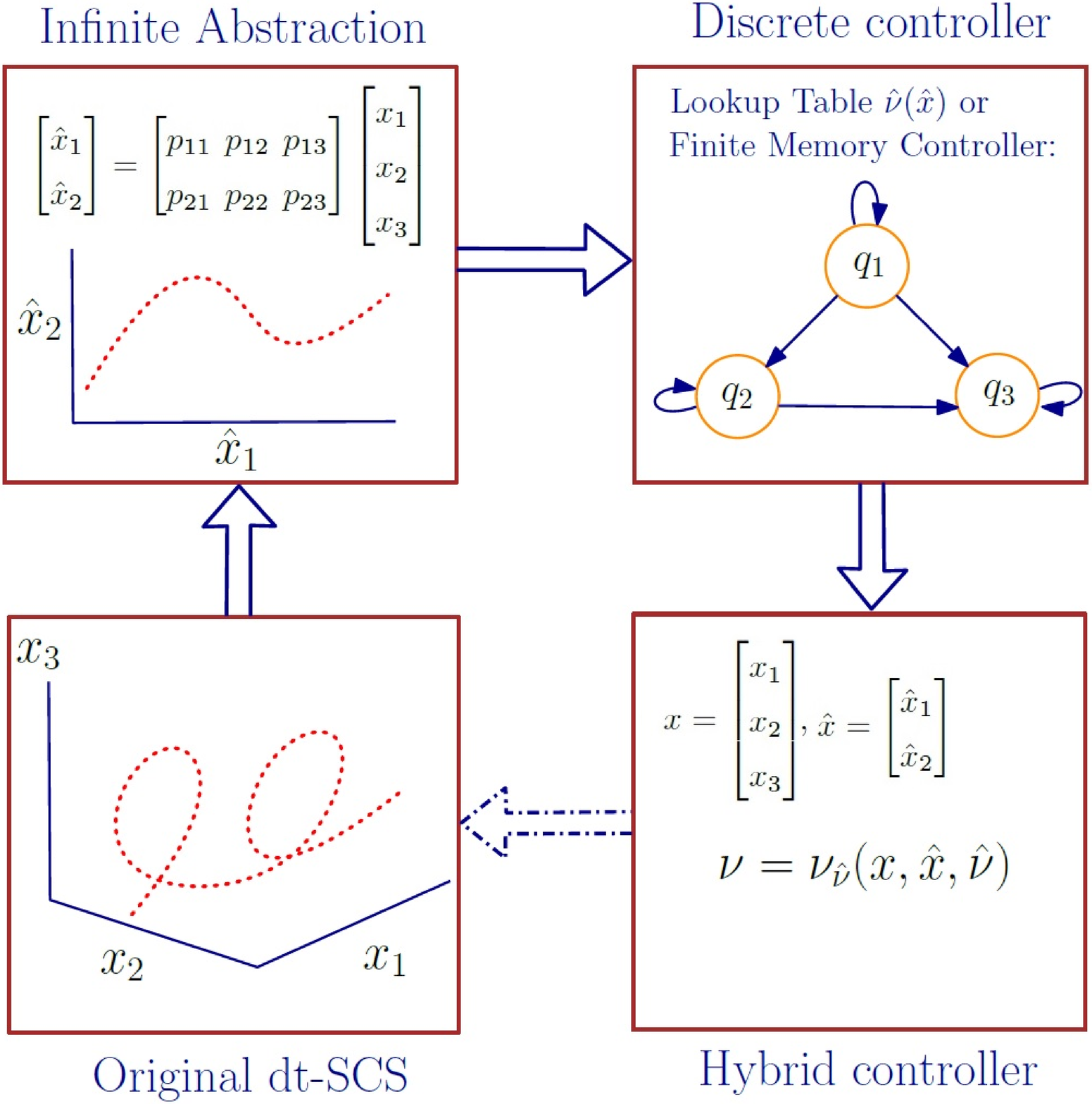} 
		\caption{\textbf{Infinite abstractions.} The original dt-SCS has a 3-dimensional state set while its abstraction has a 2-dimensional state set. This model reduction can be performed via a transformation matrix $P$ satisfying conditions~\eqref{Eq11a} and~\eqref{Eq12a}.}
		\label{Fig1_2}
	\end{center}
\end{figure}

The computational complexity associated to verifying or to synthesizing controllers for dt-SCS (and thus for SHS) models can be alleviated leveraging abstractions in two consecutive stages. 
In the first phase, the original complex systems can be abstracted by models either with simpler dynamics (e.g., linear, noiseless, etc.) or lower-dimensional state spaces (this is also known in the control literature as ``model-order reduction''~\cite{antoulas2005approximation,ionescu2015nonlinear}.
Then one can employ those simpler models (a.k.a. infinite abstractions) as a replacement of original systems, perform analysis and synthesis over those models, 
and finally refine the results back (via an interface map) over the original models. Since the mismatch between outputs of original systems and those of their infinite abstractions are formally quantified, one can guarantee that concrete systems also satisfy the specifications as abstract ones with some guaranteed error bounds. 
An example of infinite abstractions is schematically depicted in Fig.~\ref{Fig1_2}. In comparison with Fig.~\ref{Fig1}, which focuses on discretization-based techniques to obtain finite abstractions, Fig.~\ref{Fig1_2} focuses on infinite abstractions with lower-dimensional systems.  

\begin{remark}
	Infinite abstractions can take numerous shapes and forms: they can for instance be linearized versions of the original models, 
	they can be obtained via polynomial truncation, 
	they can be models with different noises (e.g., stochastic realizations~\cite{van1989stochastic}), 
	or models obtained by disregarding the noise terms \cite{zamani2014symbolic,zamani2014approximately} (cf. Theorem~\ref{IAbs} and Figure~\ref{Fig1_2}). 
	The main focus of this section is placed on infinite abstractions with lower-dimensional state spaces, which are in practice compact representations of the concrete models.   
\end{remark} 

Note that one can construct finite abstractions directly, without going through infinite abstractions first. 
However, constructing finite abstractions for high-dimensional systems can result in large, finite state spaces, 
which might not be practically viable with limited computational and memory resources. 
One of the main benefits of infinite abstractions is thus to help reducing dimensions or complexity of concrete systems, 
which can then allow leveraging finite abstractions for the reduced-order models,  
while still providing the probabilistic closeness guarantees. 

Developed earlier for continuous-time models~\cite{julius2009approximations,cA09,zamani2016approximations} and further discussed in Section \ref{CTSS}, 
the construction of infinite abstractions for discrete-time
stochastic control systems is proposed in~\cite{lavaei2017compositional} and~\cite{lavaei2018ADHSJ} and summarized in the results below. 
The abstraction framework is based on notions of stochastic simulation functions, introduced earlier (Def.~\ref{SSF}).
These functions relate output trajectories of an abstract system to those of the original one, 
such that the mismatch between the output trajectories of two systems remains within some guaranteed error bound. 
Through these stochastic simulation functions it is possible to quantify the probabilistic distance between the original stochastic system and its abstraction, 
based on the closeness in~\eqref{Pro2}. 
The aforementioned work also focuses on a class of discrete-time \emph{linear} stochastic control systems, 
as in~\eqref{linear} and further detailed next, 
and proposes a computational scheme to construct infinite abstractions together with their corresponding stochastic simulation functions.

Consider the class of discrete-time linear stochastic control system (a special instance of dt-SCS), as 
\begin{align}\label{linear}
	\Sigma\!:\left\{\hspace{-1.5mm}\begin{array}{l}x(k+1)=Ax(k)+B\nu(k)+R\varsigma(k),\\
		y(k)=Cx(k),\end{array}\right.
\end{align}
where the additive noise $\varsigma(k)$ is a sequence of independent random vectors with multivariate standard normal distributions. We use the tuple $\Sigma=\left(A,B,C,R\right)$ to refer to the class of linear systems in \eqref{linear}.
In the next theorem, we establish a formal relation between $\Sigma$ and its reduced-order model $\widehat\Sigma$, by constructing corresponding matrices $\hat A,\hat B,\hat C,\hat R$.

\begin{theorem}\label{IAbs}
	Let $\Sigma=(A,B,C,R), \widehat \Sigma=(\hat A,\hat B,\hat C,\hat R)$ be two linear dt-SCS with independent additive noises. Suppose there exist a matrix
	$K$ and a positive-definite matrix $M$ such that the following matrix inequalities
	\begin{align}\label{Eq13a}
		&C^TC\preceq M, \\
		\big((1+\pi)(A+BK&)^TM(A+BK)-M\big)\preceq -\hat\kappa M, \label{Eq14a}
	\end{align}
	hold for some constants $0<\pi$ and $0<\hat\kappa<1$. If further
	\begin{align}\label{Eq11a}
		AP&=P\hat A-BQ,\\\label{Eq12a}
		CP&=\hat C,
	\end{align}
	hold for some matrices $Q$ and $P$ of appropriate dimension, 
	then there exists a quadratic SSF $V(x,\hat x)$~\cite{lavaei2017compositional} between $\Sigma$ and  $\widehat\Sigma$ as
	\begin{align}\
		V(x,\hat x)=(x-P\hat x)^TM(x-P\hat x), \label{Eq10a}
	\end{align}
	where $P\in\R^{n\times\hat n}$ is a matrix of an appropriate dimension with $\hat n$ being the dimension of the reduced-order model $\widehat\Sigma$.
\end{theorem}

The stochastic simulation function $V(x,\hat x)$ in \eqref{Eq10a} gives a probabilistic closeness guarantee between the original dt-SCS $\Sigma$ and its infinite abstraction $\widehat\Sigma$, as per~\eqref{Pro2}.

\begin{remark}
	Condition~\eqref{Eq11a} holds as long as condition (V.18) in~\cite{zamani2017compositionalMurat} is satisfied. 
	In addition, notice that the results in Theorem~\ref{IAbs} do not impose any condition on the matrix $\hat B$, which thus can be chosen arbitrarily. 
	As an example, one can select $\hat B=\mathds{I}_{\hat n}$, which renders the abstract system $\widehat \Sigma$ fully actuated and, hence, can facilitate a subsequent synthesis task.
	
	Notice further that the matrix $\hat R$ can be also chosen arbitrarily. 
	In this case, the probabilistic closeness between two systems $\Sigma$ and $\widehat\Sigma$ can be quantified as $\lambda_2$ in~\eqref{Pro2}, where
	$$
	\psi = \textsf{Tr}\big(R^TMR+\hat R^TP^TMP\hat R\big). 
	$$
	One can readily verify that selecting $\hat R = 0$ results in a tighter relationship between the original system $\Sigma$ and its infinite abstraction $\widehat\Sigma$. 
	However, observe that this is not the case when the noises of the concrete system and of its infinite abstraction are the same, 
	as assumed in~\cite{zamani2014compositional,zamani2016approximations}, in where $\hat R$ can be chosen appropriately to minimize the error term.  
\end{remark}

The construction of infinite abstractions for dt-SCS is also discussed in~\cite{lavaei2018CDCJ}. 
The proposed approach employs the notion of stochastic storage function (a variant of the stochastic simulation function in Definition \ref{SSF}) between a concrete system and its abstraction, which allows to provide a closeness guarantee as in~\eqref{Pro2}.
This work also focuses on a specific class of discrete-time \emph{nonlinear} stochastic systems by adding $E\tilde \varphi(Fx(k))$ to~\eqref{linear} in which $E\in\mathbb R^{n\times 1}$, $F\in\mathbb R^{1\times n}$, and $\Upsilon:\R\rightarrow\R$ is the nonlinearity term
satisfying a slope restriction as
\begin{align}\label{slop}
	0\leq\frac{\Upsilon(c)-\Upsilon(d)}{c-d}\leq b,
\end{align} 
for any $c,d\in\R,c\neq d,$ for some $b\in\R_{>0}\cup\{\infty\}$,     
and proposes a construction scheme for building infinite abstractions together with their corresponding stochastic storage functions.

It is worth mentioning that the contributions in~\cite{lavaei2017compositional,lavaei2018CDCJ} do not raise any restrictions on the sources of uncertainty in the concrete and abstract systems (\emph{i.e.,}  the noise of the abstraction can be completely independent of that of the concrete system). In particular, the results provided in~\cite{lavaei2017compositional,lavaei2018CDCJ} are more general than~\cite{zamani2016approximations}, where the noises in the concrete and abstract systems are assumed to be the same, which practically means the abstraction has access to the noise of the concrete system. The results in~\cite{lavaei2017compositional,lavaei2018CDCJ} provide a closeness guarantee between output trajectories of $\Sigma$ and $\widehat \Sigma$ as in~\eqref{Pro2}.

To provide a broader context, approximations of large-scale dynamical systems in the context of model-order reduction are studied in~\cite{antoulas2005approximation} by combining system theory with numerical linear algebra. A notion of moment matching is presented in~\cite{ionescu2015nonlinear}, which discussed a family of (nonlinear) parametrized reduced-order models that achieve moment matching.

A general framework for structure-preserving model reduction of a second-order
network system based on graph clustering is studied in~\cite{cheng2017reduction}, where the dissimilarities of vertices are quantified by the
$\mathcal H_2$-norms of the transfer function discrepancies. An $\mathcal H_2$ sub-optimal model reduction for second-order network systems is proposed in~\cite{yu2019},
and an extension
is recently presented in~\cite{yu2021h_2}, in which the main objective is to find a reduced-order model that not only approximates the input-output mapping of the original
system but also preserves crucial model structure.
\begin{remark}
	Note that the model-order reduction techniques in~\cite{yu2019,yu2021h_2,cheng2017reduction} deal with models in the frequency domain, and their main goal is to establish a closeness relation between the transfer function of the original system and of its reduced-order model by providing  closeness guarantees based on the $\mathcal H_2$ norm. Since studies in the frequency domain are mainly developed for stability and input-output behaviour, handling more complex logical  properties (such as the discussed safety, reachability, etc.) via those techniques is not straightforward. In comparison, the discussed  infinite-abstraction techniques concerning models in the time domain can readily be employed to study verification and synthesis problems over  logical specifications.  
\end{remark}

{\bf Running example (continued)}. We consider the above running example, concerning a two-dimensional model, and now aim at constructing a one-dimensional infinite abstraction (\emph{i.e.,} a proper reduced-order model), by satisfying conditions~\eqref{Eq13a}-\eqref{Eq12a}. The two-dimensional room temperature regulation model is given by
	\begin{equation*}
		\Sigma\!:\left\{\hspace{-1.5mm}\begin{array}{l}{T}(k+1)=A{T}(k)+\gamma T_{h}\nu(k)+ \theta T_{E}+R\varsigma(k),\\
			y(k) = CT(k),\end{array}\right.
	\end{equation*}
	where: 
	\begin{align}\notag
		&A=\begin{bmatrix}1-2\sigma-\theta& \sigma \\ 
			\sigma & 1-2\sigma-\theta\end{bmatrix}\!\!,  T_E=[T_{e_1};T_{e_2}], T(k)=[T_1(k);T_2(k)],  \nu(k)=[\nu_1(k);\nu_2(k)],\varsigma(k)=[\varsigma_1(k);\varsigma_2(k)].
	\end{align}
	Moreover, $R = 0.01\mathds{I}_2$, $C = \mathds{1}^T_2$, $T_{e_i}=-1\,^\circ C$, $i \in\{1,2\}$, $T_h=50\,^\circ C$, $\theta = 0.4$, $\gamma = 0.5$, and $\sigma = 0.1$ (the latter is a conduction factor between the two rooms). The  goal is to construct a one-dimensional infinite abstraction $\widehat\Sigma$ from $\Sigma$ by satisfying conditions~\eqref{Eq13a}-\eqref{Eq12a}, which can be met by selecting 
	\begin{align*}
		& M = \mathds{I}_2, ~ P =  \mathds{1}_2, ~ K = \mathbf{0}_{2\times 2}, ~ Q =  \mathds{1}_2, \hat A = 25.5, ~ \hat C = 1, ~ \pi = 1, ~ \hat \kappa = 0.34.
	\end{align*}
	Then, there exists a quadratic SSF $V(x,\hat x)$ between $\Sigma$ and  $\widehat\Sigma$, as in~\eqref{Eq10a}. By taking $\hat R = 0.01$ and the initial states of the two models $\Sigma$ and $ \widehat \Sigma$ to be equal to $20$, and using the bound in~\eqref{Pro2}, one can guarantee that the distance between the outputs of $\Sigma$ and $\widehat \Sigma$ does not exceed $\varepsilon = 1$ over the time horizon $T_d=100$, with a probability of at least $95\%$, \emph{i.e.,} 
	\begin{equation*}
		\mathbb P\Big\{\Vert y(k)-\hat y(k)\Vert\le 1,\,\, \forall k\in[0,100]\Big\}\ge 0.95.
	\end{equation*}
	One can utilize the obtained reduced-order model and construct a finite abstraction for later verification and synthesis purposes - this goal will be further discussed in the next section.\hfill \qed

\section{Finite Abstractions}\label{Finite}

In the second phase of the abstraction procedure (cf. Fig.~\ref{Fig1}),
one can construct finite abstractions usually in the form of finite Markov decision processes (MDPs). 
These abstractions are approximate descriptions of (reduced-order) systems, in which each discrete state corresponds to a set of continuous states of the (reduced-order) systems. Since the obtained abstractions are finite, one can employ algorithmic machineries and existing software tools to automatically synthesize controllers, which can then be applied (refined) over the concrete models, thus enforcing complex properties, including specifications expressed as temporal logical formulae. 

A concrete model $\Sigma$ is approximated by a \emph{finite} $\widehat\Sigma$ using Algorithm~\ref{algo:MC_app}. For the sake of an easier presentation, we present the construction algorithm just for dt-SCS, however we refer the interested reader to~\cite{ZTA17,tkachev2017quantitative,SIAM17} for related, more general SHS, 
and to~\cite{ZA14,ZAG15,lavaei2019HSCC_J} for the construction of finite MDPs for a class of SHS namely stochastic \emph{switched} systems. 
To construct such a finite approximation, the state and input sets (over which one is interested to perform analysis and synthesis) of the \mbox{dt-SCS} $\Sigma$ are restricted to be compact.\footnote{This compactness assumptions can be relaxed, albeit at the cost of additional (but quantifiable) approximation errors, as discussed in \cite{bcSA14,SA15b}.} The rest of the state space can be considered as a single absorbing state. Algorithm~\ref{algo:MC_app} first constructs a finite partition of the state set $X = \cup_i \mathsf X_i$ and the input set $U = \cup_i \mathsf U_i$. Then arbitrary ``representative points'' $\bar x_i\in \mathsf X_i$ and $\bar \nu_i\in \mathsf U_i$ are selected as abstract states and inputs.
Transition probabilities in the finite MDP $\widehat\Sigma$ are computed according to \eqref{eq:trans_prob}. The output map $\hat h$ is the same as $h$ with its domain restricted to the finite set $\hat X$ (cf. Step \ref{step:output_map}) and the output set $\hat Y$ is the image of $\hat X$ under $h$ (cf. Step \ref{step:output_space}).

\begin{algorithm}[ht!]
	\caption{Approximation of a dt-SCS $\Sigma$ by a finite MDP $\widehat\Sigma$}
	\label{algo:MC_app}
	\begin{center}
		\begin{algorithmic}[1]
			\REQUIRE 
			Input dt-SCS $\Sigma=(X,U,T_{\mathsf x},Y,h)$
			\STATE
			Select finite partitions of sets $X,U$ as $X = \cup_{i=1}^{n_{\bar x}} \mathsf X_i$, $U = \cup_{i=1}^{n_{\bar\nu}} \mathsf U_i$
			\STATE
			For each $\mathsf X_i$, and $\mathsf U_i$, select single representative points $\bar x_i \in \mathsf X_i$, $\bar\nu_i \in \mathsf U_i$
			\STATE
			Define 
			$\hat X := \{\bar x_i, i=1,...,n_{\bar x}\}$ as the finite state set of MDP~$\widehat\Sigma$ with the finite input set
			$\hat U := \{\bar\nu_i, i=1,...,n_{\bar\nu}\}$
			\STATE
			\label{step:refined}
			Define the map $\Xi:X\rightarrow 2^X$ that assigns to any $x\in X$, the corresponding partition set it belongs to, \emph{i.e.,}
			$\Xi(x) = \mathsf X_i$ if $x\in \mathsf X_i$ for some $i\in\{1,2,\ldots,n_x\}$
			\STATE
			Compute the discrete \emph{transition probability matrix} $\hat T_{\mathsf x}$ for $\widehat\Sigma$ as:
			\begin{equation}
				\label{eq:trans_prob}
				\hat T_{\mathsf x} (x'|x,\nu) 
				= T_{\mathsf x} (\Xi(x')|x,\nu),
			\end{equation}
			for all $x,x'\in \hat X, \nu\in \hat U$
			\STATE
			\label{step:output_space}
			Define the output space $\hat Y := h(\hat X)$
			\STATE
			\label{step:output_map}
			Define the output map $\hat h := h|_{\hat X}$
			\ENSURE
			Output finite MDP $\widehat\Sigma = (\hat X, \hat U, \hat T_{\mathsf x}, \hat Y, \hat h)$
		\end{algorithmic}
	\end{center}
\end{algorithm}

Given a dt-SCS $\Sigma=(X,U,\varsigma,f,Y,h)$,
the finite MDP $\widehat\Sigma$ constructed in Algorithm~\ref{algo:MC_app} can be represented as
\begin{equation}
	\label{eq:abs_tuple}
	\widehat\Sigma =(\hat X,\hat U, \varsigma,\hat f, \hat Y, \hat h),
\end{equation}
where $\hat f:\hat X\times\hat U\times \mathcal{V}_\varsigma\rightarrow\hat X$ is defined as
\begin{equation}\label{Abstraction Map}
	\hat f(\hat{x},\hat{\nu},\varsigma) = \Pi_{x}(f(\hat{x},\hat{\nu},\varsigma)),	
\end{equation}
and $\Pi_x:X\rightarrow \hat X$ is the map that assigns to any $x\in X$, the representative point $\bar x\in\hat X$ of the corresponding partition set containing $x$.
The initial state of $\widehat\Sigma$ is also selected according to $\hat x_0 := \Pi_x(x_0)$, with $x_0$ being the initial state of $\Sigma$. 

The dynamical representation of the abstract finite MDP $\widehat\Sigma$ employs the map $\Pi_x:X\rightarrow \hat X$,  
which satisfies the inequality
\begin{equation}
	\label{eq:Pi_delta}
	\Vert \Pi_x(x)-x\Vert \leq \delta,\quad \forall x\in X,
\end{equation}
where $\delta:=\sup\{\|x-x'\|,\,\, x,x'\in \mathsf X_i,\,i=1,2,\ldots,n_x\}$ is the state discretization parameter.

\begin{figure*}[ht!]
	\begin{center}
		\includegraphics[width=0.33\linewidth]{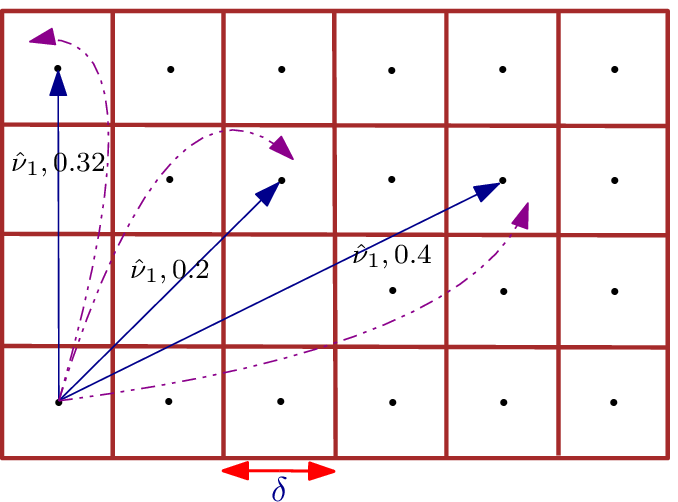} \hspace{0.7cm}
		\includegraphics[width=0.33\linewidth]{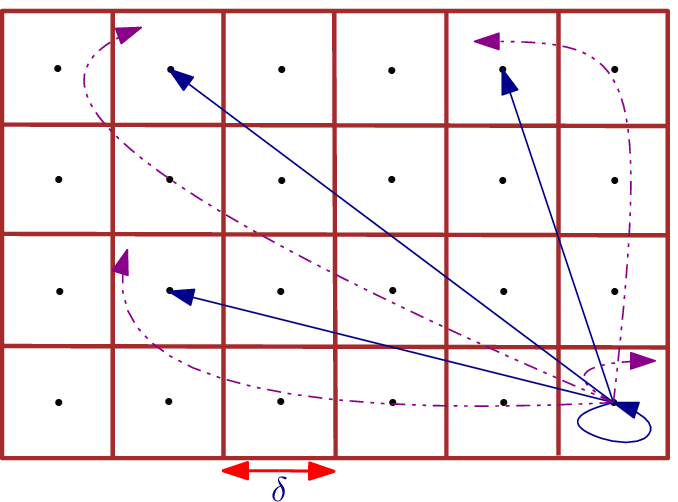} 
		\caption{Construction of finite MDPs: A grid is first overlaid on the state and input sets. The center of each cell is considered as a representative point, and the transition probability (\emph{i.e.,} probability of jumping from each representative point in a cell to all other cells) for all possible discrete inputs is computed. By repeating the process and storing the probabilities in a matrix called \emph{transition probability matrix}, the corresponding finite MDP is accordingly constructed.}
		\label{Fig4}
	\end{center}
\end{figure*}

\begin{remark}\label{rem:abs_errors}
	Observe that the state discretization parameter $\delta$ appears in the probabilistic closeness quantified in~\eqref{Pro1}-\eqref{Pro2}: 
	thus, one can decrease the error by reducing the state discretization parameter, namely by aptly refining the state partitions. 
	Notice that there is no requirement on the shape of the partition elements in constructing the finite MDPs. For the sake of an easier implementation, 
	one can for instance consider partition sets as hyper-boxes, and representative points as centers of each box (cf. Fig.~\ref{Fig4}).  
	The errors and guarantees derived above provide flexibility and have been embedded in a few software tools generating finite abstractions, cf.~\cite{FAUST15,lavaei2020AMYTISS}.  
\end{remark}

\subsection{Abstractions for Finite-Horizon Specifications}

The construction of finite abstractions of SHSs has been initially proposed in~\cite{AKLP10} and used for formal verification and synthesis. This work investigates probabilistic safety and reachability over a finite-time horizon for a general class of discrete-time SHS with control inputs. 
The proposed framework characterizes the set of initial conditions providing a certain probabilistic guarantee that the system will keep evolving within a desired `safe' region of the state
space in terms of a value function, and determines `maximally safe' Markov policies via dynamic programming over the finite abstract MDP. 
An improved gridding scheme, which is adaptive and sequential, for the abstraction and verification of stochastic processes is proposed in~\cite{SA13} (cf. Remark \ref{rem:abs_errors}). 
The abstract model is constructed as a Markov chain using an adaptive gridding algorithm that conforms to the underlying dynamics of the model and thus mitigates the curse of dimensionality unavoidably related to the partitioning procedure. 
The work focuses on the study of a particular specification (probabilistic safety or invariance, over a finite horizon) and the results are then extended to SHS models with hybrid state spaces. The closeness guarantee between the original SHS and their finite abstractions is in the form of~\eqref{Pro1}. 

The above results in general rely on Lipschitz continuity of the stochastic kernel associated with the system. The works \cite{SA12,SATAC12} extend the method for systems with discontinuous stochastic kernels and provide error bounds for inequalities of the form \eqref{Pro1}. On the other hand, if the kernel admits higher-order derivatives, 
refined computations are proposed in~\cite{SAH12}, with errors that naturally depend on higher orders of the discretization parameter $\delta$ and that can show faster convergence to zero. 

Among the many applications of these formal abstractions, 
the approach is employed to aggregate modeling and control of thermostatically controlled loads, as in~\cite{soudjani2014formal}, and in building automation~\cite{cCA18}. In the aggregation procedure, each thermostatically controlled load model in the population, in principle similar to the Running Example in this paper, is formally abstracted as a finite MDP, and the cross product of these MDPs is lumped into its coarsest (exact) probabilistic bisimulation, and can be used for predictive energy scheduling. The abstraction procedure allows for the quantification of the induced error in the form of~\eqref{Pro1}.

The construction of finite abstractions for stochastic control systems is presented in~\cite{SAM15,SAM17}. These studies investigate the problem of finite-horizon probabilistic invariance for dt-SCS by providing a closeness guarantee between two systems in the form of~\eqref{Pro1}. The proposed approach is more general than~\cite{lavaei2018ADHSJ,lavaei2017HSCC}, since the provided framework does not require original systems to be $\delta$-ISS. On the other hand, the abstraction error in~\cite{SAM15,SAM17} depends on the Lipschitz constants of the stochastic kernels associated with the system, and accordingly, it grows to infinity as the standard deviation of the noise goes to zero, which is not the case in~\cite{lavaei2018ADHSJ,lavaei2017HSCC}. 

A method to generate finite Markovian abstractions for discrete-time linear stochastic systems.
The proposed approach
proceeds by approximating the transition probabilities from one partition set
to another by calculating the probability from a single representative point in the first region.
The work employs an adaptive refinement algorithm that takes advantage of the dynamics of the system to achieve a desired error value. 
The proposed approach is similar to that of~\cite{SA13} with a closeness guarantee in the form of~\eqref{Pro1}, however here the transition probabilities are averaged over partition sets.

The construction of finite abstractions for discrete-time stochastic systems is also pursued in~\cite{lavaei2017HSCC}.
Focusing on a specific class of linear dt-SCS, these results employ notions of stochastic simulation (or storage) functions by providing a probabilistic distance between the interconnection of stochastic control subsystems and that of their finite abstractions
based on~\eqref{Pro2}.

The construction of finite MDPs for stochastic systems that are not necessarily stabilizable is presented in~\cite{lavaei2019NAHS}.
The proposed frameworks rely on a relation between a system and its finite abstraction employing a new notion called \emph{finite-step} stochastic simulation. In comparison with the existing notions of simulation functions in which stability or stabilizability of each subsystem is required, a \emph{finite-step} stochastic simulation function needs to decay only after some finite numbers of steps (rather than at each time step). This results in a less conservative approach in the sense that one can compositionally construct finite MDPs such that stabilizability of each subsystem is not necessarily required. The work in~\cite{lavaei2019NAHS} provides a closeness guarantee between output trajectories of $\Sigma$ and $\widehat \Sigma$ as per~\eqref{Pro2}.

The construction of finite abstractions for stochastic \emph{switched} systems is presented in~\cite{lavaei2019HSCC_J,lavaei2020LSS_J}.
The transition map switches between a finite set of modes and the switched system accepts multiple Lyapunov (or storage) functions with a dwell-time condition that puts a lower bound on the interval between two consecutive switching time instants. The dwell-time is deterministic and always met by the controller designed using the finite MDP. 
In particular, switching signals in those works are control inputs and the main goal is to synthesize them with a specific dwell-time, such that the output of original systems satisfies some high-level specifications, such as safety, reachability, etc. Those works utilize notions of stochastic simulation (or storage) functions and provide a closeness guarantee in the form of~\eqref{Pro2} but adapted to the switched setup.   
These works also show that under standard assumptions ensuring incremental input-to-state stability
of switched systems similar to Definition~\ref{Def:4} (\emph{i.e.,} existence of common incremental Lyapunov (or storage) functions, or multiple incremental Lyapunov (or storage) functions with some dwell-time conditions), one can construct finite MDPs for nonlinear stochastic switched systems. These results also propose an approach to construct finite MDPs together with their corresponding stochastic simulation (or storage) functions for a particular class of nonlinear stochastic switched systems whose nonlinearity $\Upsilon$ satisfies either a slope restriction similar to~\eqref{slop},
or an incremental quadratic inequality as
\begin{align}
	\begin{bmatrix}\label{quadratic_constraint}
		d_2-d_1\\\Upsilon_{p}(k,d_2)-\Upsilon_{p}(k,d_1)
	\end{bmatrix}^T\!\!\bar Q_p\begin{bmatrix}
		d_2-d_1\\\Upsilon_{p}(k,d_2)-\Upsilon_{p}(k,d_1)
	\end{bmatrix}\geq 0,
\end{align}
for all $k\in\mathbb N$, $d_1, d_2 \in \mathbb R$, for all switching modes $p\in P  = \{1,\dots, m \}$, and for all $\bar Q_p \in \mathcal{\bar Q}_p$, where $\mathcal{\bar Q}_p$ is the set of symmetric matrices referred to as ``incremental multiplier'' matrices. 
For this class of nonlinear systems, the aforementioned incremental stability property can be readily checked via matrix inequalities. The quadratic inequality in~\eqref{quadratic_constraint} is called ``incremental,'' as the difference between $d_1,d_2$ and their functions $\Upsilon_{p}(k,d_1), \Upsilon_{p}(k,d_2)$ appears in the inequality.

Abstraction-based synthesis of general MDPs using approximate probabilistic relations is proposed in~\cite{lavaei2019NAHS1}.
The abstraction framework is based on the notion of $\delta$-lifted relations, which is similar to \cite{SIAM17}, using which one can quantify the distance in probability between dt-SCS and that of their abstractions as a version of the closeness guarantee proposed in~\eqref{Pro1}. The works focus on a class of stochastic nonlinear dynamical systems and construct their (in)finite abstractions using both model order reduction and state space discretization.  

A general abstraction technique for verifying safety problems for probabilistic hybrid systems is proposed in~\cite{zhang2010safety}. Safety verification of linear discrete-time stochastic systems over bounded and unbounded time horizons is studied in~\cite{lal2020safety}. For bounded safety verification, the work reduces the problem to the satisfiability of a semidefinite programming problem, whereas for the unbounded safety verification, the paper proposes an abstraction procedure to reduce the safety problem to that of a finite graph, wherein, the nodes of the graph correspond to the regions of a partition of the state space.
A counterexample-guided abstraction refinement algorithm for a subclass of probabilistic hybrid systems, called polyhedral probabilistic hybrid systems, is proposed in~\cite{lal2019counterexample}, where the continuous dynamics are specified using a polyhedral set within which the derivatives of the continuous executions lie.

An abstraction-based reachability analysis for finite
stochastic hybrid systems is studied in~\cite{zhang2017abstraction}.  The work addresses the problem of computing the probability
of reaching a desired set in a subclass of SHS,
wherein the stochasticity arises from the randomness of the initial
distribution of continuous states, and the probabilistic transitions in
the underlying finite-state Markov chain. Hierarchical abstractions for reachability analysis of probabilistic hybrid systems are proposed in~\cite{lal2018hierarchical}, in which discrete and probabilistic
dynamics are captured using finite-state MDPs, and the continuous dynamics are modeled by annotating the states of the MDP with differential equations and inclusions. An abstraction-based framework to check probabilistic specifications of
MDPs using stochastic two-player game abstractions is proposed in~\cite{kattenbelt2009verification}. The work also proposes a four-valued PCTL semantics for the developed game abstractions. A counterexample-guided abstraction refinement technique for the automatic verification of probabilistic systems is proposed in~\cite{hermanns2008probabilistic11}. A survey on various abstraction-based
techniques of probabilistic systems is presented in~\cite{dehnert2012abstraction}.

\begin{figure}
	\centering
	\includegraphics[width=7.7cm]{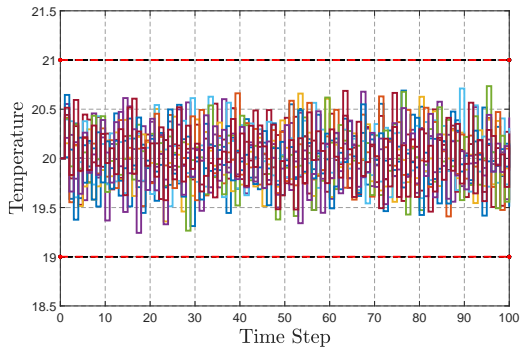}
	\caption{Closed-loop state trajectories with $10$ different noise realizations for the finite time horizon $T_d=100$.}
	\label{Sim1}
\end{figure}

\begin{figure}
	\centering
	\includegraphics[width=7.7cm]{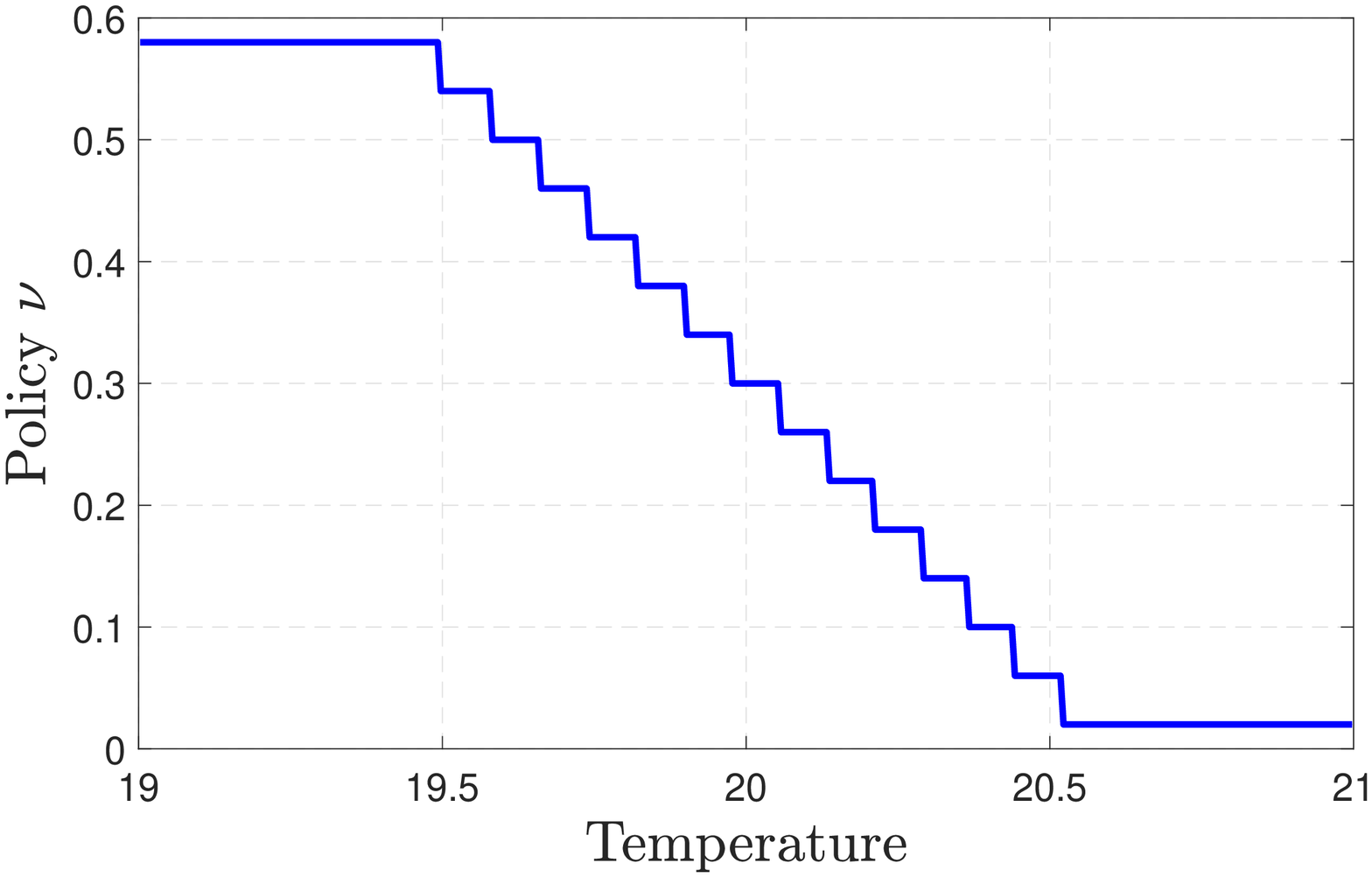}
	\caption{Synthesized policy as a function of state (room temperature).}
	\label{Sim3}
\end{figure}

{\bf Running example (continued).} We construct a finite MDP from the model in~\eqref{RE} according to Algorithm~\ref{algo:MC_app}, with the state discretization parameter $\delta = 0.005$. By taking the initial states of the two models $\Sigma$ and $ \widehat \Sigma$ to be equal to $20$, and using the proposed bound in~\eqref{Pro2}, one can guarantee that the distance between the outputs of $\Sigma$ and $\widehat \Sigma$ does not exceed $\varepsilon = 0.5$ over the time horizon $T_d=100$, with a probability of at least $98\%$, \emph{i.e.,} 
\begin{equation}
	\label{eq:guarantee}
	\mathbb P\Big\{\Vert y(k)-\hat y(k)\Vert\le 0.5,\,\, \forall k\in[0,100]\Big\}\ge 0.98.
\end{equation}
Let us now synthesize a controller, where $U = [0,0.6]$, for $\Sigma$ via its finite abstraction $\widehat \Sigma$, such that for $\Sigma$ the temperature of the room remains in the safe region $[19,21]$.  
This is attained by employing the software tool \textsf{AMYTISS}~\cite{lavaei2020AMYTISS}.
The synthesized policy of the room  as a function of state is illustrated in Fig.~\ref{Sim3}. 
Closed-loop state trajectories describing the dynamics of the room temperature over the finite-time horizon $T_d=100$, under $10$ different noise realizations, are illustrated in Fig.~\ref{Sim1}. 
We remark that the synthesized concrete policy for this example is simply chosen as $\nu = \hat \nu$, which is a special case of the interface function discussed in Remark~\ref{Interface}.

In order to better understand the provided probabilistic bound in~\eqref{eq:guarantee}, we also run Monte Carlo simulation of $10000$ runs. One expects that the distance between the outputs of $\Sigma$ and $\widehat \Sigma$ is always less than or equal to $0.1$ with a probability at least of $98\%$, and indeed this bound is easily matched in practice. Indeed, we expected the empirical outcomes to be tighter, due to the conservative nature of Lyapunov-like techniques (simulation functions) and associated error bounds.  

We now further elaborate the running example and quantify other closeness bounds discussed in Section~\ref{SSR}. One can compute the closeness bound $\lambda_1$ in~\eqref{Pro1} with $\delta = 0.005$, $T_d=100$, $\sigma = 0.6$, $\mathscr{H} = \dfrac{2|a|}{\sigma\sqrt{2\pi}} = 0.39$ (cf. Remark~\ref{Lipschitz}) as 
\begin{equation}\label{eq:guarantee_1}
	|\PP(\Sigma_{\hat \nu}\vDash\varphi) - \PP (\widehat\Sigma_{\hat \nu}\vDash\varphi)|\le 0.19.
\end{equation}
According to Remark~\ref{Lipschitz}, one can quantify $\mathscr{\bar H} = \dfrac{2(|a| + |\gamma T_{h}|)}{\sigma\sqrt{2\pi}} = 17.02$, and accordingly, compute the closeness bound $\bar \lambda_1$ in~\eqref{Pro1_2} as $8.51$.
Similarly, one can readily compute the proposed closeness bound in~\eqref{Pro1_1} as $17.02$.
As can be observed, the obtained closeness error bounds from~\eqref{Pro1_2}-\eqref{Pro1_1} are vacuous and even the closeness bound in~\eqref{eq:guarantee_1} is more conservative than the one from~\eqref{eq:guarantee}. This issue is expected and the main reason is that closeness bounds~\eqref{Pro1}-\eqref{Pro1_1} do not require the concrete model to be $\delta$-ISS (which is the case in~\eqref{Pro2}) but rather only the Lipschitz continuous. In addition, the closeness bounds~\eqref{Pro1}-\eqref{Pro1_1} do not require the original and abstract systems to share the same source of stochasticity (which is the case in~\eqref{Pro2}), but this comes at the cost of providing more conservative closeness guarantees. One can ready improve the closeness bounds from~\eqref{Pro1}-\eqref{Pro1_1} using an abstraction refinement approach. \hfill\qed

\subsection{Abstractions for Infinite-Horizon Specifications}

The construction of finite Markov chains for discrete-time stochastic models with continuous state spaces and their use to verify infinite-horizon properties (\emph{e.g.,} safety and reachability specifications) is proposed in~\cite{tkachev2011infinite,cTA12,cTA12b,TA14}.  
The proposed approaches employ notions of stochastic bisimulation functions and provide a lower bound for infinite-time probabilistic invariance 
(cf. equations~\eqref{Pro4} and~\eqref{Pro5})  
by decomposing this property into a finite-time reach-avoid together with an infinite-time invariance 
around absorbing sets (cf. Definition~\ref{ABS} below) over the state space of the model. 

A quantitative abstraction-based controller synthesis for SHS is discussed in~\cite{tmka2013} and later extended in~\cite{TMKA17}. The problem is reformulated as an optimization of a probabilistic reachability property over a product process (known as product automaton), which is obtained from the model of the specification and that of the system. The work develops a discretization procedure, which results in a standard synthesis problem over Markov decision processes with history-independent Markov policies, 
with errors of the form \eqref{Pro1} and \eqref{Pro1_2}), respectively.

The satisfaction probability of infinite-horizon properties is theoretically investigated in~\cite{TA14}. 
Extending to control-dependent models in~\cite{TMKA17}, 
it is shown that the satisfaction probability depends on the existence of \emph{absorbing sets}, as defined next. 
\begin{definition}\label{ABS}
	The set $\mathcal A\in\mathcal B (X)$ is called (weakly) absorbing if there exists a randomized selector $\bar \mu$ such that for all $x\in\mathcal A$, it holds that $\bar\mu(U\,\big|\,x)=1$ and	
	\begin{align}\notag
		\int_U T_{\mathsf x} (\mathcal A\,\big|\,x,\nu)\bar\mu(\mathsf{d}\nu\,\big|\,x)=1.
	\end{align} 
	We say that the set $\mathcal A\in\mathcal B (X)$ is simple if it does not have non-empty (weakly) absorbing subsets.
\end{definition}

Employing Definition~\ref{ABS}, the following theorem is proposed in~\cite{TMKA17}.
\begin{tcolorbox}[enhanced,
	standard jigsaw,
	boxrule=0.5pt,
	opacityback=0,]
	\begin{theorem}\label{ABS1}
		The infinite-horizon safety probability for a continuous-space dt-SCS and a compact safe set $\mathcal S$ is equal to zero over the entire set if and only if the safe set $\mathcal S$ does not contain any absorbing sets. 
	\end{theorem}
\end{tcolorbox}

If the underlying system is a finite MDP, the simple absorbing sets in Definition~\ref{ABS} are bottom strongly connected components (BSCCs) of the MDP, 
and computing these BSCCs is straightforwardly done by graph search algorithms. 
Conversely, no computational method is proposed in the literature for finding absorbing sets of continuous systems. Motivated by Definition~\ref{ABS} and Theorem~\ref{ABS1}, we present the following open problem, which would allow to expand beyond the results in~\cite{TA14,TMKA17}.

\begin{resp}
	\begin{open}
		Given a dt-SCS with continuous-state space, compute its absorbing sets, or to compute over- and under-approximations of such absorbing sets within a-priori precision.  
	\end{open}
\end{resp}\medskip

An alternative approach to handle infinite-horizon specifications is to employ interval MCs or interval MDPs, as discussed in the next subsection. 

\subsection{Abstractions as Interval Markov Models}\label{IMDPs}

The classical finite-state Markov models seen in the previous sections are not the only possible architecture for abstractions: 
uncertain Markov models can as well be employed for this task, and indeed they have been in particular employed to construct finite abstractions that are capable of satisfying infinite-time horizon properties in a more natural manner than standard Markov models. 
Uncertain Markov models have been studied under different but related perspectives and semantics: 
\cite{Junges2020Thesis} provides an overview of existing approaches, 
and in particular focuses on models where the uncertainty is described parametrically, 
where probabilities are symbolic expressions rather than concrete values. 
\cite{Junges2020Thesis} discusses the parameter synthesis problem for the analysis of this class of Markov models.  
Alternatively, 
when the probabilities of transition between states belong to intervals, we can use interval Markov chains (IMC) and interval Markov decision processes (IMDP).  

\begin{figure}[ht]
	\begin{center}
		\includegraphics[width=3.7cm]{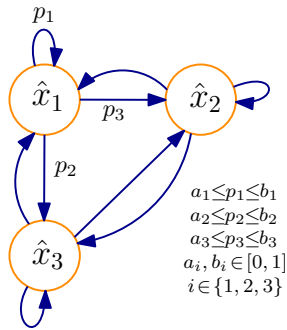}
		\caption{Example of an IMC.}
		\label{Fig9}
	\end{center}
\end{figure}

The definition of IMDP is similar to finite MDP as in Algorithm~\ref{algo:MC_app} with a tuple  $\widehat\Sigma_{\text{I}} = (\hat X, \hat U, \hat T_{\mathsf x_1}, \hat T_{\mathsf x_2}, \hat Y, \hat h)$ where the exact transition probabilities are not known but are bounded above and below as $\hat T_{\mathsf x_1} \leq \hat T_{\mathsf x} \leq \hat T_{\mathsf x_2}.$ An IMC is schematically depicted in Figure~\ref{Fig9}.

It is worth mentioning that constructing IMCs/IMDPs can be more complicated compared to standard MCs/MDPs since one needs to provide both lower and upper bounds for probabilities of transitions among partition sets, by solving max-min optimization problems. 
However, one can mitigate the construction complexity if the system has a property called \emph{mixed-monotonicity} and if the noise of the system has some nice properties.

\begin{definition}\label{Mix}
	A function $f: X\rightarrow X$ is called mixed monotone if there exists a decomposition
	function $g: X\times X\rightarrow X$ satisfying~\cite{smith2008global,coogan2015efficient}
	\begin{itemize}
		\item $\forall x\in X\!: f(x) = g(x,x),$
		\item $\forall x_1,x_2,z \in X\!: x_1 \leq x_2$ implies $g(x_1,z) \leq g(x_2,z),$
		\item $\forall x,z_1,z_2 \,\in X\!: z_1 \leq z_2$ implies $g(x,z_2) \leq g(x,z_1).$
	\end{itemize}
\end{definition}

Mixed monotonicity generalizes the notion of monotonicity in dynamical systems, which is recovered when $g(x, z) = f(x)$ for all $x, z$.

Consider the discrete-time stochastic system $x(k+1)=f(x(k)) + \varsigma(k)$, where $f(\cdot)$ is mixed monotone and entries of the noise $\varsigma(\cdot)$ are independent with unimodal distributions. Then, an IMC as abstraction of this system can be computed without the need for the optimization required in the computation of $ \hat T_{\mathsf x_1}, \hat T_{\mathsf x_2}$~\cite{dutreix2019specification}.

Specification-guided verification and abstraction refinement for mixed-monotone stochastic systems against omega-regular specifications are proposed in~\cite{dutreix2019specification}. The article presents a procedure to compute a finite-state interval-valued Markov chain abstraction of discrete-time
mixed-monotone stochastic systems subject to additive noise,   
given a rectangular partition of the state space. An algorithm is proposed for performing verification
against omega-regular properties in IMCs that aims to compute bounds on the probability of satisfying a specification from any initial state of the IMC, in the form of~\eqref{Pro4}. This is achieved by solving a reachability problem on sets of so-called ``winning and losing'' components in the Cartesian product between the IMC and a Rabin automaton representing the original specification. 

The results of~\cite{dutreix2019specification} have been recently extended to the controller synthesis problem for discrete-time, continuous-state stochastic systems, 
under omega-regular specifications~\cite{dutreix2020abstraction}. 
The work presents a synthesis algorithm for optimizing the probability that a discrete-time
stochastic switched system with a finite number of modes satisfies an omega-regular
property. The approach relies on a finite-state abstraction of the underlying
dynamics in the form of a bounded-parameter Markov decision process
arising from a finite partition of the model's domain, with errors in the form of~\eqref{Pro4}. 
Such Markovian abstractions allow for a range of probabilities of transitions between states for each selected action representing a mode of the original system. The proposed framework decomposes the synthesis into a qualitative problem, where the so-called greatest permanent winning or losing components of the product automaton are created.

\begin{remark}
	The results in~\cite{dutreix2019specification} and \cite{dutreix2020abstraction} leverage the mixed-monotonicity property (cf. Definition~\ref{Mix}) of the deterministic part of the map $f$ by assuming that (i) the stochasticity is additive, (ii) the distribution of the noise is unimodal, and (iii) noises of different states are independent from each other. This observation leads to the following open problem. 
\end{remark}
\begin{resp}
	\begin{open}
		Provide a suitable definition of mixed-monotonicity for stochastic systems based on their stochastic kernels and investigate which classes of systems satisfy that property. An initial investigation is provided in Subsection~\ref{MM-SHSs}. 
	\end{open}
\end{resp}\medskip

An abstraction framework for mapping a discrete-time stochastic system to an IMC and mapping a switched dt-SCS to a bounded-parameter Markov decision process (BMDP) is proposed in~\cite{lahijanian2015formal}.
The work constructs model checking algorithms for IMCs and BMDPs against PCTL formulae to find sets of initial states that \emph{definitely}, \emph{possibly}, and \emph{never} satisfy a given specification. It also develops an algorithm for BMDPs that synthesizes a policy maximizing the probability of satisfaction, and further proposes an adaptive refinement algorithm that exploits the dynamics of the system and the geometry of the partition to increase the precision of the solution. The work proposes a closeness guarantee in the form of~\eqref{Pro4}.

Approximate abstractions of dt-SCS with interval MDPs are proposed in~\cite{zacchia2018approximate}. The abstraction leverages the semantics of IMDPs and the standard notion of approximate probabilistic bisimulation.
The resulting model presents a smaller one-step bisimulation error, in the form of equation~\eqref{Pro1} or~\eqref{Pro1_2}, when compared to a Markov chain abstraction. The work outlines a method to perform probabilistic model checking, and shows that the computational complexity of the new method is comparable to that of standard abstractions based on approximate probabilistic bisimulations.

A constructive procedure for obtaining a finite abstraction of a discrete-time SHS is proposed in~\cite{AIB11}, 
with errors as in equation~\eqref{Pro4} but usable over infinite horizons.  
Similar to the finite abstractions discussed above, the procedure consists of a partition of the state space of the system which depends on a controllable parameter. Given proper continuity assumptions on the model, the approximation errors introduced by the abstraction procedure are explicitly computed and it is shown that they can be tuned through the parameter of the partition.  

An efficient abstraction framework for formal analysis and control synthesis of a class of discrete-time SHS with linear dynamics is developed in~\cite{cauchi2019efficiency}.  
The work constructs IMDPs and focuses on temporal logic specifications over both finite- and infinite-time horizons.
A strategy that maximizes the satisfaction probability of the given specification is synthesized over the IMDP and mapped to the underlying SHS. In contrast to existing formal approaches, which are by and large limited to finite-time properties and rely on conservative over-approximations, the article shows that the exact abstraction error can be computed as a solution of convex optimization problems and can be embedded into the IMDP abstraction. This is later used in the synthesis step over both bounded- and unbounded-time properties, mitigating the known state-space explosion problem but at the cost of lack of convergence guarantees.  

The contribution in~\cite{Badetal22} presents a planning method for models with unknown disturbances, which computes a controller providing probabilistic guarantees on safely reaching a target. The continuous system is abstracted into an IMDP, adapting tools from the scenario approach~\cite{campi2018introduction} to compute probably approximately correct (PAC) assertions. The obtained IMDP is robust against uncertainty in the transition probabilities, and the tightness of the probability intervals can be controlled through the number of samples. Verification techniques are used to provide guarantees on the IMDP, and compute a controller for which these guarantees carry over to the concrete system. 

\begin{resp}
	\begin{open}
		The discussed results in the setting of IMCs/IMDPs
		by and large provide a guarantee in the form of~\eqref{Pro4}. 
		In particular, the satisfaction probability computed over the IMDPs gives a lower bound for the probability of satisfaction over the original system.
		Quantify instead the distance between the probability of satisfactions over the two systems in the form of~equation~\eqref{Pro1} or~\eqref{Pro1_2}. 
	\end{open}
\end{resp}\medskip

\section{Discretization-Free Verification \& Synthesis}\label{CBC}

As discussed in the previous sections, discretization-free approaches can prevent the curse of dimensionality arising in the construction of finite abstractions.  
In this section we discuss discretization-free approaches based on (control) barrier certificates that have been proposed in recent years.  
Work in this domain follows theoretical~\cite{prajna2007framework} and computational~\cite{cAAEGP21} contributions,
which however have been developed for continuous-time models.
We first formally define control barrier certificates in the context of this work.  

\begin{definition} \label{cbc}
	Consider a dt-SCS $\Sigma= (X,U,\varsigma,f, Y, h)$ with sets $X_0, X_u \subseteq X$ that are respectively initial and unsafe sets of the system. A function $\mathds B:X \rightarrow \mathbb{R}_{\geq 0}$ is called a control barrier certificate (CBC) for $\Sigma$ if there are constants $\eta,\beta\in\R_{\geq 0}$ with $\beta > \eta$ such that
	\begin{align}\label{sys2}
		&\mathds B(x) \leq \eta,\quad\quad\quad\quad\quad\!\forall x \in X_{0},\\\label{sys3}
		&\mathds B(x) \geq \beta, \quad\quad\quad\quad\quad\!\forall x \in X_{u},
	\end{align}  
	and $\forall x\in X$, $\exists u\in U$, such that
	\begin{align}\label{cbceq}
		\mathbb{E}&\Big[\mathds B(f(x,u,\varsigma)) \,\,\big|\,\, x, u\Big]\leq \max\Big\{\kappa\mathds B(x), c\Big\},
	\end{align}
	for constants $0<\kappa\leq 1$ and $c\in\R_{\geq 0}$.
\end{definition}

\begin{remark}
	Note that the existential quantifier for the condition in~\eqref{cbceq} implies the existence of a feedback controller for a model satisfying the conditions. 
\end{remark}

Employing Definition~\ref{cbc}, one can propose an upper bound on the probability that the dt-SCS in~\eqref{Def:1} reaches an unsafe region over a finite time horizon, as presented in the next theorem. Note that the requirement $\beta > \eta$
is needed in order to propose meaningful probabilistic bounds. Corollary \ref{Kushner3} and the subsequent remark discuss the choice of the constant $c$ in the statement above. 

\begin{theorem}\label{Kushner}
	Consider a dt-SCS $\Sigma= (X,U,\varsigma,f, Y, h)$ and a CBC\, $\mathds B$ for $\Sigma$. Then the probability that the solution process of $\Sigma$ starts from any initial state $x(0) \in X_0$ and reaches $X_u$ under the policy $\nu(\cdot)$ (associated with the CBC $\mathds B$) within the time interval $[0,T_d]$ is bounded by $\bar\delta$, namely 
	\begin{tcolorbox}[enhanced,
		standard jigsaw,
		boxrule=0.5pt,
		opacityback=0,
		]
		\begin{equation} \label{eqlemma2}
			\PP\Big\{x(k)\in X_u ~~\text{for some}~~k\in[0,T_d] \,\, \big| x(0)\in X_0 \Big\} \leq \bar\delta,
		\end{equation}
		where if ~~	$0<\kappa < 1$:
		\begin{equation}\label{Kushner1}
			\bar\delta:=  \begin{cases} 
				1-(1-\frac{\eta}{\beta})(1-\frac{c}{\beta})^{T_d}, & ~~~ \text{if } \beta \geq \frac{c}{{\kappa-1}}, \\
				(\frac{\eta}{\beta})\kappa^{T_d}+(\frac{c}{(1-{\kappa})\beta})(1-{\kappa}^{T_d}), & ~~~\text{if } \beta< \frac{c}{{\kappa-1}}, \\
			\end{cases}
		\end{equation}
		whereas if ~~	$0<\kappa \le 1$: 
		\begin{equation}\label{Kushner2}
			\bar\delta:= \frac{\eta + cT_d}{\beta}.
		\end{equation}
	\end{tcolorbox}
	The upper bound proposed in~\eqref{Kushner1} is less conservative than that of~\eqref{Kushner2} in the sense that \eqref{Kushner1} yields a tighter probabilistic bound.  On the other hand, the proposed bound in~\eqref{Kushner2} is more general, since there may not exists a $\kappa$ strictly less than one satisfying condition~\eqref{cbceq} for many classes of models and dynamics.
\end{theorem}

The results in Theorem~\ref{Kushner} provide upper bounds on the probability that the models reach unsafe regions within a \emph{finite} time horizon. One can generalize the proposed results to an \emph{infinite} time horizon, provided that the constant $c = 0$, as stated in the following corollary.

\begin{corollary}\label{Kushner3}
	Let $\Sigma= (X,U,\varsigma,f, Y, h)$ be a dt-SCS and suppose $\mathds B$ is a CBC for $\Sigma$ with $c = 0$ in~\eqref{cbceq}. Then the probability that the trajectory of $\Sigma$ starts from any initial state $x(0)\in X_0$ and reaches $X_u$ under the policy $\nu(\cdot)$ is bounded by
	\begin{tcolorbox}[enhanced,
		standard jigsaw,
		boxrule=0.5pt,
		opacityback=0,
		]
		\begin{equation}\label{Kushner4}
			\PP\Big\{x(k)\in X_u ~~\text{for some}~~ k\ge 0\,\,\big|\,\,  x(0) \Big\} \leq \frac{\eta}{\beta}.
		\end{equation}
	\end{tcolorbox}
\end{corollary}

\begin{remark} 
	Note that a CBC $\mathds B$ satisfying condition~\eqref{cbceq} with $c = 0$ is a non-negative supermartingale~\cite[Chapter I]{1967stochastic}.  Although the supermartingale property on $\mathds B$ allows one to provide probabilistic guarantees for infinite-time horizons via Corollary~\ref{Kushner3}, it is restrictive in the sense that a supermartingale $\mathds B$ may not exist \cite{steinhardt2012finite,Pushpak2019}. One may therefore employ a more general $c$-martingale type condition as in~\eqref{cbceq} that does not require such an assumption at the cost of providing probabilistic guarantees for finite-time horizons.
\end{remark}

Note that the computation underlying CBC does not generate an abstract model, and accordingly it does not rely on any similarity relation and closeness error as presented in Definition~\ref{CG}.
Instead, one can employ Definition~\ref{cbc} together with Theorem~\ref{Kushner3} and directly compute an upper bound on the probability that a dt-SCS reaches an unsafe region in a finite-time horizon, much alike~\eqref{Pro4}. 

\begin{remark}
	Note that the verification problem is a special case of the synthesis one, in which the main goal is to verify that the property of interest is satisfied by means of some lower bound on the probability. The statements above can be accordingly tailored by changing the quantifier `\,$\exists$' in~\eqref{cbceq} to `\,$\forall$', and the results follow. 
\end{remark}

\subsection{Computation of CBC and of Control Policies}
\label{computation}
In  this subsection, we discuss suitable methods to search for CBCs and to synthesize corresponding control policies.
We study two different approaches based on (i) sum-of-squares (SOS) optimization and on (ii) counter-example guided inductive synthesis (CEGIS)~\cite{Pushpak2019}.

\subsubsection{Sum-of-Squares Optimization Problems}
\label{sossubsec}

We reformulate conditions~\eqref{sys2}-\eqref{cbceq} as an SOS optimization problem~\cite{Parrilo2003}, where a CBC is restricted to be a non-negative polynomial that can be written as a sum of squares of different polynomials. To do so, the following assumption is required.

\begin{assumption} \label{assumeSOS}
	The dt-SCS $\Sigma$ has a continuous state set $X \subseteq \mathbb{R}^{n}$, and continuous input set $U \subseteq \mathbb{R}^{m}$. Moreover, the vector field $f: X \times U\times \mathcal V_{\varsigma} \rightarrow X$
	is a polynomial function of the state $x$ and of the input $u$. 
	Sets $X$ and $U$ are bounded semi-algebraic sets (i.e., they can be represented by the intersection of polynomial inequalities).
\end{assumption}

Under Assumption \ref{assumeSOS}, one can reformulate conditions~\eqref{sys2}-\eqref{cbceq} as an SOS optimization problem to search for a polynomial CBC $\mathds B$ and a polynomial controller $\nu(\cdot)$ for the dt-SCS $\Sigma$. The following lemma provides the SOS formulation.
\begin{lemma}\label{sos}
	Suppose Assumption \ref{assumeSOS} holds and sets $X_{0}$, $X_{u}$, $X$, and $U$ can be defined as $X_{0}=\{x \in \mathbb{R}^{n} \mid g_{0}(x) \geq 0\}$, $X_{u}=\{x \in \mathbb{R}^{n} \mid g_{u}(x) \geq 0\}$, $X=\{x \in \mathbb{R}^{n} \mid g(x) \geq 0\}$, $U=\{u \in \mathbb{R}^{m} \mid g_\nu(x) \geq 0\}$, where $g_0,g_u$, $g$ and $g_\nu$ are vectors of polynomials and inequalities are intended element-wise. Suppose for a given dt-SCS $\Sigma$, there exists a sum-of-squares polynomial $\mathds B(x)$, constants $\eta,\beta, \bar c \in \R_{\geq 0}$, with $\beta > \eta$, $0<\bar \kappa <1$, vectors of sum-of-squares polynomials $\lambda_{0}(x), \lambda_{u}(x)$, $\lambda(x,u), \lambda_\nu(x,u),$ and polynomials $\lambda_{\nu_{j}}(x)$ corresponding to the $j^{\text{th}}$ input in $u=(u_{1},u_{2},...,u_{m}) \in U \subseteq \mathbb{R}^{m}$ of appropriate dimensions, such that the following expressions are sum-of-squares polynomials:
	\begin{align} \notag
		-&\mathds B(x)-\lambda^{T}_{0}(x)g_{0}(x)+\eta\\\notag
		&\mathds B(x)-\lambda^{T}_{u}(x)g_{u}(x)-\beta\\\label{new}
		-&\mathbb{E}\Big[\mathds B(f(x,u,\varsigma)) \mid x,u\Big]+\bar\kappa\mathds B(x)+\bar c-\sum_{j=1}^{m}(u_{j}-\lambda_{\nu_{j}}(x))-\lambda^{T}(x,u)g(x) -\lambda_\nu^{T}(x,u)g_\nu(x).
	\end{align}
	Then $\mathds B(x)$ is a CBC satisfying conditions~\eqref{sys2}-\eqref{cbceq} and $u=[\lambda_{\nu_{1}}(x);\dots;\lambda_{\nu_{m}}(x)]$, is the corresponding controller of the dt-SCS $\Sigma$, 
	where 
	\begin{align}\notag
		&{\kappa} =1 -(1-{\pi} )(1 - \bar\kappa),\quad c=\frac{\bar c}{\pi(1 - \bar\kappa)},
	\end{align}
	with $0<\pi<1$.
\end{lemma}

For such computations, one can readily employ existing software tools available in the literature such as \textsf{SOSTOOLS} \cite{papachristodoulou2013sostools}, 
together with a semidefinite programming (SDP) solver \cite{sturm1999using,yurtsever2021scalable}.

\subsubsection{Counter-Example Guided Inductive Synthesis}

One can find a CBC with a given parametric form, \emph{e.g.,} a polynomial, by utilizing satisfiability modulo theories (SMT) solvers such as Z3 \cite{de2008z3}, dReal \cite{gao_-complete_2012} or MathSat \cite{cimatti_mathsat5_2013}.
The counter-example guided inductive synthesis (CEGIS) \cite{solar2006combinatorial} scheme can compute CBC for finite input sets, 
and it does not require any restrictions on underlying dynamics beyond what required by the SMT solver of choice. 
One can employ the following lemma and reformulate conditions~\eqref{sys2}-\eqref{cbceq} as a satisfiability modulo theory problem, as follows. 
\begin{lemma}
	Consider the dt-SCS $\Sigma$. Suppose there exists a function $\mathds{B}(x)$, constants $\eta,\beta,c \in \mathbb{R}_{\geq 0}$, and $0<\kappa<1$ such that 		
	\begin{align*} 
		\Theta(x) = \bigwedge_{x \in X_{0}}  \hspace{-0.5em} (\mathds{B}(x)  \hspace{-0.2em} \leq  \hspace{-0.2em} \eta)  
		\hspace{-0.5em} \bigwedge_{x \in X_{u}}  \hspace{-0.5em} (\mathds{B}(x) \hspace{-0.2em}\geq \hspace{-0.2em} \beta)\bigwedge_{x \in X}\bigvee_{u \in U} \left(\mathbb{E}\Big[\mathds{B}(f(x,u,\varsigma)) \big| x, u\Big]\right. \nonumber 
		\left.\leq \max\Big\{{\kappa}\mathds{B}(x),c \Big\}\right),
	\end{align*}		
	where the index sets of conjunctions and disjunctions are possibly infinite. Then $\mathds{B}(x)$ is a CBC satisfying conditions~\eqref{sys2}-\eqref{cbceq}.
\end{lemma}

Note that in the CEGIS approach, SMT solvers are employed to compute the CBC $\mathds{B}(x)$ given a finite set $\overline{X} \subset X$ of data samples. If $\neg \Theta(x)$ has no feasible solution, this implies that $\mathds{B}(x)$ is a true CBC. However, if $\neg \Theta(x)$ is feasible for some $\overline{x} \in X$, then $\overline{x}$ is a counter example. In this case, data samples should be updated to $\overline{X} = \overline{X} \cup \overline{x}$ and coefficients of the barrier should be recomputed iteratively until $ \neg \Theta(x)$ becomes infeasible. The control policy corresponding to the true CBC  would be the sequence of inputs from the finite input set $U$ that renders $\Theta(x)$ feasible.

\begin{remark}
	The computational complexity in the construction of finite MDPs as in Algorithm~\ref{algo:MC_app} grows exponentially with the dimension of the state set. 
	In contrast, in the case of sum-of-squares optimization, the computational complexity depends on both the degree of the polynomials and the number of state variables. 
	It is shown that for fixed degree of the polynomials, the required computation grows polynomially with the dimension~\cite{wongpiromsarn2015automata}. 
	Hence, we expect this technique to be more scalable than discretization-based approaches to study specific problems, such as safety analysis. 
	The CEGIS approach~\cite{Pushpak2019} has a bottleneck that resides with the SMT solver, 
	and it is difficult to provide any analysis on the computational complexity due to its iterative nature and lack of completeness (termination) guarantees.
\end{remark}

\subsubsection{Related Work on Barrier Certificates}

Within this line of work, 
the synthesis of invariants to study probabilistic safety of infinite-state models (probabilistic programs) is discussed in~\cite{chakarov2013probabilistic}. The proposed analysis employs concentration inequalities and martingales theory.

Finite-time safety verification of stochastic nonlinear systems using barrier certificates is proposed in~\cite{steinhardt2012finite}.  
The work considers the problem of bounding the probability of failure (defined as leaving a given bounded region of the state space) over a finite-time horizon for continuous-time continuous -space stochastic nonlinear systems. 
The proposed approach searches for exponential barrier functions (\emph{e.g.,} $\mathds B(x,k) = e^{\frac{1}{2}x^TM(k)x} - 1$)
that provide bounds using $c$-martingale type conditions as in~\eqref{cbceq}, however in continuous time.

Probabilistic safety verification of systems using barrier certificates is proposed in~\cite{huang2017probabilistic}. 
The paper considers stochastic hybrid systems where the dynamics are represented as polynomial relations (equalities and inequalities) over the system variables and where random variables denote the initial discrete mode.
The proposed approach guarantees the safety over an infinite-time horizon.
Control barrier certificates for a class of stochastic nonlinear systems against safety specifications are discussed in~\cite{liu2018adaptive}. The proposed scheme provides probabilistic safety guarantees by reasoning over the stability properties of the model. 

Temporal logic verification of stochastic systems via barrier certificates is proposed in~\cite{jagtap2018temporal}. The goal is to find a lower bound on the probability that a complex temporal logic property is satisfied by finite trajectories of the system (cf.~\eqref{Kushner2}), in the spirit of~\eqref{Pro4}. 
Considering properties expressed as safe LTL formulae, 
the proposed approach relies on decomposing the negation of the specification into a union of sequential reachabilities and then using barrier certificates to compute upper bounds for those reachability probabilities. The results of~\cite{jagtap2018temporal} are recently extended in~\cite{Pushpak2019} to provide a formal synthesis framework for stochastic systems. The extended work distinguishes uncountable and finite input sets in the computation of control barrier certificates, using respectively SOS optimization and the CEGIS approach.

A controller synthesis framework  for stochastic control systems based on control barrier functions is also provided in~\cite{clark2019control}. The paper considers both complete information systems, in which the controller has access to the full system information, as well as incomplete information systems where the state must be reconstructed from noisy measurements. In the complete information case, it formulates barrier functions that leads to sufficient conditions for safety with probability $1$. However, in order to provide infinite-time horizon guarantees, this result requires that the control barrier functions exhibit supermartingale property, which presupposes stochastic stability and vanishing noise at the equilibrium point of the system. This approach is only applicable to systems with unbounded input sets and it does not provide any probabilistic guarantee $1$ when the input set is bounded.
In the incomplete information case, it formulates barrier functions that take an estimate from an extended Kalman filter (cf. \cite{BB04,BB07} for this notion in the context of SHS) as input, and derives bounds on the probability of safety as a function of the asymptotic error for the filter. The results in \cite{Niloofar_LCSS20} study formal synthesis of control policies for partially observed jump-diffusion systems (affected by both Poisson processes and Brownian motions)
against complex logic specifications. Given a state estimator, the results in \cite{Niloofar_LCSS20} synthesize control policies providing (potentially maximizing)
lower bounds on the probabilities that the trajectories of
the partially observed jump-diffusion systems satisfy some
complex specifications expressed by deterministic finite
automata as in Definition \ref{DFA_Original}.

A methodology for safety verification of non-stochastic systems using barrier certificates is proposed in~\cite{prajna2005necessity}. 
Using the concepts of convex duality and density functions, the paper presents a converse statement for barrier certificates, showing that the existence of a barrier certificate is also necessary for safety. 
The results are then extended in~\cite{wisniewski2015converse} to more general classes of dynamical systems:
in particular,~\cite{wisniewski2015converse} proves converse barrier certificate theorems for a class of structurally stable dynamical systems. 
Dovetailing on these recent results, we present the following challenge, which is later generalized further.

\begin{resp}
	\begin{open}
		There is in general no a-priori guarantee on the existence of barrier certificates for a given SHS. In particular, Definition~\ref{cbc} provides a set of \emph{sufficient} conditions for the existence of CBC.
		One interesting direction as a future work is to investigate \emph{necessary and sufficient} conditions for the existence of control barrier certificates for SHS.
	\end{open}
\end{resp}\medskip

\begin{resp}	
	\begin{open}
		Develop computational techniques to construct CBC for general, nonlinear dt-SCS (not only polynomial-type models).  
	\end{open}	
\end{resp}\medskip

{\bf Running example (continued)}. The regions of interest in this example are considered as $X \in [1,50]$, $X_{0} \in [19.5,20]$, and $X_{u} = [1,17]\cup [23,50]$. The main goal is to find a CBC for the system, for which a safety controller is synthesized maintaining the temperature of the room in a comfort zone $[17,23]$.

\begin{figure}
	\centering
	\includegraphics[width=7.7cm]{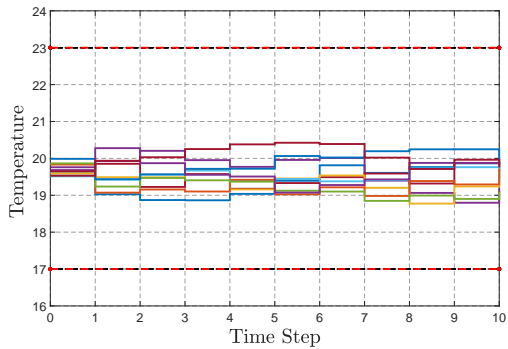}
	\caption{Closed-loop state trajectories with $10$ different noise realizations for the finite-time horizon $T_d=10$.}
	\label{Barrier_1}
\end{figure}

We employ software tool \textsf{SOSTOOLS}~\cite{papachristodoulou2013sostools} and the SDP solver \textsf{SeDuMi}~\cite{sturm1999using} to compute CBC as described in Definition~\ref{cbc}. We compute
CBC of an order $2$ as $\mathds B(T) = 0.86043T^2-33.78116T+331.57433$ and the corresponding safety controller as $\nu(T) = -0.0120155T+0.9$. Furthermore, the corresponding parameters in Definition~\ref{cbc} satisfying conditions~\eqref{sys2}-\eqref{cbceq} are quantified as $\eta = 0.13, \beta = 4.4, \kappa = 0.99,$ and $c = 99\times 10^{-4}$.

By employing Theorem~\ref{Kushner}, one can guarantee that the temperature of the room starting from the initial set $X_{0} = [19.5,20]$ remains in the safe set $X_u =[17,23]$ during the time horizon $T_d=10$ with a probability at least $95\%$, \emph{i.e.,}
\begin{equation}\label{threshold}
	\mathbb{P}\Big\{x(k)\in X_u ~\text{for some}~ k\in[0,T_d] \,\, \big|\,\, a\Big\}\ge 0.95\,.
\end{equation}
Closed-loop state trajectories with $10$ noise realizations are illustrated in Figure~\ref{Barrier_1}. \hfill\qed

\subsection{Analysis of SHS with Optimization-based Methods}\label{Optimization}
For the sake of context and of completeness, we should mention literature dealing with alternative, 
discretization-free techniques for the analysis of SHS, which are mainly based on optimization approaches. 
Obviously the nature of these approaches is quite different than those presented so far, 
and indeed they tend to focus on other objectives, such as stability. 
However, in some instances, they might also accommodate some basic forms of specifications: for instance, safety might be able to be asserted for models that are shown to be stable. 
We should however remark that, in general, these alternative approaches are essentially different in nature than the techniques at the core of this survey.

Lyapunov-based conditions for stability and recurrence for a class of stochastic hybrid systems are presented in~\cite{teel2013lyapunov}, where solutions are not necessarily unique, either due to nontrivial overlap of the flow and jump sets, a set-valued jump map, or a set-valued flow map. Different notions of stability for \emph{stochastic} hybrid systems including Lyapunov, Lagrange, asymptotic stability, and recurrence analysis are overviewed in~\cite{teel2014stability}. A
moment closure technique for stochastic chemically reacting systems based on derivative-matching, which closes the moment equations by approximating higher-order moments as nonlinear functions of lower-order moments is investigated in~\cite{singh2010approximate}.
A moment-based analysis for a class of SHS, so-called linear time-triggered SHS, is presented in~\cite{soltani2017moment}.
The approach relies on embedding a Markov chain based on phase-type processes to model timing of events, and showing that the resulting system has closed moment dynamics.

A shrinking-horizon model predictive control (MPC) scheme for discrete-time linear systems with signal temporal logic (STL) specification constraints is proposed in~\cite{farahani2018shrinking}. 
The control objective is to maximize a function under the restriction that a given STL specification is satisfied with high probability against stochastic uncertainties.
An MPC problem for a discrete-time linear system constrained to satisfy a co-safe LTL is studied in~\cite{gol2015temporal}. 
An overview of the main developments in the area of stochastic model predictive control (SMPC), together with potential perspectives for future research,  can be found in~\cite{mesbah2016stochastic}.

A Lagrangian technique to compute under-and-over approximations of target tube problem, a more generalized version of the finite-time horizon reach-avoid problem, for discrete-time nonlinear systems is proposed in~\cite{gleason2021lagrangian}. The proposed Lagrangian technique eliminates the necessity to grid the state, input, and disturbance spaces allowing for increased scalability and faster computation.
A stochastic reachability problem, which maximizes the
probability that the state remains within time-varying state constraints (\emph{i.e.,} a ``target tube"), despite
bounded control inputs is studied in~\cite{vinod2021stochastic}: the work proposes sufficient conditions under which the reach set
is closed, compact and convex by providing an under-approximative interpolation technique for reach set.
A linear programming approach for stochastic reach-avoid problems is proposed in~\cite{kariotoglou2017linear}, 
where the objective is to synthesize a control policy to maximize the probability of reaching a target set at a given time, while staying in a safe set at all prior times. 
A class of stochastic reachability problems with state constraints from an optimal control perspective and set characterization for diffusions
is proposed in~\cite{esfahani2016stochastic}.

\begin{remark}
	Let us again comment on the comparison between abstraction-based techniques and SMPC, 
	bearing in mind the discussed differences between the two classes of problems. 
	Since abstraction-based approaches rely on discretizing state and input sets, 
	they can readily handle any type of nonlinearity in models, 
	and the spatial sets (characterizing specifications of interest) can be non-convex. 
	In comparison, SMPC would be very challenging with nonlinear dynamics or non-convex constraints. 
	Furthermore, whilst in principle SMPC may ensure a form of invariance via its recursive feasibility feature, 
	it is not straightforward to enforce more complicated, high-level logical properties as optimization constraints in SMPC.
	In conclusion, SMPC is considered as an open and challenging problem~\cite{mesbah2016stochastic} that, in general, cannot be easily utilized to provide the formal guarantees on verification and controller synthesis for complex SHS which are the main focus of this survey paper.
\end{remark}	

\section{Temporal Logic Verification and Synthesis}\label{TLVS}

In this section, we discuss how one can perform verification and synthesis for stochastic hybrid systems over interesting requirements, 
such as safety, reachability, or even more complex specifications encompassed by temporal logic or omega-regular languages.
In presenting work at the interface between control theory and formal methods, 
we mainly focus on LTL and PCTL properties (or related expressions as automata) in this survey for the sake of better readability, 
thus leaving the survey of results on different temporal requirements to bibliographical pointers.

Let us start with basic specifications, expressed over DFAs: 
a quantitative, abstraction-based controller synthesis for SHS is proposed in~\cite{tmka2013}.  
The problem is first reformulated as an optimization of a probabilistic reachability property over a product process obtained from the model for the specification and the model of the system. The article develops a discretization procedure leading into standard dynamic programming problems over finite MDPs with history-independent Markov policies. Errors are in the form of equation~\eqref{Pro1_2}.
A similar controller design scheme for stochastic hybrid systems is also provided in~\cite{kamgarpour2013control}.
As a generalization, 
an optimal control synthesis approach defined over general  discrete-time Markov decision processes is proposed in~\cite{tkachev2017quantitative} in which the probability of a given event is optimized:
it is shown that the optimization over a wide class of LTL and $\omega$-regular properties can be reduced to the solution of one of two fundamental problems: reachability and repeated reachability.

A policy refinement scheme for dt-SCS via approximate similarity relations based on $\delta$-lifting is proposed by~\cite{SIAM17} by providing a closeness guarantee similar to~\eqref{Pro1}. 
In particular, given safety properties over the concrete system, the work constructs an epsilon-perturbed specification over the abstract model whose probability of satisfaction gives a lower bound for the probability of satisfaction in the concrete domain with some quantified error bounds in the form of~\eqref{Pro1}. 
The work is then generalized in~\cite{haesaert2018temporal} to a larger class of temporal properties ((bounded) probabilistic reachability and other temporal logic specifications) and in~\cite{HS19} to synthesize policies for a robust satisfaction of specifications.

\begin{figure}[ht]
	\begin{center}
		\includegraphics[width=6.5cm]{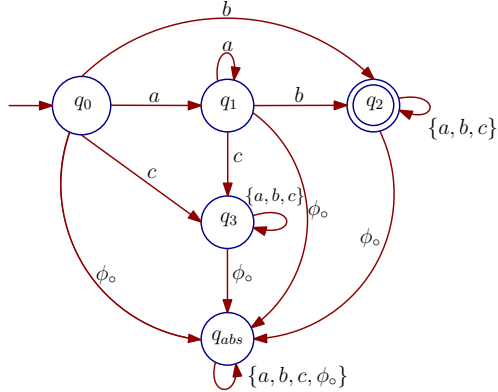}
		\caption{Modified DFA $\mathcal {\hat A}_{\varphi}$ of the specification $(a\,\until b)$.}
		\label{DFA_Modified}
	\end{center}
\end{figure}

Policy synthesis with respect to co-safe linear temporal logic for stochastic control systems is proposed in~\cite{lavaei2018CDCJ} in which it is discussed how synthesized policies for abstract systems can be refined back to original models while providing guarantees on the probability of satisfaction. 
All the results in~\cite{SIAM17,haesaert2018temporal,HS19,lavaei2018CDCJ} quantify a probabilistic distance between the original system and its epsilon-perturbed abstraction, 
as a version of closeness guarantee proposed in~\eqref{Pro1}.  
We should highlight that, given the DFA $\mathcal A$ in Definition~\ref{DFA_Original}, the epsilon-perturbed specification in~\cite{SIAM17,haesaert2018temporal,HS19,lavaei2018CDCJ} corresponds to a new DFA $\mathcal{\hat A}_{\varphi}=(\bar Q_{\ell},q_0,\mathsf{\bar\Sigma}_{\textsf{a}},F_{\textsf{a}},\bar{\trans})$
in which one absorbing location $q_{\textsf{abs}}$ and one letter $\phi_\circ$ are added as
$\bar Q_{\ell}:=Q_{\ell} \cup \{q_{\textsf{abs}}\}$
and $\mathsf{\bar\Sigma}_{\textsf{a}}:=\mathsf{\Sigma}_{\textsf{a}} \cup \{\phi_\circ\}$. The initial and accept locations are the same with $\mathcal A_{\phi}$.
The transition relation is defined, $\forall q \in \bar Q_{\ell}, \forall a \in \mathsf{\bar\Sigma}_{\textsf{a}}$, as
\begin{equation*}
	\bar{\trans} (q, a):=
	\begin{cases}
		\trans (q,a)& \text{if}~ q \in Q_{\ell}, a \in \mathsf{\Sigma}_{\textsf{a}},\\
		q_{\textsf{abs}} &\text{if}~ a = \phi_\circ, q \in \bar Q_{\ell}, \\
		q_{\textsf{abs}} & \text{if}~ q = q_{\textsf{abs}}, a \in \mathsf{\bar\Sigma}_{\textsf{a}}.
	\end{cases}
\end{equation*}
In other words, an absorbing state $q_{\textsf{abs}}$ is added and all states will jump to this absorbing state with label $\phi_\circ$.
As an example, the modified DFA of the reach-avoid specification in Figure \ref{DFA} is plotted in Figure~\ref{DFA_Modified}.

Temporal logic verification and synthesis of stochastic systems via control barrier certificates against a fragment of linear temporal logic, \emph{i.e.,} safe LTL, over finite traces are presented in~\cite{jagtap2018temporal,Pushpak2019}. Those results got extended to $\omega$-regular specifications in \cite{Lavaei_AUT21}.

Forward stochastic reachability analysis for uncontrolled linear systems with affine (bounded or unbounded) disturbance is presented in~\cite{vinod2017forward}.   
The proposed method utilizes Fourier transforms to efficiently compute the forward stochastic reach probability measure (density) and the forward stochastic reach set.
Underpinned by the same technique, an under-approximation of the
stochastic reach-avoid probability
for high-dimensional linear stochastic systems is presented in~\cite{vinod2017scalable} while providing guarantees of the type is equations~\eqref{Pro4} and~\eqref{Pro5}. 
The proposed framework exploits fixed control sequences parameterized by the initial condition (an open-loop control policy) to generate the under-approximation. For Gaussian disturbances, the under-approximation can be obtained using existing efficient algorithms by solving a convex optimization problem.
The work in \cite{vinod2018scalable} proposes a scalable algorithm to construct a polytopic under-approximation of the terminal hitting time stochastic reach-avoid set, for the verification of high-dimensional linear stochastic systems with arbitrary stochastic disturbance. The existence of a polytopic under-approximation is proved by characterizing sufficient conditions under which the stochastic reach-avoid set and the proposed open-loop under-approximation are compact and convex.    

A framework for analyzing probabilistic safety and reachability problems for discrete-time SHS, in scenarios where system dynamics are affected by competing agents, is proposed in~\cite{kamgarpour2011discrete}. The provided framework considers a zero-sum game formulation of the probabilistic reach-avoid problem, in which the control objective is to maximize the probability of reaching a desired subset of the
hybrid state space, while avoiding an unsafe region, subject to the worst-case behavior of a rational adversary.
The results are then extended in~\cite{ding2013stochastic}
to demonstrate how the proposed results can be specialized to address the safety problem, by computing the minimal probability that the system state reaches an unsafe subset of the state space. 

\begin{figure*}[ht!]
	\begin{center}
		\includegraphics[width=0.7\linewidth]{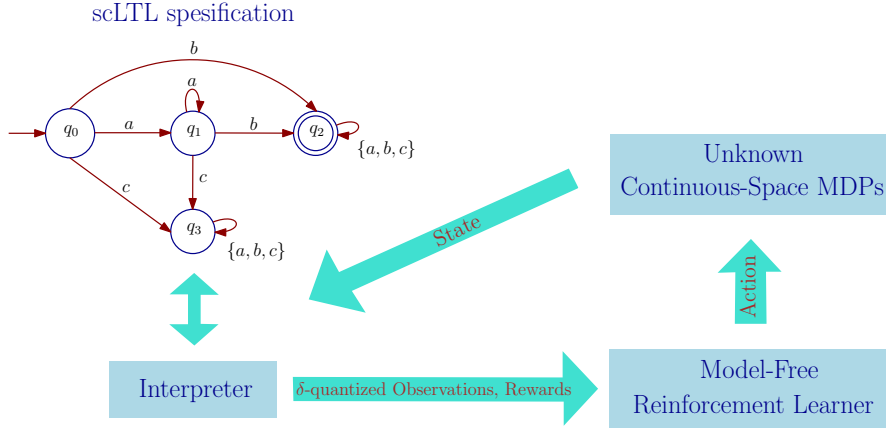} \hspace{0.5cm}
		\caption{Model-free reinforcement learning is employed
			by a DFA corresponding to an \textsc{scLTL}
			objective. In particular,
			the $\delta$-quantized observation set of the dt-SCS $\Sigma$ is used by an \emph{interpreter} process to
			compute a run of the DFA.
			When the run reaches a final state, the interpreter gives the reinforcement learner a positive
			reward and the training episode terminates.
			Any converging reinforcement learning
			algorithm over such $\delta$-quantized
			observation set is guaranteed to maximize the probability of satisfaction of the scLTL
			objective and converge to an optimal strategy over the unknown dt-SCS $\Sigma$.}
		\label{Fig10}
	\end{center}
\end{figure*}

Under-approximation of  finite-time horizon, stochastic reach-avoid sets for discrete-time stochastic nonlinear systems is discussed in~\cite{gleason2017underapproximation} via Lagrangian methods. The article utilizes the concept of target-tube reachability to define robust reach-avoid sets that are parameterized by the target set, safe set, and the set which the disturbance is drawn from. The proposed framework unifies two existing Lagrangian approaches to compute these sets, and establishes that there exists an optimal Markov control policy for the robust reach-avoid sets.
The results characterize a subset of the disturbance space whose corresponding robust reach-avoid set for a given target and safe set is a guaranteed underapproximation of the stochastic reach-avoid level set of interest.
Although the proposed method is conservative, it does not rely on a grid, implying scalability features that now hinge on geometrical computations.  

\cite{LA18} investigate multi-objective optimal control for dt-SHS with safety as a priority, by means of a \emph{lexicographic} approach that priorities the safety as a constraint to be met prior to optimizing a given reward. The paper in \cite{Sofie_LCSS20} proposes an abstraction framework for computation of policies to satisfy multiple specifications with different priorities by encoding them in a multi-objective framework. A tutorial covering multi-objective probabilistic model checking, to analyze trade-offs between several different quantitative properties, is in~\cite{forejt2011automated}. 

A computational framework for the automatic deployment of a robot with sensor, actuator noise and temporal logic specifications is proposed in~\cite{lahijanian2011temporal}. The work models the motion of the robot in the environment as a finite-state MDP and translates the motion specification to a formula of probabilistic computation tree logic. There are alternative approaches that deploy similar results on robotics applications modelled via  stochastic systems~\cite{lacerda}.

We discuss briefly the formal synthesis over SHS via learning and data-driven approaches in Subsection~\ref{Data} as an open research direction. 
Here, 
to conclude this section, 
we only present a few limited recent works with a focus on temporal logic verification and synthesis via learning approaches. 
A reinforcement learning framework for controller synthesis of finite MDPs with unknown transition probabilities against LTL objectives with a proof of convergence is proposed in~\cite{hasanbeig2019certified,hahn2019omega}.
A key feature of the proposed techniques is the compilation of $\omega$-regular properties into limit deterministic B\"uchi automata (LDBA), 
instead of the Rabin automata that are standard with MDPs.
If the dt-SCS is finite, theoretical guarantees are provided on the convergence of the RL algorithm to an optimal policy, maximizing the satisfaction probability.

A model-free reinforcement learning scheme to synthesize policies for unknown continuous-space dt-SCS is proposed in~\cite{lavaei2020ICCPS}. 
The proposed approach is schematically illustrated in Figure~\ref{Fig10}: the properties of interest for the system belong to so-called syntactically co-safe linear temporal logic formulae, and the synthesis requirement is to maximize the probability of satisfaction within a given bounded time horizon. 
The work provides control strategies maximizing the probability of satisfaction over unknown continuous-space dt-SCS while providing probabilistic closeness guarantees in the form of~\eqref{Pro1}.
Similarly based on the scheme in Figure~\ref{Fig10}, extensions to continuous spaces and $\omega$-regular properties are studied in \cite{lcnfq,kazemi2020formal}, as well as in  \cite{hasanbeig2020drl,CHXAK21} by means of deep neural nets generalizers, but without providing optimality guarantees for the synthesized policies when refined over unknown dt-SCS.

\begin{resp}
	\begin{open}
		Provide (approximate) optimality guarantee for learning-based approaches that compute a controller for dt-SCS to satisfy any given LTL specification.
	\end{open}
\end{resp}\medskip

\section{Compositional Techniques}\label{Network}

\begin{figure*}[ht]
	\begin{center}
		\includegraphics[width=0.33\linewidth]{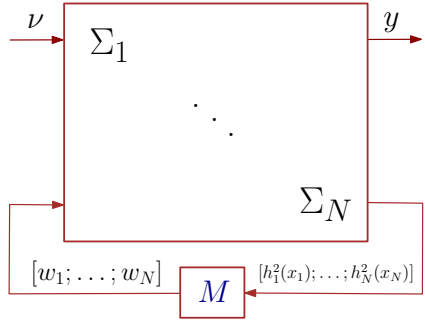} 
		\caption{An interconnected dt-SCS with external input and output signals $\nu$ and $y$, respectively. 
			Note that in so-called small-gain settings discussed later, the interconnection matrix $M$ is a permutation matrix that results in an element-wise interconnection constraint (\emph{i.e.,} $\forall i,j\in \{1,\dots,N\},i\neq j\!:w_{ij}=h^2_{ji}(x_j)$).}
		\label{Fig3}
	\end{center}
\end{figure*}

\begin{figure*}[ht]
	\begin{center}
		\includegraphics[width=0.6\linewidth]{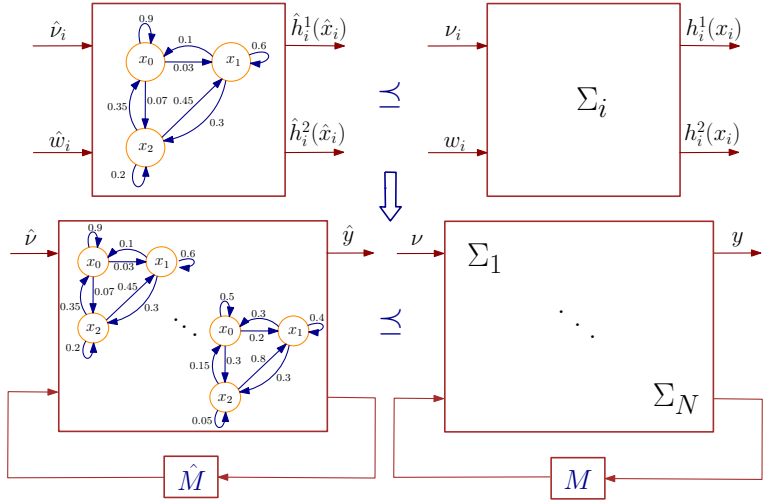} 
		\caption{Representation of compositionality results. Here, we denote $\widehat\Sigma\preceq\Sigma$ if there exists an SSF $V$ from $\widehat\Sigma$ to $\Sigma$ (cf. Definition~\ref{SSF}).}
		\label{Fig5}
	\end{center}
\end{figure*} 

It is of interest to extend the techniques introduced in previous sections to interconnected models, 
or to models with specially structured dynamics or coupling between variables.  
Moreover, the construction of (in)finite abstractions for large-scale stochastic hybrid systems in a monolithic manner suffers severely from the curse of dimensionality. 
To mitigate this issue, one promising solution is to consider a large-scale model as an interconnected system composed of several smaller subsystems. 
Compositional techniques are specifically suitable to tackle these problems, and are broadly studied in this section.  
We overview results on compositional frameworks for the construction of (in)finite abstractions for interconnected systems using abstractions of smaller subsystems.  

We first define stochastic control \emph{subsystems} next. 
The term ``internal'' is employed for inputs and outputs of subsystems that are affecting each other in the interconnection: an internal output of a subsystem affects an internal input of another subsystem downstream. The term ``external'' instead is utilized to denote (exogenous) inputs and outputs that are not employed for the construction of the interconnection. 
\begin{definition}\label{Def:10}
	A discrete-time stochastic control subsystem (dt-SCS) is described by the tuple
	\begin{equation}\label{Eq:10}
		\Sigma=(X,U,W,\varsigma,f,Y^1,Y^2, h^1, h^2),
	\end{equation}
	where: 
	\begin{itemize}
		\item $X\subseteq \mathbb R^n$ is a Borel space as the state space of the subsystem;
		\item $U\subseteq \mathbb R^m$ is a Borel space as the \emph{external} input space of the subsystem;
		\item $W\subseteq \mathbb R^{\bar p}$ is a Borel space as the \emph{internal} input space of the subsystem; 
		\item $\varsigma$ is a sequence of i.i.d. random variables from a sample space $\Omega$ to the measurable space $(\mathcal{V}_\varsigma,\mathcal F_\varsigma)$; 
		\begin{equation*}
			\varsigma:=\{\varsigma(k):(\Omega,\mathcal F_\Omega)\rightarrow (\mathcal{V}_\varsigma,\mathcal F_\varsigma),\,\,k\in\N\}; 
		\end{equation*}
		\item $f:X\times U\times W\times \mathcal{V}_{\varsigma} \rightarrow X$ is the transition map;
		\item  $Y^1\subseteq \mathbb R^{q^1}$ is a Borel space as the \emph{external} output space of the subsystem;
		\item  $Y^2\subseteq \mathbb R^{q^2}$ is a Borel space as the \emph{internal} output space of the subsystem;
		\item  $h^1:X\rightarrow Y^1$ is a measurable function as the \emph{external} output map that takes a state $x\in X$ to its \emph{external} output $y^1 = h^1(x)$;
		\item  $h^2:X\rightarrow Y^2$ is a measurable function as the  \emph{internal} output map that takes a state $x\in X$ to its \emph{internal} output $y^2 = h^2(x)$.
	\end{itemize}
	
\end{definition}
Properties of the interconnected system are specified over external outputs, as in Definition~\ref{Def:10}, 
and the synthesis objective is to control external inputs in order to satisfy desired properties over external outputs; 
whereas internal signals are utilized for the sake of interconnections amongst subsystems.  
We are now well equipped to define an interconnected dt-SCS.

\begin{definition}\label{Def:11}
	Consider $N\in\N_{\geq1}$ stochastic control subsystems $\Sigma_i=(X_i,U_i,W_i,\varsigma_i, f_i, Y^1_{i},Y^2_{i},h^1_{i},h^2_{i})$, $\forall i\in\{1,\ldots,N\}$, and a matrix $M$ of appropriate dimensions, defining the coupling between these subsystems. 
	The interconnection of  $\Sigma_i$, $i\in\{1,\ldots,N\}$, is the dt-SCS $\Sigma=(X,U,\varsigma, f, Y,h)$, denoted by
	$\mathcal{I}(\Sigma_1,\ldots,\Sigma_N)$, such that $X:=\prod_{i=1}^{N}X_i$,  $U:=\prod_{i=1}^{N}U_i$, $f:=\prod_{i=1}^{N}f_{i}$, $Y:=\prod_{i=1}^{N}Y^1_{i}$, and $h=\prod_{i=1}^{N}h^1_{i}$, subjected to the following interconnection constraint:
	\begin{align}\notag
		[{w_{1};\ldots;w_{N}}]=M[{h^2_{1}(x_1);\ldots;h^2_{N}(x_N)}].
	\end{align} 
\end{definition} 
An interconnected dt-SCS based on Definition~\ref{Def:11} is schematically depicted in Fig.~\ref{Fig3}.

In this section, we discuss two different compositional approaches based on small-gain and dissipativity conditions. Small-gain and dissipativity techniques have been traditionally employed in the context of stability analysis for networks of interconnected systems~\cite{dashkovskiy2010small,arcak2016networks}. Suppose $\Sigma$ is an interconnected dt-SCS with $N$ stable subsystems $\Sigma_1,\ldots,\Sigma_N$. Under some small-gain or dissipativity conditions, one can ensure that the composed network $\Sigma=\mathcal{I}(\Sigma_1,\ldots,\Sigma_N)$ is also stable. A similar idea can be utilized here in the setting of similarity relations, using which one can establish a formal relation between an interconnected system $\Sigma$ and its abstraction $\widehat\Sigma$, based on relations between subsystems and their corresponding abstractions. We present the semantics of compositionality techniques in the following. 

{\bf Semantics of compositionality techniques.} Consider an interconnected dt-SCS $\Sigma=\mathcal{I}(\Sigma_1,\ldots,\Sigma_N)$, and assume proper relations between subsystems  $\Sigma_i$ and their corresponding abstractions $\widehat \Sigma_i$ in the sense of Definition~\ref{SSF}. 
Under some compositionality conditions,
one can construct  an overall relation between the two interconnected systems $\widehat \Sigma=\widehat {\mathcal{I}}(\widehat \Sigma_1,\ldots,\widehat\Sigma_N) \footnote{Interconnection topology in the abstract domain can be constructed similar to the interconnection topology of the concrete domain (see \emph{e.g.,}~\cite[Section VI]{lavaei2018ADHSJ}).}$
and $\Sigma$, based on the relations between the subsystems and their abstractions. 

Compositionality conditions based on dissipativity approaches are in the form of LMI that can be readily checked via semidefinite programming (SDP) solvers such as \textsf{SeDuMi}~\cite{sturm1999using}. For small-gain reasoning, we distinguish the corresponding compositionality conditions based on so-called sum-type and  $\max$-type small-gain approaches. In particular, in sum-type small-gain approach, the second condition of the stochastic simulation function (SSF) is in the form of~\eqref{eq:V_dec}, and the overall SSF is a weighted sum of SSF of subsystems. Accordingly, one deals with a spectral radius of some matrix that needs to be strictly less than one as the compositionality condition. In contrast, in the $\max$-type small-gain approach, the upper bound in~\eqref{eq:V_dec} is in the max form and the overall SSF is based on the maximum of SSF of subsystems. We refer the interested readers for more details on the compositionality conditions in (sum and $\max$-type) small-gain and dissipativity approaches respectively to~\cite{lavaei2017compositional,lavaei2018ADHSJ} and~\cite{lavaei2018CDCJ}. It will be further discussed in this section that the closeness guarantees are of the type in~\eqref{Pro2}.

Compositionality results have been schematically depicted in Fig.~\ref{Fig5}. As illustrated, if there exists a local stochastic simulation function between each original subsystem and its corresponding finite MDP, one can construct an overall stochastic simulation function between the original interconnected system and its interconnected finite abstraction provided that some compositionality conditions are satisfied.  

\begin{remark}
	Note that the proposed compositionality results based on sum-type
	small-gain approaches~\cite{lavaei2017compositional} require linear growth
	on gains
	of subsystems (cf.~\cite[Assumption 1]{lavaei2017compositional})
	and provide an additive overall error (i.e., the error of the interconnected abstraction is linear combination of errors of abstractions of subsystems). 
	In contrast, the $\max$-type
	small-gain approaches~\cite{lavaei2018ADHSJ} are more general, since they do not require any linearity assumption on gains of subsystems and the overall error is the maximum error of abstractions of subsystems. Both errors provide closeness guarantees of the type in~\eqref{Pro2}. 
	On the other hand, checking the compositionality condition in the sum-type small-gain is much easier than the $\max$-type one, since it is based on the spectral radius of some matrix that can be easily checked.
\end{remark}

\begin{figure}[h!]
	\begin{center}
		\includegraphics[width=0.53\linewidth]{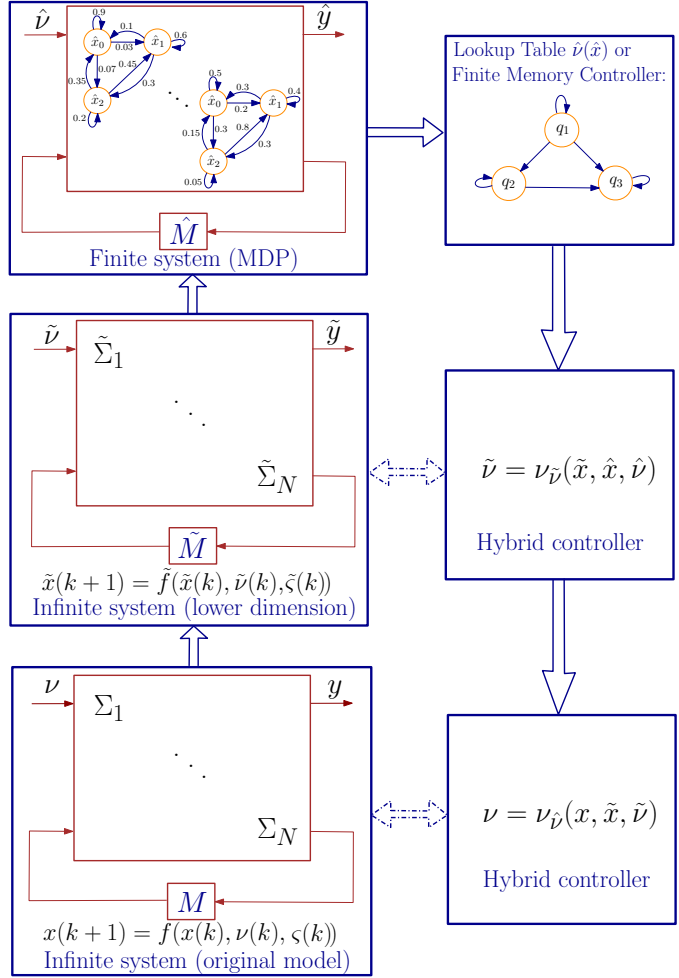} 
		\caption{Compositional techniques based on abstractions.}
		\label{Fig2}
	\end{center}
\end{figure}

Compositional techniques based on infinite and finite abstractions have been schematically illustrated in Fig.~\ref{Fig2}.

A compositional reasoning methodology for the design of \emph{finite} systems with stochastic and/or non-deterministic aspects is proposed in~\cite{delahaye2011probabilistic}. The work focuses on models
of assume/guarantee contracts for stochastic systems, in which the contract allows to distinguish hypotheses made on a system (the guarantees) from those made on its environment (the assumptions). An automated technique for assume-guarantee
style checking of strong simulation between a system and a specification,
both expressed as non-deterministic \emph{finite} labeled probabilistic transition
systems is presented in~\cite{komuravelli2012assume}.

Compositional construction of finite abstractions for stochastic control systems is presented in~\cite{SAM15,SAM17}. These results investigate the finite-horizon probabilistic invariance for dt-SCS
and provide a closeness guarantee between two systems in the form of~\eqref{Pro1}. The compositionality framework is based on finite \emph{dynamic Bayesian networks (DBNs)} and the results exploit the structure of the underlying Markov process to compute the abstraction separately for each dimension and discuss how factor graphs and the sum-product algorithm for DBNs can be utilized to solve the finite-horizon probabilistic invariance problem.

Compositional construction of \emph{infinite} abstractions for interconnected dt-SCS is proposed in~\cite{lavaei2017compositional} via sum-type small-gain conditions. The abstraction framework is based on notions of SSF in Definition \ref{SSF}, using which one can quantify the probabilistic distance between original interconnected stochastic control systems and their abstractions based on the closeness type in~\eqref{Pro2}. A compositional scheme for constructing \emph{infinite} abstractions based on dissipativity approaches is presented in~\cite{lavaei2018CDCJ}. The proposed scheme employs the interconnection matrix and joint dissipativity-type properties of subsystems and their abstractions described by a notion of stochastic storage functions~\cite[Definition 3.1]{lavaei2018CDCJ}.

Compositional construction of both infinite and finite abstractions via $\max$-type small-gain conditions is discussed in~\cite{lavaei2018ADHSJ}. The proposed overall error is computed based on maximum errors of subsystems. The articles in~\cite{lavaei2018ADHSJ} employ a variant of notions of SSF in Definition \ref{SSF}) and provide the probabilistic distance between the interconnection of stochastic control subsystems and that of their (in)finite abstractions based on~\eqref{Pro2}. The proposed framework also leverages the $\delta$-ISS property of original systems as in Definition~\ref{Def:4} and provides an approach to construct finite MDPs of the concrete models (or their reduced-order versions).

Compositional construction of finite abstractions for stochastic control systems is also presented in~\cite{lavaei2017HSCC} but using dissipativity conditions. This work provides a closeness in the form of~\eqref{Pro2} and proposes an approach to construct finite MDPs for the general setting of discrete-time nonlinear SCS satisfying a
passivity-like property, whereby one can construct finite MDPs by selecting a suitable discretization of the input and state sets. Moreover, for linear dt-SCS, the aforementioned property boils down to a matrix inequality. 

Compositional construction of finite MDPs for networks of not necessarily stabilizable stochastic systems is presented in~\cite{lavaei2019NAHS} respectively via \emph{relaxed} dissipativity approaches. The proposed framework relies on a relation between each subsystem and its finite abstraction employing a new notion of simulation functions, called \emph{finite-step} stochastic simulation functions~\cite[Definition A.4]{lavaei2019NAHS}. In comparison with the existing notions of simulation functions in which stability or stabilizability of each subsystem is required, a \emph{finite-step}  stochastic simulation function needs to decay only after some finite numbers of steps instead of at each time step. This relaxation results in a \emph{less conservative} version of small-gain or dissipativity conditions.

Compositional construction of finite abstractions for networks of classes of stochastic hybrid systems, namely, stochastic \emph{switched} systems, is presented in~\cite{lavaei2019HSCC_J,lavaei2020LSS_J} via respectively small-gain and dissipativity approaches. These contributions utilize notions of stochastic simulation (or storage) functions and provide a closeness guarantee in the form of~\eqref{Pro2}, however adapted to switched models. 

Compositional abstraction-based synthesis of dt-SCS using approximate probabilistic relations is proposed in~\cite{lavaei2019NAHS1}. The abstraction framework is based on the notion of $\delta$-lifted relations, using which one can quantify the distance in probability between the interconnected dt-SCS and that of their abstractions as a version of closeness guarantee proposed in~\eqref{Pro1}.
Those results provide some matrix (in)equality conditions for simultaneous existence of relations incorporating the structure of the network. It is shown that the unified compositional scheme is less conservative than the two-step consecutive procedure that independently constructs infinite and finite abstractions (\emph{e.g.,}~\cite{lavaei2017HSCC,lavaei2018CDCJ}).

Results on compositional multi-objective synthesis  for finite probabilistic models are proposed in~\cite{kwiatkowska2013compositional}, and for continuous-space models in~\cite{Sofie_LCSS20} but with in a monolithic manner. However, those approaches are not, to the best of our knowledge,  applicable or in general computationally tractable for large-scale SHS. Therefore, we raise the following open problem.

\begin{resp}
	\begin{open}
		Develop a compositional, multi-objective synthesis framework for continuous-space SHS, and display its applicability to classes of real-life large-scale problems.
	\end{open}
\end{resp}\medskip

Although the proposed compositional frameworks for constructing finite abstractions can mitigate the state-space explosion problem, the curse of dimensionality may still arise at the level of single subsystems. 
As discussed in Section~\ref{CBC}, an alternative direction is to employ \emph{control barrier functions} as a discretization-free technique for controller synthesis of complex stochastic systems. However, searching for control barrier certificates for large-scale systems can be also computationally expensive. Consequently, developing compositional techniques for constructing control barrier functions is a promising solution to alleviate this complexity.
Compositional construction of control barrier certificates for large-scale stochastic control systems is presented in~\cite{Lavaei_TAC20} and \cite{Lavaei_AUT21}. The proposed compositional methodologies are based on a notion of \emph{control sub-barrier certificates}, enabling one to construct control barrier certificates of interconnected systems by leveraging some $\max$-type small-gain or dissipativity-type compositionality conditions, respectively.
Compositional construction of control barrier certificates for discrete-time stochastic switched systems accepting multiple barrier certificates with some dwell-time conditions is also proposed in~\cite{Amy_LCSS20}. 

The results in \cite{Niloofar_TCNS21} also propose a compositional framework for the synthesis of safety controllers for
networks of partially-observed discrete-time stochastic control systems (a.k.a. continuous-space POMDPs).
The proposed framework is based on a notion of so-called local control
barrier functions computed for subsystems in two different ways. In the first scheme, no prior knowledge of
estimation accuracy is needed. The second framework utilizes a probability bound on the estimation accuracy
using a notion of so called stochastic simulation functions. In both proposed schemes, sufficient
small-gain type conditions are derived in order to compositionally construct control barrier functions for interconnected
POMDPs using local barrier functions computed for subsystems. Leveraging compositionality results, the
constructed control barrier functions enable computing lower bounds on the probabilities that the interconnected POMDPs avoid certain unsafe regions in finite-time horizons.

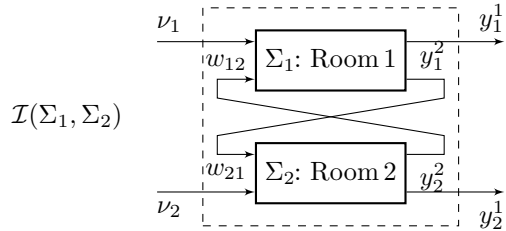
\begin{figure}[ht]
	\begin{tikzpicture}[>=latex']
	\tikzstyle{block} = [draw, 
	thick,
	rectangle, 
	minimum height=.8cm, 
	minimum width=1.5cm]
	
	\node at (-3.5,-0.75) {$\mathcal{I}(\Sigma_1,\Sigma_2)$};
	
	\draw[dashed] (-1.7,-2.2) rectangle (1.7,.7);
	
	\node[block] (S1) at (0,0) {$\Sigma_1\!\!:$ Room \!1};
	\node[block] (S2) at (0,-1.5) {$\Sigma_2\!\!:$ Room \!2};
	
	\draw[->] ($(S1.east)+(0,0.25)$) -- node[very near end,above] {$y^{1}_1$} ($(S1.east)+(1.3,.25)$);
	\draw[<-] ($(S1.west)+(0,0.25)$) -- node[very near end,above] {$\nu_{1}$} ($(S1.west)+(-1.3,.25)$);
	
	\draw[->] ($(S2.east)+(0,-.25)$) -- node[very near end,below] {$y^{1}_2$} ($(S2.east)+(1.3,-.25)$);
	\draw[<-] ($(S2.west)+(0,-.25)$) -- node[very near end,below] {$\nu_{2}$} ($(S2.west)+(-1.3,-.25)$);
	
	\draw[->] 
	($(S1.east)+(0,-.25)$) -- node[very near end,above] {$\!\!\!y^{2}_1$} 
	($(S1.east)+(.5,-.25)$) --
	($(S1.east)+(.5,-.5)$) --
	($(S2.west)+(-.5,.5)$) --
	($(S2.west)+(-.5,.25)$) -- node[very near start,below] {$~~w_{21}$}
	($(S2.west)+(0,.25)$) ;
	
	\draw[->] 
	($(S2.east)+(0,.25)$) -- node[very near end,below] {$\!\!\!y^{2}_2$} 
	($(S2.east)+(.5,.25)$) --
	($(S2.east)+(.5,.5)$) --
	($(S1.west)+(-.5,-.5)$) --
	($(S1.west)+(-.5,-.25)$) -- node[very near start,above] {$~~w_{12}$}
	($(S1.west)+(0,-.25)$) ;
	
	\end{tikzpicture}
	\caption{Interconnection of two rooms $\Sigma_1$ and $\Sigma_2$.}
	\label{system}
\end{figure}

\smallskip

{\bf Running example (continued)}. We present the dynamics of~\eqref{RE} based on a network of two rooms, as illustrated in Figure~\ref{system}. The evolution of the temperatures $T_i$ can be described by~\eqref{RE} in which $A$ is a matrix with diagonal elements $\bar a_{ii}=(1-2\sigma-\theta-\gamma\nu_{i}(k))$, $i\in\{1,2\}$, and off-diagonal elements $a_{1,2}=a_{2,1}=\sigma$. Parameter $\sigma = 0.1$ is the conduction factor between the rooms. Furthermore, $ T(k)=[T_1(k);T_2(k)]$,  $\nu(k)=[\nu_1(k);\nu_2(k)]$, $\varsigma(k)=[\varsigma_1(k);\varsigma_2(k)]$, $T_E=[T_{e_1};T_{e_2}]$, and $R = 0.3\mathds{I}_2$. By considering the individual rooms $\Sigma_i$ as
\begin{align}\label{subs}
	\Sigma_i:\left\{\hspace{-0.5mm}\begin{array}{l}T_{i}(k+1)=\bar a_{ii}T_{i}(k)+\gamma T_{h} \nu_i(k)+D_i w_i(k)+\theta T_{ei}+0.3\varsigma_i(k),\\
		y_i(k)=T_{i}(k),\\
	\end{array}\right.
\end{align}
one can readily verify that $\Sigma=\mathcal{I}(\Sigma_1,\Sigma_2)$ where $D_i = \sigma$, and $w_i(k) = y^2_{i-1}(k)$ for any $i\in\{1,2\}$ (with $y^2_0 = y^2_2$). One can also establish a quadratic stochastic simulation
function between $\Sigma_i$ and $\widehat \Sigma_i$ in the form of $S_i(T_i,\hat T_i)=(T_i-\hat T_i)^2$ satisfying~\cite[conditions (III.1), (III.2)]{lavaei2018ADHSJ} with $\alpha_{i}(s)=s^2$, $\kappa_i(s)=0.99s$, $\rho_{\mathrm{int}i}(s)=0.97s^2$, $\rho_{\mathrm{ext}i}(s)=0$, $\forall s\in \mathbb R_{\ge0}$, and $\psi_i = 6.06\,\delta_i^2$, for any $i\in\{1,2\}$.

\begin{figure}
	\centering
	\includegraphics[width=7.7cm]{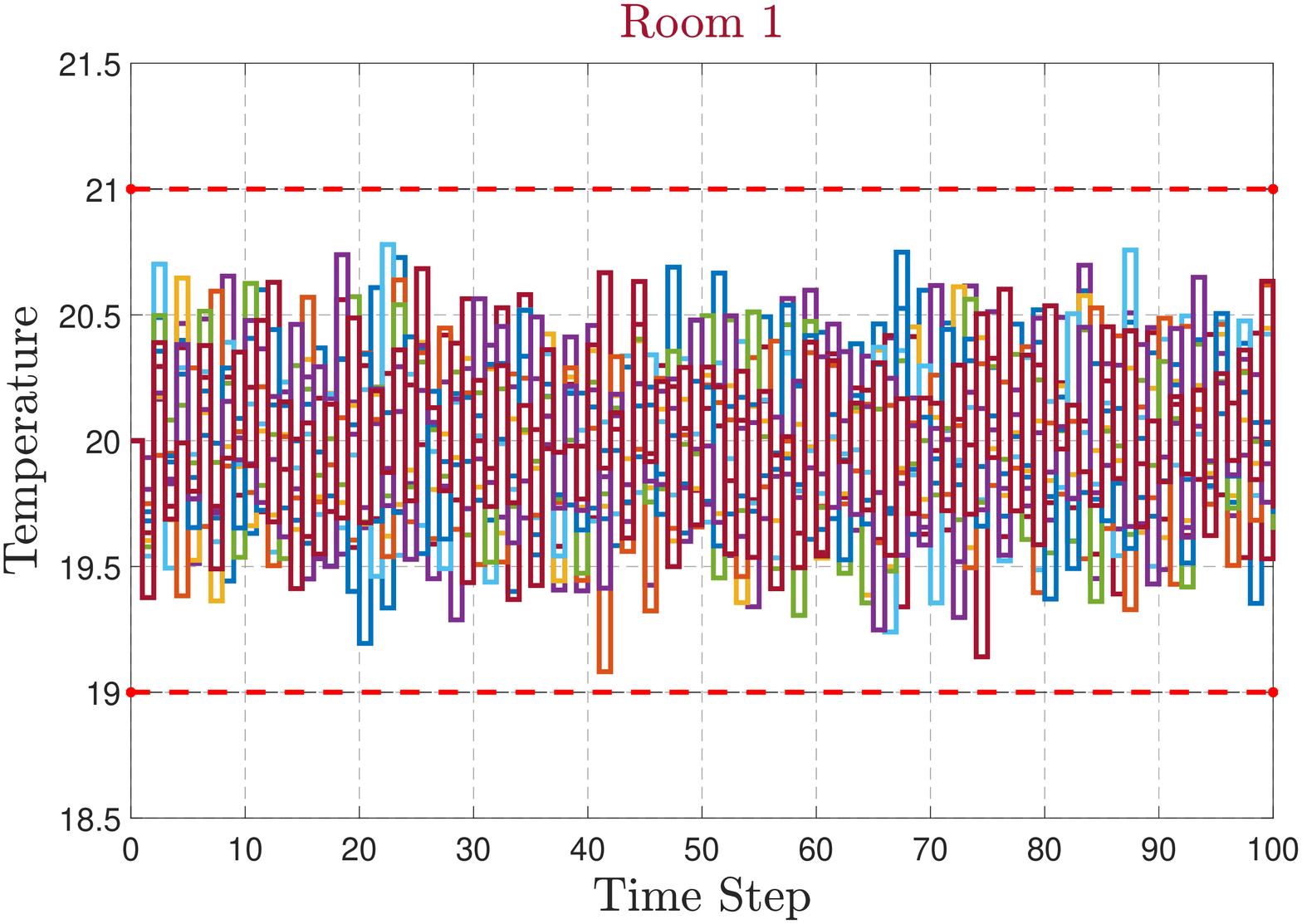}\hspace{-7mm}	
	\includegraphics[width=7.7cm]{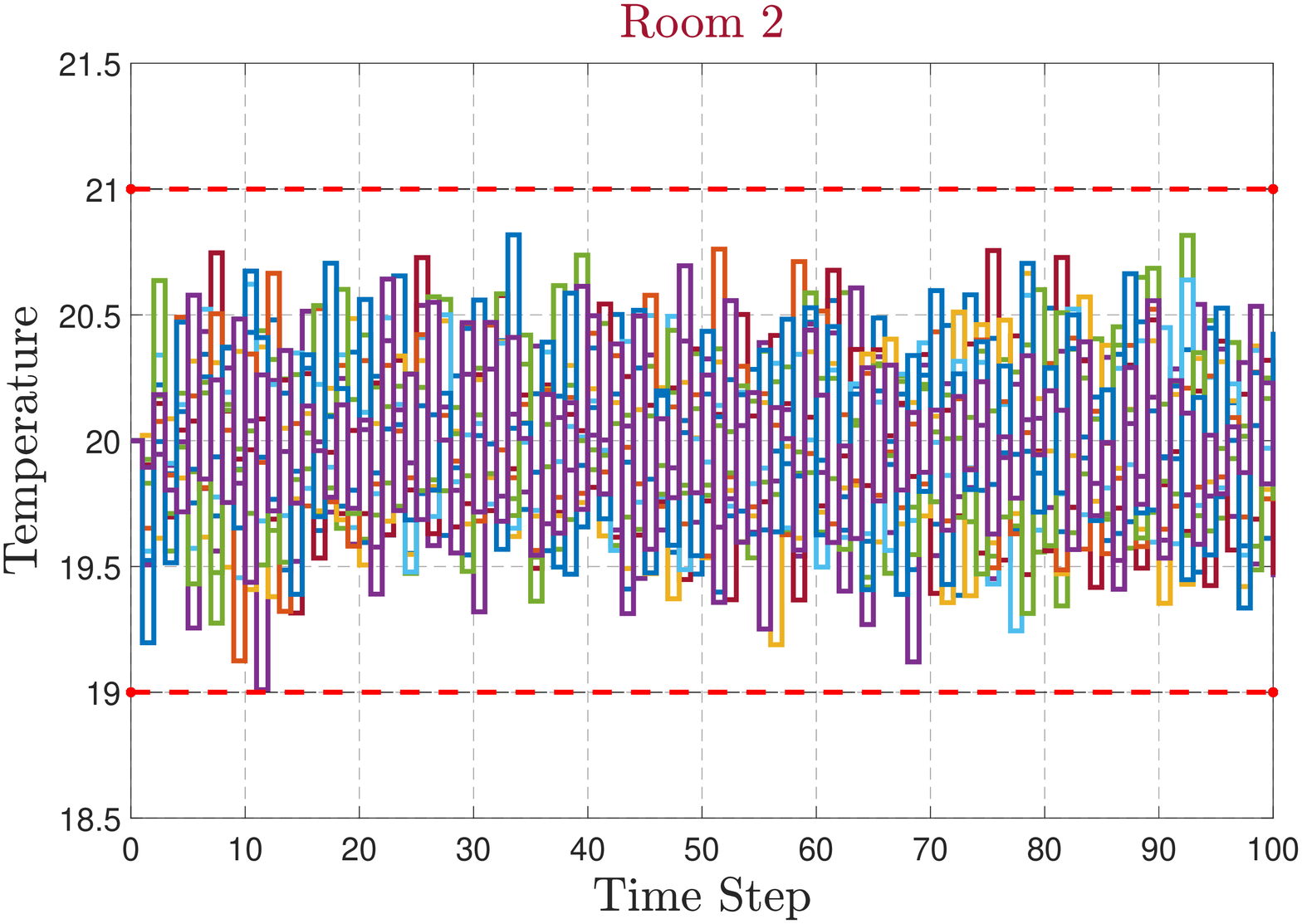}
	\caption{Closed-loop state trajectories of Room 1 (top) and Room 2 (bottom) with  $10$ different noise realizations for the finite-time horizon $T_d=100$.}
	\label{Simulation}
\end{figure}

Now we need to check the small-gain condition for the interconnected system~\eqref{RE}. The small-gain condition~\cite[equation (V.2)]{lavaei2018ADHSJ} is readily satisfied if \begin{align}\label{SGC}
	\kappa_{12}.\kappa_{21} <1,
\end{align}
where  $\kappa_{ij}$ for any $i,j\in\{1,2\}, i \neq j$, is defined as $\kappa_{ij}(s)=\rho_{\mathrm{int}i}(\alpha_j^{-1}(s))$.
Since $\kappa_{12}=\kappa_{21}=0.97$, the small-gain condition~\eqref{SGC} is simply satisfied. Hence, $V(T,\hat T)=\max_{i} (T_i-\hat T_i)^2$ for any $i\in\{1,2\}$ is a stochastic simulation function from the interconnected system $\widehat\Sigma$ to $\Sigma$.\footnote{We should highlight that condition \eqref{SGC} is similar
	to what proposed in \cite{zames} in the context of stability verification of feedback interconnection of two linear systems.} 

By taking the \emph{state} discretization parameter $\delta = 0.005$, the initial states of the interconnected systems $\Sigma$ and $ \widehat \Sigma$ as $20\mathds{1}_{2}$, and using~\cite[ineqality (III.3)]{lavaei2018ADHSJ}, one can guarantee that the distance between outputs of $\Sigma$ and $\widehat \Sigma$ will not exceed $\varepsilon = 0.5$ during the time horizon $T_d=100$ with a probability of at least $98\%$, in the form of~\eqref{Pro2}, \emph{i.e.,}
\begin{equation}
	\label{eq:guarantee5}
	\mathbb P\Big\{\Vert y(k)-\hat y(k)\Vert\le 0.5,\,\, \forall k\in[0,100]\Big\}\ge 0.98\,.
\end{equation}
Let us now synthesize a controller for $\Sigma$ via its finite abstraction $\widehat \Sigma$ such that the controller maintains the temperature of any room in the comfort zone $[19,21]$. We design a local controller for the abstract subsystem $\widehat \Sigma_i$, and then refine it back to the subsystem $\Sigma_i$ using an \emph{interface} map. Consequently, the overall controller for the interconnected system  $\Sigma$ would be a vector such that each of its components is the controller
for subsystems  $\Sigma_i$. We employ the software tool \textsf{AMYTISS}~\cite{lavaei2020AMYTISS} to synthesize controllers for $\Sigma_i$. Closed-loop state trajectories of two rooms with $10$ different noise realizations are illustrated in Figure~\ref{Simulation}. The simulations show that none of $10$ trajectories violates the specification, which is in accordance with the theoretical guarantee \eqref{eq:guarantee5}. As discussed in Section~\ref{Finite}, if one employs our designed controllers and run Monte Carlo simulations of the closed-loop model, the distance between outputs of $\Sigma$ and $\widehat \Sigma$ will likely be empirically closer than $0.5$ with the same probability as in~\eqref{eq:guarantee5}. This is as expected, in view of the conservative nature of formal guarantees provided using Lyapunov-like techniques (simulation functions).  

\begin{figure}
	\centering
	\includegraphics[width=5cm]{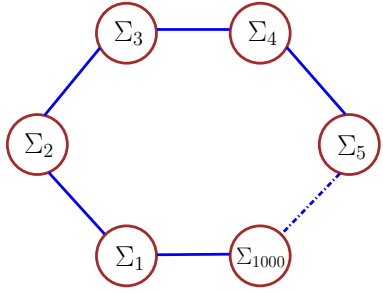}
	\caption{A circular interconnection for a network of $1000$ rooms.}
	\label{Fig6}
\end{figure}

We now increase the number of rooms to $n = 1000$ and interconnect them in a circular fashion, as depicted Figure~\ref{Fig6}. In this case, $A$ in~\eqref{RE} is a matrix with diagonal elements $\bar a_{ii}=(1-2\sigma-\theta-\gamma\nu_{i}(k))$, $i\in\{1,\ldots,n\}$, off-diagonal elements $\bar a_{i,i+1}=\bar a_{i+1,i}=\bar a_{1,n}=\bar a_{n,1}=\sigma$, $i\in \{1,\ldots,n-1\}$, and all other elements are identically zero. Moreover,  $\sigma$ is a conduction factor between pairs of room $i \pm 1$ and $i$, $T(k)=[T_1(k);\ldots;T_n(k)]$,  $\nu(k)=[\nu_1(k);\ldots;\nu_n(k)]$, $ \varsigma(k)=[\varsigma_1(k);\ldots;\varsigma_n(k)]$, $T_E=[T_{e1};\ldots;T_{en}]$, $R = 0.3\mathds{I}_n$. 
Considering the individual rooms $\Sigma_i$ as~\eqref{subs}, one can readily verify that $\Sigma=\mathcal{I}(\Sigma_1,\ldots,\Sigma_N)$ where $D_i = [\sigma;\sigma]^T$, and $w_i(k) = [y^2_{i-1}(k);y^2_{i+1}(k)]$ (with $y^2_0 = y^2_n$ and $y^2_{n+1} = y^2_1$).

We set the state discretization parameter $\delta = 0.005$, and initial states of the interconnected systems $\Sigma$ and $ \widehat \Sigma$ as $20\mathds{1}_{1000}$. Using the proposed bound in~\cite[ineqality (III.3)]{lavaei2018ADHSJ}, one can guarantee that the distance between outputs of $\Sigma$ and $\widehat \Sigma$ will not exceed $\varepsilon = 0.5$ during the time horizon $T_d=100$ with the probability at least $98\%$, \emph{i.e.,}
\begin{equation}\label{eq:guarantee1}
	\PP\Big\{\Vert y(k)-\hat y(k)\Vert\le 0.5,\,\, \forall k\in[0,100]\Big\}\ge 0.98.
\end{equation}

\begin{figure}
	\centering
	\includegraphics[width=7.7cm]{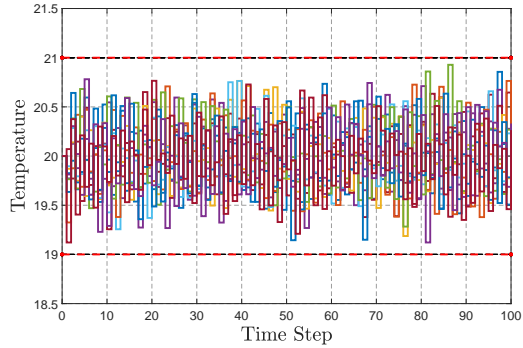}
	\caption{Closed-loop state trajectories of a representative room with $10$ different noise realizations, for the network of $1000$ rooms.}
	\label{Sim2}
\end{figure}

We employ \textsf{AMYTISS}~\cite{lavaei2020AMYTISS} and synthesize a controller compositionally for $\Sigma$ via the abstraction $\widehat \Sigma$ such that the controller maintains the temperature of any room in the comfort zone $[19,21]$. Closed-loop state trajectories of a representative room with $10$ different noise realizations are illustrated in Figure~\ref{Sim2} for a finite-time horizon $T_d=100$. \hfill\qed

\section{Continuous-Time Stochastic Hybrid Systems}\label{CTSS} 

Foundations of continuous-time SHS can be traced back to the work on piecewise-deterministic Markov models in \cite{d1993}, 
which was extended to diffusion processes in \cite{DBLP:conf/hybrid/HuLS00}. 
An early survey of work can be found in \cite{DBLP:conf/adhs/PolaBLB03}. 
In this survey, we present selected results for the continuous-time setting, categorized according to the different topics discussed in the previous sections.

{\bf Notations.} We assume that for continuous-time processes, the triple $(\Omega,\mathcal F_{\Omega},\mathds{P}_{\Omega})$ denotes a probability space endowed with a filtration $\mathbb{F} = (\mathcal F_s)_{s\geq 0}$ satisfying the standard conditions of completeness and right-continuity~\cite{oksendal2013stochastic}. In addition, we denote by $(\mathbb W_s)_{s \ge 0}$ a ${\textsf b}$-dimensional $\mathbb{F}$-Brownian motion. We now define continuous-time stochastic control systems which are studied in this section.

\begin{definition}
	A continuous-time stochastic control system (ct-SCS) is characterized by the tuple
	\begin{align}\label{eq:ct-SCS}
		\Sigma=(X,U,\mathcal U,f,\sigma,Y,h),
	\end{align}
	where:
	\begin{itemize}
		\item $X\subseteq \mathbb R^n$ is the state space of the system;
		\item $U\subseteq \mathbb R^m$ is the input space of the system;
		\item $\mathcal U$ is a subset of the sets of all $\mathbb{F}$-progressively measurable processes taking values in $\mathbb R^m$;  
		\item $f:X\times U\rightarrow X$ is the drift term which is globally Lipschitz continuous: there exist constants $\mathscr{L}_x, \mathscr{L}_\nu \in  \mathbb R_{\ge0}$ such that $\Vert f(x,\nu)-f(x',\nu')\Vert \leq \mathscr{L}_x\Vert x-x'\Vert+\mathscr{L}_\nu\Vert \nu-\nu'\Vert$ for all $x,x' \in X$, and for all $\nu,\nu' \in U$;
		\item $\sigma: \mathbb R^n \rightarrow \mathbb R^{n\times \textsf b}$ is the diffusion term which is globally Lipschitz continuous with the Lipschitz constant $\mathscr{L}_\sigma$;
		\item  $Y\subseteq \mathbb R^{q}$ is the output space of the system;
		\item  $h:X\rightarrow Y$ is the output map.
	\end{itemize}
\end{definition}
A continuous-time stochastic control system $\Sigma$ satisfies
\begin{align}\label{sys1}
	\Sigma\!:\left\{\hspace{-1.5mm}\begin{array}{l}\mathsf{d}x(t)=f(x(t),\nu(t))\,\mathsf{d}t+\sigma(x(t))\,\mathsf{d}\mathbb W_t,\\
		y(t)=h(x(t)),\\
	\end{array}\right.
\end{align}
$\mathds P$-almost surely ($\mathds P$-a.s.) for any $\nu \in \mathcal U$, where stochastic processes $x:\Omega \times \mathbb R_{\ge 0}\rightarrow X$ and $y:\Omega \times \mathbb R_{\ge 0}\rightarrow Y$ are called the \emph{solution process} and the \emph{output trajectory} of $\Sigma$, respectively. We also employ $x_{a \nu}(t)$ to denote the value of the solution process at time $t\in\mathbb R_{\ge 0}$ under an input trajectory $\nu$ from an initial condition $x_{a \nu}(0)= a$ $\mathds P$-a.s., where $a$ is a random variable that is $\mathcal F_0$-measurable. We also denote by $y_{a \nu}$ the \emph{output trajectory} corresponding to the \emph{solution process} $x_{a \nu}$.

{\bf Stochastic Similarity Relations.} Here, we first define the notion of stochastic simulation functions (SSF) for \emph{continuous-time} stochastic control systems as a counterpart of Definition~\ref{SSF}.

\begin{definition}\label{SSF1}
	Consider two ct-SCS
	$\Sigma =(X,U,\mathcal U,f,\sigma,Y,h)$ and
	$\widehat\Sigma =(\hat X,\hat U,\hat{\mathcal U},\hat f,\hat\sigma,\hat Y,\hat h)$.
	A twice-differentiable function $V:X\times\hat X\to\R_{\ge0}$ is
	called a \emph{stochastic simulation function} (SSF) from $\widehat\Sigma$  to $\Sigma$ if
	\begin{itemize}
		\item $\exists \alpha\in \mathcal{K}_{\infty}$ such that
		\begin{eqnarray}\notag
			\forall x\in X,\forall \hat x\in\hat X,\quad \alpha(\Vert h(x)-\hat h(\hat x)\Vert^{\bar q})\le V(x,\hat x),
		\end{eqnarray}
		\item $\forall x\in X,\hat{x}\in\hat X,\hat{u}\in\hat U$, $\exists u\in U$ such that
		\begin{align*}
			\mathcal{L}V(x,\hat{x})
			\leq-\kappa (V(x,\hat{x}))
			+\rho_{\mathrm{ext}}(\Vert\hat u\Vert^{\bar q})+\psi,
		\end{align*}
		for some $\kappa \in \mathcal{K}_{\infty}$, $\rho_{\mathrm{ext}} \in \mathcal{K}_{\infty}\cup \{0\}$, and $\psi \in\R_{\ge 0}$, where $\bar q\in\mathbb N_{\ge 1}$ denoting the moment of a random variable and $\mathcal{L} V$ is the infinitesimal generator of the stochastic process acting on the function $V$~\cite{oksendal2013stochastic}, defined as
		\begin{align*}
			\mathcal{L} V(x,\hat{x})=\partial_x Vf(x,u) +\partial_{\hat x} Vf(\hat x,\hat u)+\frac{1}{2}\mathsf{Tr}(\sigma(x)\sigma(x)^T\partial_{x,x}V)+ \frac{1}{2}\mathsf{Tr}(\hat\sigma(\hat x)\hat\sigma(\hat x)^T\partial_{\hat x,\hat x}V).
		\end{align*}
	\end{itemize}
\end{definition}
In the next theorem, we present a result on the closeness in \emph{expectation (moment)} of the difference between output trajectories of original continuous-time systems $\Sigma$ and their corresponding abstractions $\widehat \Sigma$, as proposed in~\cite{zamani2014symbolic}. This relates to \emph{condition (iii)} in Definition \ref{CG}, which is the last of the four closeness results that have been introduced in this article.

\begin{theorem}
	Let $\Sigma$ be a continuous-time SCS and $\widehat\Sigma$ be its abstraction.
	Suppose there exists a stochastic simulation function $V:X\times\hat X\to\R_{\ge0}$ from $\widehat\Sigma$  to $\Sigma$ as in Definition~\ref{SSF1}.
	For any input trajectory $\hat\nu(\cdot)\in\mathbb{\hat U}$ that preserves the Markov property for the closed-loop $\widehat\Sigma$, and for any random variables $a$ and $\hat a$ as the initial states of $\Sigma$ and $\widehat \Sigma$, respectively, one can construct an input trajectory $\nu(\cdot)\in\mathbb{U}$ for $\Sigma$ through an interface function associated with $V$ (cf. Def.~\ref{SSF}),
	such that:		
	\begin{tcolorbox}[enhanced,
		standard jigsaw,
		boxrule=0.5pt,
		opacityback=0,]
		\begin{equation}\label{Pro3}
			\EE\Big[\Vert y_{a\nu}(t)-\hat y_{\hat a \hat\nu}(t)\Vert^{\bar q}\Big]\leq	\lambda_4,\quad \forall t\in\mathbb R_{>0},
		\end{equation}
	\end{tcolorbox}
	where, $$\lambda_4 := \beta (\EE[V(a,\hat a)],t) +\rho_{\mathrm{ext}}(\EE[\Vert\hat \nu\Vert^{\bar q}_\infty])+c,$$ 
	with $\beta\in \mathcal{KL}, \rho_{\mathrm{ext}}\in\mathcal{K}_{\infty}, c\in\R_{>0}$, 
	and $\bar q\in\mathbb N_{\ge 1}$ denoting the moment of a random variable.
\end{theorem}
\begin{remark}
	Note that one can leverage the bound in~\eqref{Pro3} together with the Markov inequality~\cite{oksendal2013stochastic} to provide a lower bound on the probability of satisfaction of logic specifications
	for which satisfiability is only concerned at single time instances (e.g., reachability). The new bound is similar to~\eqref{Pro2}, but the supremum appears outside of the probability operator~\cite{zamani2014symbolic}. As such, note that this reasoning is not useful for providing probabilistic bounds on satisfaction of general logic specifications, such as those involving the always (\,$\square$) or until (~\!$\until$) operators, since it does not capture the joint distributions across different time instants.
	Then, one can conclude that the bound in~\eqref{Pro3} implies~\eqref{Pro2} and accordingly~\eqref{Pro1} and~\eqref{Pro4}.
\end{remark}

The characterization and computation of probabilistic bisimulations of SCS is discussed in~\cite{cA09}. This work proposes sufficient conditions for the existence of a stochastic simulation function based on the use of contractivity analysis (a variant of incremental stability) for probabilistic systems. The results are then extended in~\cite{cA10} to probabilistic simulations between two SCS that are additionally endowed with switching and resetting behaviors. 
A notion of stochastic simulation function for continuous-time stochastic systems is discussed in~\cite[Definition 3.2]{zamani2016approximations}. 

{\bf Infinite abstractions.} The construction of infinite abstractions for a class of continuous-time SHS was initially proposed in~\cite{julius2009approximations}. The approximation framework is based on stochastic simulation functions and the work provides a closeness guarantee in the form of continuous-time counterpart of~\eqref{Pro2}
but for infinite-time horizons (\emph{i.e.,} $0\le k<\infty$). For the class of jump linear stochastic systems and linear stochastic hybrid automata, the article shows that the computation of stochastic simulation functions can be cast as a linear matrix inequality (LMI) problem. A method for verifying
continuous-time SHS using the Mori-Zwanzig model reduction is proposed in~\cite{wang2016verifying}, where properties are specified as Metric Interval Temporal Logic (MITL) formulas. MITL is an extension of LTL that deals with models with dense time. 
By partitioning the state space of the continuous-time SHS and computing the optimal transition rates between partitions, the work provides
a procedure to both reduce a continuous-time SHS to a continuous-time
Markov chain (CTMC), and the associated MITL formulae
defined on the continuous-time SHS to MITL specifications on the CTMC.

{\bf Finite abstractions.} Construction of symbolic models (\emph{i.e.,} finite abstractions) for incrementally stable stochastic control and switched systems is proposed in \cite{zamani2014symbolic} and~\cite{ZAG15}, respectively. The underlying switched systems in \cite{ZAG15} have a probabilistic evolution over a continuous domain and control-dependent discrete dynamics over a finite set of modes. Both papers
constructively derive approximately equivalent (bisimilar) symbolic models of stochastic systems. The result in \cite{ZAG15} provides two different symbolic abstraction techniques: one requires state space discretization, but the other one does not require any space discretization which can be potentially more efficient than the first one especially when dealing with higher dimensional stochastic switched systems.
Construction of symbolic models for \emph{randomly switched} stochastic system is studied in~\cite{zamani2014approximately}. The proposed framework is based on approximate bisimilar relations and leverages some incremental stability assumption over randomly switched stochastic systems to establish the relation between original systems and their corresponding finite symbolic models. All those aforementioned results provide a closeness bound between the original system and its bisimilar finite abstraction based on the one proposed in~\eqref{Pro3}. 

{\bf Control barrier certificates.} Discretization-free approaches based on barrier certificates for safety verification of continuous-time stochastic hybrid systems are initially proposed in~\cite{prajna2007framework}. The article leverages the supermartingale property and quantifies an upper bound on the probability that a trajectory of the system ever reaches a given unsafe set (over an infinite time horizon) as proposed in~\eqref{Kushner4}. For polynomial-type systems, barrier certificates can be constructed using convex optimization, which is computationally tractable.  

Verification and control for finite-time safety of stochastic systems via barrier functions are discussed in~\cite{santoyo2019verification}. The proposed certificate condition includes a state-dependent bound on the infinitesimal generator, allowing for tighter probability bounds. Moreover, for stochastic systems where the drift dynamics are affine in the control input, the paper proposes a method for synthesizing a polynomial state-feedback controller that achieves a specified safety probability.

Control barrier functions for stochastic systems under process and measurement noise have been proposed in~\cite{clark2020control}. 
The article first considers the case where the system state is known at each time step, and presents a construction that guarantees almost sure safety. 
It then extends the approach to models with incomplete state information, where the state must be estimated: it is shown that the proposed certificates ensure safety with probability $1$ when the state estimate is within a given bound of the true state, which can be achieved using an Extended Kalman Filter\footnote{This and other particle filters have been developed for SHS in~\cite{BB04,BB07}.} when the system is linear or the process and measurement noises are sufficiently small. 

Synthesis for stochastic systems with partial state information via control barrier functions is discussed in~\cite{NiloofarIFAC2020}. Given an estimator with a probabilistic guarantee on its accuracy, the paper proposes an approach to compute a controller providing a lower bound on the probability that the trajectories of the stochastic control system remain safe over a finite time horizon (similar to~\eqref{Kushner2}). This work does not require a supermartingale property on the control barrier functions, and in particular it does not require any stability assumption on the model.
The results of this article are generalized in~\cite{Niloofar_LCSS20} in which no a-priori knowledge about the estimation accuracy is needed. Besides, the class of properties is extended to those expressed by nondeterministic finite automata (NFA), and the dynamics are also generalized to partially-observed jump-diffusion systems. 

{\bf Compositional techniques.} Compositional construction of infinite abstractions (in particular, reduced-order models) for a class of SHS is proposed in~\cite{zamani2016approximations} using sum-type small-gain conditions. The class of systems includes both jump linear stochastic systems and linear stochastic hybrid automata. The work employs stochastic simulation functions to quantify an error between the interconnection of stochastic hybrid subsystems and that of their approximations, in the form of~\eqref{Pro3} (but in the continuous-time setting). It also focuses on a specific class of SHS, namely jump linear stochastic systems, and proposes a constructive scheme to determine approximations together with their corresponding stochastic simulation functions.

Compositional construction of finite abstractions for stochastic control systems is presented in~\cite{Mallik2017CDC}. The proposed framework is based on a notion of (approximate) disturbance bisimulation relation, which results in a closeness guarantee in the form of~\eqref{Pro3} (but in the continuous-time setting). Given
any SCS satisfying a stochastic version of a $\delta$-ISS property and a positive error bound, the article shows how to construct a finite-state transition
system (whenever existing) which is disturbance-bisimilar to the given stochastic control system.

Compositional construction of finite MDPs for networks of continuous-time stochastic systems via $\max$ small-gain conditions is proposed in~\cite{AmyJournal2020}. The proposed framework leverages stochastic simulation functions to relate continuous-time stochastic systems with their discrete-time (in)finite counterparts. In order to propose the construction procedure for finite abstractions, the paper first introduces infinite abstractions as time-discretized versions of original continuous-time stochastic hybrid systems (as a middle step) since the finite abstractions are constructed from the discrete-time counterparts. The work quantifies the distance in probability between original continuous-time stochastic hybrid systems and their discrete-time (finite or infinite) abstractions at sampling times in the form of~\eqref{Pro2}. It also constructs finite abstractions together with their corresponding stochastic simulation functions for a particular class of  \emph{nonlinear} SHS. Although the original models in~\cite{ZAG15,zamani2014approximately,Mallik2017CDC} are stochastic, the constructed abstractions are finite, non-stochastic labeled transition systems. However, the finite abstractions in~\cite{AmyJournal2020} are finite MDPs and potentially less conservative for the sake of controller synthesis and satisfaction probabilities.

Compositional construction of infinite abstractions for networks of stochastic hybrid systems under randomly switched topologies are proposed in~\cite{awan2018compositional}. The proposed framework leverages the interconnection topology, switching randomly between $\mathcal P$ different interconnection topologies (it is modelled by a Markov chain), and the joint dissipativity-type properties of subsystems and their abstractions. The abstraction itself is a stochastic hybrid system (possibly with a lower dimension) and can be used as a substitute of the original system in the controller design process. The work provides a closeness guarantee based on $k$ moments similar to~\eqref{Pro3}. 

Compositional construction of infinite abstractions of interconnected stochastic hybrid systems via dissipativity theory is discussed in~\cite{awan2019dissipativity}. The proposed results leverage a notion of stochastic simulation function
in which the supply rate has its own dynamics. The stochastic noises and jumps in the concrete subsystem and its abstraction do not need to be the same. For a class of nonlinear stochastic hybrid subsystems with an incremental quadratic inequality on the nonlinearity, a set of matrix (in)equalities is established to facilitate the construction of their abstractions together with the corresponding stochastic storage functions. The article quantifies a formal error between the output behaviors of the original system and the ones of its infinite abstractions in the form of~\eqref{Pro3}. 

Compositional construction of control barrier functions for networks of continuous-time stochastic systems is presented in~\cite{AmyIFAC12020}. The proposed scheme is based on notions of \emph{pseudo-barrier functions} (similar to Definition~\ref{cbc} but computed over subsystems), using which one can synthesize state-feedback controllers for interconnected systems  enforcing safety specifications over a finite time horizon. This work leverages sum-type small-gain conditions to compositionally construct control barrier functions for interconnected systems based on the corresponding pseudo-barrier functions computed for subsystems. Then, using the constructed control barrier functions, it quantifies upper bounds on the probability that an interconnected system reaches certain \emph{unsafe} regions in a finite time horizon, similar to Theorem~\ref{Kushner} but in the continuous-time setting. The work also employs a systematic technique based on the SOS optimization program to search for pseudo-barrier functions of subsystems, while synthesizing safe  controllers.

{\bf Temporal logic verification and synthesis.} 
Early work on SHS is in \cite{DBLP:conf/hybrid/GaoLQ06} and \cite{DBLP:conf/hybrid/KoutsoukosR06}, which focused on probabilistic reachability analysis. Stochastic reachability analysis of hybrid systems is also studied in~\cite{DBLP:conf/hybrid/BujorianuL03,bujorianu2012stochastic}, and this line of work continues to these days \cite{DBLP:journals/tac/WisniewskiBS20,cosentino2021gridfree}.

A reachability analysis problem for an aircraft conflict prediction, modelled as a stochastic hybrid system, is studied in~\cite{prandini2008application} in which a switching diffusion
is presented to predict the future positions of an aircraft given a flight plan. The work
proposes a numerical algorithm for estimating
the probability that the aircraft either enters an unsafe region or closely approaches to another aircraft.

A probabilistic approach for control of continuous-time linear stochastic systems subject to LTL formulae over a set of linear predicates in the state of the system is presented in~\cite{lahijanian2009probabilistic}. The article defines a polyhedral partition of the state space and a finite collection of controllers, represented
as symbols, and constructs a finite MDP. By utilizing an algorithm resembling LTL model checking, it determines a run satisfying the formula in a corresponding Kripke structure. A sequence of control actions in the MDP is determined to maximize the probability of following the run. 

Measurability and safety verification of a class of SHS are discussed in~\cite{franzle2011measurability} in which the continuous-time behaviour is given by differential equations, but discrete jumps are chosen by probability distributions. In this work, non-determinism is also supported, and it is exploited in an abstraction and evaluation method that establishes safe upper bounds on reachability probabilities. 

An optimal control problem for continuous-time stochastic systems subject to objectives specified in a fragment of metric interval temporal logic specifications, a temporal logic with real-time constraints, is presented in~\cite{fu2015computational}. The work proposes a numerical method for computing an optimal policy with which the given specification is satisfied with  maximal probability in the discrete approximation of the underlying stochastic system. It is shown that the policy obtained in the discrete approximation converges to the optimal one for satisfying the specification in the continuous or dense-time semantics, as the discretization becomes finer in both state and time.

Motion planning for continuous-time stochastic processes via dynamic programming is discussed in~\cite{esfahani2015motion}. The work studies stochastic motion planning problems which involve a controlled process, with possibly discontinuous sample paths, visiting certain subsets of the state-space while avoiding others in a sequential fashion. A weak dynamic programming principle (DPP) is proposed that characterizes the set of initial states which admit a control enabling the process to execute the desired maneuver with probability no less than some pre-specified value. The proposed DPP comprises auxiliary value functions defined in terms of discontinuous payoff functions. 

Reachability analysis for continuous-time stochastic hybrid systems with no resets is proposed in~\cite{laurenti2017reachability} in which continuous dynamics described by linear stochastic differential equations. For this class of models, the article studies reachability (and dually, safety) properties on an abstraction defined in terms of a discrete-time and finite-space Markov chain, with provable error bounds. The paper provides a characterization of the uniform convergence of the time discretization of stochastic processes with respect to safety properties, and this allows to provide a complete and sound numerical procedure for reachability and safety computation over the stochastic systems. 

An approach for the automated synthesis of safe and robust proportional-integral-derivative (PID) controllers for SHS is proposed in~\cite{shmarov2017automated}. The work considers hybrid systems with nonlinear dynamics (Lipschitz continuous ordinary differential equations) and random parameters, and synthesizes PID controllers such that the resulting closed-loop systems satisfy safety and performance constraints given as probabilistic bounded-reachability properties. The proposed technique leverages satisfiability modulo theories (SMT) solvers over reals and nonlinear differential equations to provide formal guarantees that the synthesized controllers satisfy such properties. These controllers are also robust by design since they minimize the probability of reaching an unsafe set in the presence of random disturbances.

Automated synthesis of controllers with formal safety guarantees for nonlinear control systems with noisy output measurements, and stochastic disturbances is recently presented in~\cite{shmarov2019automated}. The proposed method derives  controllers such that the corresponding closed-loop system, modeled as a sampled-data stochastic control system, satisfies a safety specification with probability above a given threshold. If the obtained probability is not above the threshold, the approach expands the search space for candidates by increasing the controller degree.

A theoretical and computational synthesis framework for safety properties of continuous-time continuous-space switched diffusions is proposed in~\cite{laurenti2020formal}. The work provides an appropriate discrete abstraction in the form of an uncertain Markov model that captures all possible behaviors of the system. This is achieved through a discretization of both time and space domains, each introducing an error such that the errors are formally characterized and represented as uncertain transition probabilities in the abstraction model. It also provides a robust strategy that optimizes a safety property over the abstraction. This strategy is computed by considering only the feasible transition probability distributions, preventing the explosion of the error term and resulting in achievable bounds for the safety probability. This robust strategy is mapped to a switching strategy for the system consisting of switched diffusions with the guarantee that safety probability bounds also hold for this system in the form of~\eqref{Pro4}.

{\bf Stability and Optimal control.} 
In this survey we do not delve into the broad issues of stability analysis and optimal control synthesis for SHS, 
which have been widely investigated over the past two decades. 
For stability, we point the reader to the survey in \cite{teel2014stability}, 
whereas for optimal control we refer the interested reader to the books~\cite{blom2006stochastic,cassandras2018stochastic} (which covers seminal work 
\cite{d1993,borkar93}) and to the work of \cite{DBLP:conf/cdc/PakniyatC16}.
It is worth mentioning that formal verification and synthesis have a tight connection to \emph{optimal control} by expressing logic specifications as a part of constraints in the optimization problem \cite{LA18,Sofie_LCSS20}. 
Finally, let us mention the qualitative connection with stochastic MPC~\cite{mesbah2016stochastic}, discussed in more detail elsewhere in this survey. 

\section{SHS Simulation and Statistical Model Checking}\label{SIM_SHS}

In this section, we study simulation-based analysis of stochastic hybrid systems, and also encompass work on statistical model checking (SMC). 
Early work deals with the development of filtering algorithms that are applicable to SHS~\cite{BB04,BB07}.  
Similarly, sequential Monte Carlo simulations and the use of Petri Nets are discussed in~\cite{blom2007free}, 
and employed in collision risk estimation of free flight operations. 
The proposed results are applicable to rare-event estimation over complex models.   
Safety risk analysis of an air traffic management operation over a SHS is similarly discussed in~\cite{blom2013modelling}. 
\cite{bouissou} puts forward techniques for efficient Monte Carlo simulation of stochastic hybrid systems.

So far, we have widely discussed formal verification and synthesis of SHS in which a closeness guarantee between original SHS and their abstractions is formally provided in four different forms, as elaborated in Definition~\ref{CG}. 
For simulation-based analysis instead, new guarantees are provided. Suppose $(\hat x_i)^{\bar N}_{i=1}$ are $\bar N$ i.i.d. sampled data from a set $\Omega$. Simulation-based guarantees are presented in two-layer probabilities~\cite{calafiore2006scenario}, as follows: 
\begin{equation}\label{CC}
	\PP\Big\{(\hat x_i)^{\bar N}_{i=1} \in \Omega\!: \PP \big\{\bar y_{a\nu} \not\models \varphi \big\} \leq \bar\varepsilon \Big\}\ge \bar\beta,
\end{equation}
where $\varphi$ is the property of interest, $\bar y_{a\nu}$ is any given random output trace,
and $\bar\varepsilon, \bar\beta \in [0,1]$ are respectively a threshold and confidence level. As a comparison with the studied approaches in the previous sections, the formal guarantee there comes with only one layer probability similar to~\eqref{eqlemma2}. If the confidence level $\bar\beta$ is increased to one,
the chance constrained problem~\eqref{CC} can be understood to be similar to~\eqref{eqlemma2}.

A statistical model checking (SMC) algorithm to verify stochastic properties with unbounded until is presented in~\cite{sen2005statistical}. The algorithm is based on Monte Carlo simulation of the model and hypothesis testing of the samples, as opposed to sequential hypothesis testing. Statistical model checking for synthesizing policies on stochastic models including finite MDPs is presented in~\cite{henriques2012statistical}. The proposed framework develops an algorithm that resolves nondeterminism probabilistically, and then uses multiple rounds of sampling and reinforcement learning to provably improve resolutions of nondeterminism with respect to satisfying a bounded linear temporal logic (BLTL) property. The proposed algorithm thus reduces an MDP to a fully probabilistic Markov chain on which SMC may be applied to give an approximate estimation of the probability of the BLTL property.

A numerically rigorous Monte Carlo approach for computing probabilistic reachability in hybrid systems subject to random and nondeterministic parameters is proposed in~\cite{shmarov2016probabilistic}.  Instead of standard simulation, the work employs $\delta$-complete SMT procedures, which enables formal reasoning for
nonlinear systems up to a user-definable numerical precision. Monte Carlo
approaches for probability estimation assume that sampling is possible
for the real system at hand, however when using $\delta$-complete simulations,
one instead samples from an over-approximation of the random quantities at hand. The article introduces a Monte Carlo-SMT approach for
computing probabilistic reachability confidence intervals that are both statistically and numerically rigorous. 
A survey on statistical model checking is provided in~\cite{agha2018survey}, which covers SMC algorithms, techniques, and tools, while emphasizing limitations and tradeoffs between precision and scalability. 

A multilevel Monte Carlo method for statistical model checking of continuous-time stochastic hybrid systems is proposed in~\cite{soudjani2017multilevel}. 
The provided approach relies on a 
sequence of discrete-time stochastic processes whose executions approximate
and converge weakly to that of the original continuous-time SHS with respect to the satisfaction of a property of interest.
With focus on bounded-horizon reachability, the paper casts the model
checking problem as the computation of the distribution of an exit time,
which is in turn formulated as the expectation of an indicator function. 
This latter computation involves estimating discontinuous functionals,
which reduces the bound on the convergence rate of the Monte Carlo
algorithm. The work then proposes a smoothing step with tuneable precision and formally
quantifies the error in the mean-square sense, which is composed of smoothing error, bias, and variance.

A confidence bound for statistical model checking of probabilistic hybrid systems is proposed in~\cite{ellen2012confidence}. The work presents an approximation algorithm based on confidence intervals obtained from sampling which allow for an explicit trade-off between accuracy and computational effort. Although the algorithm gives only approximate results in terms of confidence intervals, it is still guaranteed to converge to the exact solution. A Bayesian approach to
statistical model-checking of discrete-time Markov chains with respect to continuous stochastic logic specifications is presented in~\cite{lal2020bayesian}. 
Related to this approach, \cite{bcMA20} formally integrates the use of sequential Monte Carlo techniques for approximate Bayesian inference with (Bayesian) statistical model checking.

In conclusion, statistical model checking approaches appear to be suitable for verification goals, whereas they appear to be less efficient when synthesis is in order. The latter objective  could be considered as an open problem for future research. In addition, most of the proposed SMC results are suitable for \emph{finite-time} horizons. More precisely, in the setting of SMC approaches in \emph{infinite-time} horizons, the proposed results require some strong assumptions that are not in general satisfiable by SHS. More emphasis on infinite-horizon properties via SMC can be cast as another future research direction.

\section{Software Tools}\label{RST}

In this section, we discuss software tools for verification and synthesis, as well as simulation,  of stochastic hybrid systems. This is a growing and fast-pacing area, thus we focus on existing tools at the time of writing, emphasizing their architectures and relating them to the underlying theory, as well as presenting relevant open-science initiatives in this area. 

\subsection{The \textsf{Modest Toolset}}	
\textsf{Modest Toolset}~\cite{hartmanns2014modest} performs modelling and analysis for hybrid, real-time, distributed and stochastic systems. At its core are models of networks of stochastic hybrid automata (SHA), which combine nondeterministic choices, continuous system dynamics, stochastic decisions and timing, and real-time behaviour, including nondeterministic delays. The \textsf{Modest Toolset} is a modular framework, supporting as input the high-level Modest modelling language and providing a variety of analysis backends for various special cases of SHA. Many existing automata-based formalisms are special cases of SHA.

\subsection{\textsf{SReach}}
\textsf{SReach}~\cite{wang2015sreach} solves probabilistic, bounded-time reachability problems for two classes of models: (i) nonlinear hybrid automata with parametric uncertainty, and (ii) probabilistic hybrid automata with additional randomness on both transition probabilities and variable resets. 
Standard approaches to reachability analysis for linear hybrid systems require numerical solutions of large optimization problems, 
which become practically infeasible for systems involving both nonlinear dynamics and stochasticity. 
\textsf{SReach} instead encodes stochasticity by using a set of random variables, and combines $\delta$-complete decision procedures and statistical tests to solve $\delta$-reachability problems. Compared to standard simulation-based methods, \textsf{SReach} supports non-deterministic branching and allows one to increase the coverage of performed simulations. 

\subsection{\textsf{ProbReach}}
\textsf{ProbReach}~\cite{shmarov2015probreach1} is a statistical model checking tool that studies bounded-time reachability and other quantitative properties.
It handles SHS with random continuous quantities encompassing model parameters or initial conditions that are chosen within an initial set and which remain unchanged throughout the system evolution. For continuous dynamics, \textsf{ProbReach} can analyze any Lipschitz-continuous differential equations with stochastic parameters. Given an SHS with random continuous quantities and an arbitrarily small $\epsilon > 0$, \textsf{ProbReach} returns an interval of size not larger than $\epsilon$ containing the exact bounded-reachability probability. This result is guaranteed to
be numerically sound, \emph{e.g.,} free from floating-point inaccuracies. 
The introduction of discrete random parameters to the system will not affect the guarantees provided by \textsf{ProbReach},  
however if the model features only discrete random parameters, then these guarantees do not hold: 
this happens because probability distributions over discrete random parameters are not continuous, hence an arbitrary precision cannot be provided any longer. Introducing nondeterministic continuous parameters affects the guarantees the tool provides, as well: this happens because nondeterministic parameters do not have any probability measure:  
in this case, \textsf{ProbReach} computes an enclosure that is guaranteed to contain all the possible reachability probabilities. 
In general, such an enclosure may have size larger than $\epsilon > 0$. \textsf{ProbReach} employs a validated integration procedure to obtain a partition over the random continuous quantities in such a way that the guarantees described above hold. This partition is then used to enclose the probabilistic outcome by computing under- and over-approximations.

\subsection{\textsf{SReachTools}}
\textsf{SReachTools}~\cite{vinod2019sreachtools} is an open-source Matlab toolbox for performing stochastic reachability of linear, potentially time-varying, discrete-time systems that are perturbed by a stochastic disturbance. 
More precisely, this tool addresses the problem of stochastic reachability of a target tube, which also encompasses terminal-time (hitting) problems, reach-avoid problems, and related viability problems (not discussed in this survey).  
The stochastic reachability of a target tube problem maximizes the likelihood that the state of a stochastic system will remain within a collection of time-dependent target sets for a given time horizon, while respecting system dynamics and utilizing inputs within a bounded control domain. 
\textsf{SReachTools} implements several algorithms based on convex optimization, computational geometry, and Fourier transforms, to efficiently compute over- and under-approximations of stochastic reach sets.  
\textsf{SReachTools} can be employed to perform probabilistic verification of closed-loop systems, 
and can also perform controller synthesis via open-loop or affine state-feedback controllers. 

\subsection{\textsf{HYPEG}}

A statistical simulator for hybrid Petri nets with general transitions, called HYPEG, is presented in~\cite{pilch2017hypeg}. It combines discrete and continuous
components with a possibly large number of random variables, whose stochastic behavior follows arbitrary probability distributions. HYPEG employs time-bounded discrete-event simulation and well-known statistical model checking techniques to verify properties, including time-bounded reachability.

\subsection{\textsf{Mascot-SDS}}

\textsf{Mascot-SDS}~\cite{majumdar2020symbolic} is an open-source tool for synthesizing controllers with formal correctness guarantees for discrete-time dynamical systems in the presence of stochastic perturbations. Mascot-SDS is written in C++, and is an extension of Mascot~\cite{hsu2018multi}.  
The tool supports \emph{infinite-horizon} control specifications for stochastic dynamical systems and computes over- and under-approximations of the set of states that satisfy a given  specification with probability one.
The current version of the tool is developed for ``always eventually" specifications, namely for specifications dealing with ``infinitely often'' ($\omega$-regular) requirements.

\subsection{\textsf{Level-Set Toolbox}}

	The \textsf{Level-Set Toolbox}~\cite{mitchell2007toolbox} is a software package for solving time-dependent Hamilton-Jacobi partial differential equations (PDEs) in the Matlab programming environment. Level set methods are often used for simulation of dynamic implicit surfaces in graphics, fluid and combustion simulations, image processing, and computer vision. Hamilton-Jacobi and related PDEs arise in fields such as control, robotics, differential games, dynamic programming, mesh generation, stochastic differential equations, financial mathematics, and verification.
	All source code for the toolbox is provided as plain text in the Matlab m-file programming language. The toolbox is designed to allow quick and easy experimentation with level set methods, although it is not by itself a level set tutorial and so should be used in combination with the existing literature. The \textsf{Level-Set Toolbox} has been in particular used for the analysis of stochastic models  in~\cite{park2014hybrid,sprinkle2005using,ding2012methods,choi2022computation}.

\subsection{\textsf{FAUST}~$^{\mathsf 2}$}
\textsf{FAUST}$^{\mathsf 2}$ \cite{FAUST15} generates formal abstractions for continuous-space discrete-time Markov processes defined over uncountable (continuous) state spaces, and performs verification and synthesis for safety and reachability specifications. The abstract model is formally put in a relationship with the concrete model via a user-defined maximum threshold on the approximation error introduced by the abstraction procedure.  \textsf{FAUST}$^{\mathsf 2}$ allows exporting the abstract model to well-known probabilistic model checkers, such as PRISM~\cite{kwiatkowska2002prism} or Storm~\cite{DBLP:conf/cav/DehnertJK017}. Alternatively, it can handle internally the computation of PCTL properties (\emph{e.g.,} safety or reachability) over the abstract model. It also allows refining the outcomes of the verification procedures over the concrete model in view of the quantified and tuneable error, which depends on the concrete dynamics and on the given PCTL formula. 

\subsection{\textsf{StocHy}}
\textsf{StocHy} \cite{StocHy19} performs quantitative analysis of discrete-time stochastic hybrid systems. 
The tool allows to (i) simulate the SHS evolution over a given time horizon; and to automatically construct finite abstractions of the SHS.
Abstractions are then employed for (ii) formal verification or (iii) control synthesis satisfying safety and reachability specifications. 
The tool is implemented in C++ and employs manipulations based on vector calculus, using sparse matrices, the symbolic construction of probabilistic kernels, and
multi-threading. \textsf{StocHy} allows for modular modelling, and has separate simulation, verification and synthesis engines which are implemented as independent libraries. This allows for libraries to be readily used and for extensions to be easily built. 

\subsection{\textsf{AMYTISS}}
\textsf{AMYTISS}~\cite{lavaei2020AMYTISS} is developed in C++/OpenCL for designing correct-by-construction controllers of large-scale discrete-time stochastic control systems. 
\textsf{AMYTISS} natively supports both additive and multiplicative noises with different distributions including normal, uniform, exponential, and beta.
This software tool provides scalable parallel algorithms that allow to (i) construct finite MDPs from discrete-time stochastic control systems, and (ii) synthesize controllers satisfying complex logic properties including safety, reachability, and reach-avoid specifications. \textsf{AMYTISS} employs high-performance computing platforms and cloud-computing services to alleviate the effects of the state-explosion problem.
This tool improves performances over computation time and memory usage by parallel execution over different heterogeneous computing platforms including CPUs, GPUs and hardware accelerators (\emph{e.g.,} FPGAs).
\textsf{AMYTISS} significantly reduces the memory usage by setting a  probability threshold $\gamma \in [0,1]$ to control how many partition elements around the mean of the system should be stored. 
Such an approximation allows controlling the sparsity of the columns of $\hat T_{\mathsf x}$ (transition probability matrix of constructed finite MDP).
\textsf{AMYTISS} also proposes another technique that further reduces the required memory for computing $\hat T_{\mathsf x}$, named  \emph{on-the-fly abstraction} (OFA). 
In OFA, computing and storing the probability transition matrix $\hat T_{\mathsf x}$ are skipped. Instead the required entries of $\hat T_{\mathsf x}$ on-the-fly are computed as they are needed for the synthesis part via the standard dynamic programming. This reduces the required memory for $\hat T_{\mathsf x}$ but at the cost of repeated computation of their entries in each time step from $1$ to a finite-time horizon $T_d$. \textsf{AMYTISS} has been successfully applied to some large-scale applications including autonomous vehicles.

\subsection{The ARCH Initiative} 
The ARCH competition aims at providing an updated point of reference on the current state of the art in the area of models for hybrid systems, together with the currently available tools and frameworks for performing formal verification and optimal policy synthesis. 
The initiative further provides a set of benchmarks aiming to push forward the development of current and future tools. 
To provide a fair and comprehensive comparison of results, which also allows tools designed for multi-core architectures to highlight their capabilities, the competition is performed via a centralized execution of the benchmarks. To establish further trustworthiness of the results, and to bolster related \emph{open science} initiatives, the code describing the benchmarks together with the code used to compute the results are also published in a public server. The tools compete based on different aspects including implementation languages, class of models, platforms, algorithms, specifications, type of stochasticity, type of distributions, type of disturbances, etc.
Presentation and discussion of outcomes of yearly benchmarking competitions on tools for formal verification and policy synthesis of stochastic models (and in particular SHS) are provided in~\cite{abate2018arch,abate2019arch,abate2020arch,abate2021arch}.
We refer the interested readers to~\cite[Table 2]{abate2020arch} for more details on recent results of competitions.

\section{Directions for Open Research}\label{CORD}

In this subsection, we present and discuss a few open topics that can be taken up as future research initiatives.

\subsection{Formal Analysis of SHS via Learning and Data-Driven Approaches}\label{Data}

We discuss a few results on formal synthesis of SHS via learning and data-driven approaches, which is still considered as an open direction. 
A deterministic policy gradient algorithm for reinforcement learning
with continuous actions is presented in~\cite{silver14}. The framework introduces an off-policy actor-critic algorithm that learns a deterministic target policy from an
exploratory behaviour policy. It shows that the deterministic
policy gradient can be estimated much more efficiently than the usual stochastic policy gradient especially in high-dimensional action spaces. However, the results do not provide any quantitative guarantee on the optimality of synthesized policies for original MDPs.

A model-free reinforcement learning framework of $\omega$-regular objectives for finite Markov decision processes is proposed in~\cite{hahn2019omega,hasanbeig2019certified}. The $\omega$-regular properties are compiled into limit-deterministic  B\"{u}chi automata (LDBA) instead of the traditional Rabin automata; this choice sidesteps difficulties that have marred previous proposals.
\cite{hahn2019omega} presents a constructive reduction from the almost-sure satisfaction of $\omega$-regular objectives to an almost-sure reachability problem, and learns how to control an unknown model so that the chance of satisfying the objective is maximized.
\cite{hasanbeig2019certified} exploits the structure of the LDBA and shapes a synchronous reward function on-the-fly, so that an RL algorithm can synthesize a policy resulting in traces that maximize the probability of satisfying the linear temporal property.  
The approach in \cite{Bozkurt2020ControlSF} proposes a reward scheme associated to the given specification that requires two discounting factors in the reinforcement learning algorithm and provides a condition on these discounting to guarantee convergence of the learned policy to the optimal policy. 
These three approaches can be applied with off-the-shelf reinforcement learning algorithms to compute optimal strategies from the sample paths of the finite MDP. 
Extensions to continuous-space (and -actions) models are investigated in~\cite{lavaei2020ICCPS,kazemi2020formal} and in \cite{lcnfq,hasanbeig2020drl,CHXAK21}, as surveyed in Section~\ref{TLVS}. The contribution in \cite{cHAGW21} extends this setup to multi-agent cooperative games. The results in \cite{lcnfq,hasanbeig2020drl,CHXAK21,cHAGW21} are only empirically illustrated for continuous-state MDPs and do not provide any theoretical guarantees.

A data-driven verification approach under signal temporal logic constraints is proposed in~\cite{salamati2020data,salamati2020data_J}. As the dynamics are parameterized and partially unknown, the framework collects data from the system and employs Bayesian inference techniques to associate a confidence value to the satisfaction of the property. The results combine both data-driven and model-based techniques in order to have a two-layer probabilistic reasoning over the behavior of the system: one layer is related to the stochastic noise inside the system and the next layer is related to the noisy data collected from the system. Approximate algorithms are also provided for computing the confidence for linear dynamical systems.

A data-driven technique for satisfying temporal properties on unknown stochastic processes with continuous spaces is recently presented in~\cite{kazemi2020formal}. The proposed framework is based on reinforcement learning that is used to compute sub-optimal policies that are finite-memory and deterministic. The work addresses properties expressed by LTL and uses their automaton representation to give a path-dependent reward function maximized via the RL algorithm. It also develops theoretical foundations characterizing the convergence of the learned policy to the optimal one in the continuous space. To improve the performance of the learning on the constructed sparse reward function, the paper proposes a learning procedure based on a sequence of labelling functions obtained from the positive normal form of the LTL specification. This procedure is utilized to guide the RL algorithm towards the optimal policy. It is shown that the proposed approach can provide guaranteed lower bounds for the optimal satisfaction probability.

\subsection{Formal Analysis of Partially-Observed SHS}

With a few mentioned exceptions, most of the surveyed work on automated verification and synthesis of SHS assumes complete state information. However, in many real applications we do not have access to full information. There have been a limited work on formal synthesis of partially-observed SHS. 
An early formulation is put forward in \cite{cDAT13}, which characterizes the safety problem measure-theoretically and develops an application in air traffic management. 
Reachability analysis of partially observable discrete-time SHS is proposed in~\cite{lesser2014reachability}.
A dynamic programming recursion is also developed for the solution of the equivalent perfect information problem, proving that the recursion is valid, an optimal solution exists, and results in the same solution as to the original problem.

A finite-state approximation for safety verification and control of partially observable SHS is presented in~\cite{lesser2015finite,lesser2016approximate}. The papers solve a dynamic program over the finite state approximation to generate a lower bound to the viability probability, using a point-based method that generates samples of the information state. The proposed approach produces approximate probabilistic viable sets and synthesizes a controller to satisfy safety specifications. It also provides error bounds and convergence results, assuming additive Gaussian noise in the continuous-state dynamics and observations. Computing probabilistic viable sets for partially observable systems using truncated Gaussians and adaptive gridding is presented in~\cite{lesser2015computing}.

Verification of uncertain POMDPs using barrier certificates is discussed in~\cite{ahmadi2018verification}. A class of POMDPs is considered with uncertain transition and/or observation probabilities in which the uncertainty takes the form of probability intervals.
Given an uncertain POMDP representation of the system, the main goal is to propose a method for checking whether the system will satisfy an optimal performance, while not violating a safety requirement. A policy synthesis in multi-agent POMDPs via discrete-time barrier functions to enforce safety is proposed in~\cite{ahmadi2019safe}. The method is implemented online by a sequence of one-step greedy algorithms as a standalone safe controller or as a safety-filter given a nominal planning policy. Verification of partial-information probabilistic systems using counterexample-guided refinements is studied in~\cite{giro2012verification}.

A perception-aware point-based value iteration for POMDPs is presented in~\cite{ghasemi2018perception}. The approach avoids combinatorial expansion over the action space from the integration of planning and perception decisions, through a greedy strategy for observation selection that minimizes an information-theoretic measure of the state uncertainty. The article develops a point-based value iteration algorithm that incorporates this greedy strategy to pick perception actions for each sampled belief point in each iteration. A sequential decision making process using POMDPs is studied in~\cite{wu2019cost}. The work aims to find strategies that actively interact with the system, and observe its reactions so that the true model is determined efficiently and with high confidence.

Synthesis of stochastic systems with partial state information via control barrier functions is proposed in~\cite{NiloofarIFAC2020,Niloofar_LCSS20,Niloofar_TCNS21}, as surveyed in Section \ref{CTSS}.

\begin{resp}
	\begin{open}
		Formal synthesis of POMDPs (even with finite set of states) is a hard problem and the available methods are not scalable. Developing scalable algorithmic techniques for POMDPs to make the synthesis problem more tractable is a potential future research direction.
	\end{open}
\end{resp}\medskip

\subsection{Secure-by-Construction Controller Synthesis} 
Security-related attacks are increasingly becoming pervasive in safety-critical
applications, such as autonomous vehicles, implantable and wearable medical devices, smart systems and infrastructures. 
While most of the well-known attacks---such as
vehicle hacking, 
pacemaker and Implantable Cardioverter Defibrillator (ICD)
attacks~\cite{halperin2008pacemakers,HijackInsulin11}---exploit unencrypted wireless
communication, such 
attacks can be readily guarded against by following well-established
cryptographic protocols.
On the other hand, security
vulnerabilities related to information leaks via side-channels may be impossible to mitigate without requiring
a non-trivial modification to control software, as the side-channels are products of the interaction of the
embedded control software with its physical environment (which clearly represents a ``hybrid'' feature). Furthermore, the presence of wide variety of physical variables (such as temperature, electro-magnetic emissions, velocity and so on) in these control systems expose corresponding attack surfaces to the intruder and render those systems even more vulnerable than traditional digital software/hardware systems. The source of stochasticity in those systems is either due to the noisy environment or the measurement noise of intruders' observations. Hence, SHS are a good modeling framework for studying those security vulnerabilities. We refer the interested readers to the recent vision paper in \cite{liu2022securebyconstruction} explaining in detail  different security notions for hybrid systems, some interesting examples, and initial  verification and controller synthesis approaches suitable for them.

While the controller synthesis approach for SHS has been heavily investigated for safety
requirements as discussed in details in the previous sections, the secrecy requirements in SHS are often verified in a post facto manner after the design of 
controllers. Hence, if the system leaks information beyond an acceptable range,
the controller needs to be redesigned incurring very high verification and validation costs.
The secure-by-construction synthesis approach advocates a paradigm shift in the development of safe and secure
SHS by proposing a controller design scheme which generalizes
existing correct-by-construction synthesis methods by considering security
properties, simultaneously to the safety ones discussed elsewhere in this survey, during the design phase.

\begin{resp}
	\begin{open}
		The {\it correct-by-construction} controller synthesis approaches for SHS provide embedded control software from high-level safety requirements
		in an automated and formal manner. Proposing {\it secure-by-construction} controller synthesis schemes which generalize the
		correct-by-construction ones by integrating security
		requirements with the safety ones in the controller synthesis
		phase is a potential future research direction.
	\end{open}
\end{resp}\medskip

\subsection{(Mix)-monotonicity of SHS}\label{MM-SHSs}
As discussed in Subsection~\ref{IMDPs}, the construction of IMC/IMDP can be much more complex than standard abstractions based on MC/MDP, 
since one needs to compute lower and upper bounds for the probabilities of transition between states, 
rather than computing just a single number as in standard MCs/MDPs.
Under some assumptions (\emph{e.g.,} additive stochasticity, unimodal noise distribution, independent noises affecting different states), 
contributions in~\cite{dutreix2018efficient,dutreix2019specification,dutreix2020abstraction} utilize the mix-monotonicity property of the deterministic part of the map $f$ and propose an approach to compute those lower and upper bounds in an efficient way. Further research in this direction is deemed worthy of attention.

\subsection{Compositional Construction of IMCs/IMDPs}

Since constructing IMCs/IMDPs is more complex than standard abstractions, as discussed in Subsection~\ref{IMDPs}, 
a promising approach to mitigate the related computational complexity is to develop compositional techniques: 
these might allow constructing IMCs/IMDPs of high-dimensional systems based on IMCs/IMDPs of smaller subsystems.  

\subsection{Compositional Controller Synthesis for SHS}

In this survey, we mainly discussed different compositional approaches for the construction of (in)finite abstractions for networks of stochastic systems. 
Potential future work concerns the investigation of compositional controller synthesis for stochastic hybrid systems. 
In particular, given a specification over the interconnected system, it is of interest to find a formal relation between the satisfaction probabilities provided by local controllers for individual subsystems, as well as the optimal satisfaction probability for the specification on the monolithic (overall) system. 

\subsection{Extensions and Development of Software Tools} 

Developing efficient software tools based on theoretical and algorithmic results is essential for the practical use of automated verification and synthesis of SHS. 
Most of the tools discussed in Section~\ref{RST} are developed for discrete-time models. Software tool \textsf{Level-Set Toolbox}~\cite{mitchell2007toolbox} can allow for the analysis of continuous-time models. Although software tools \cite{RZ2,KZ,KZ3} are developed for formal controller synthesis of non-stochastic control systems, they can be readily utilized for the proposed results for incrementally stable SHS in \cite{zamani2014symbolic,zamani2014bisimilar,ZAG15,Mallik2017CDC}. A future  direction is to develop more general and scalable software tools for \emph{continuous-time stochastic hybrid systems}. 
Moreover, developing software tools to handle \emph{infinite-horizon specifications} is another unmet  extension. 
Finally, there is no tool at the moment that handles the construction of finite MDPs \emph{compositionally}: 
further developing software tools for compositional purposes is therefore of interest. 

A comprehensive and up-to-date discussion about different software tools together with their potential directions of extension can be found in~\cite{abate2020arch} and available at \href{https://bit.ly/3nGechr}{https://bit.ly/3nGechr}. 

\section{Closing Discussion}

In this article, we have provided the first survey of work on automated formal verification and control synthesis of stochastic hybrid systems (SHS). 
We have focused on most recent and sharpest results, and for the sake of a clear and streamlined presentation we have presented selected analysis methods, applications, and results in detail, and instead briefly overviewed alternative approaches. 
We have distinguished approaches as discretization-based and -free, and have investigated four different closeness guarantees between a concrete SHS model and its abstractions.  
We have discussed different problems including stochastic similarity relations, infinite and finite abstractions, the use of control barrier certificates, temporal logic verification and synthesis, compositional techniques, continuous-time stochastic models, data-drive approaches, and finally overviewed existing software tools that implement the discussed approaches.  
Throughout this survey, we have also added the discussion of a few open problems. 

We hope that this survey article provides an introduction to the foundations of SHS, towards an easier understanding of many challenges and existing solutions related to formal verification and control synthesis of these models, together with the associated software tools.

\bibliographystyle{alpha}
\bibliography{biblio}

\end{document}